\documentclass[12pt]{iopart}
\usepackage{tabularx}
\usepackage{graphicx}
\usepackage{cancel}
\usepackage{amssymbols}
\newcommand{\Lied}{\mathcal{L}}
\DeclareGraphicsExtensions{.jpeg,.pdf}
\begin{document}
\title{Temperature and Polarization Patterns in Anisotropic Cosmologies}
\author{Rockhee Sung$^{1,2}$ and Peter Coles$^{1}$}
\address{$^{1}$ School of Physics \& Astronomy, Cardiff University, Queen's
Buildings, 5 The Parade, Cardiff CF24 3AA, United Kingdom\\
$^2$ Cosmology and Gravity Group, University of Cape Town,
Rondebosch 7701, Cape Town, Republic of South Africa}

\ead{\mailto{Rockhee.Sung@astro.cf.ac.uk},
\mailto{Peter.Coles@astro.cf.ac.uk}}

\begin{abstract}
We study the coherent temperature and polarization patterns produced
in homogeneous but anisotropic cosmological models. We show results
for all Bianchi types with a Friedman-Robertson-Walker limit (i.e.
Types I, V, VII$_{0}$, VII$_{h}$ and IX) to illustrate the range of
possible behaviour. We discuss the role of spatial curvature, shear
and rotation in the geodesic equations for each model and establish
some basic results concerning the symmetries of the patterns
produced. We also give examples of the time-evolution of these
patterns in terms of the Stokes parameters $I$, $Q$ and $U$.
\end{abstract}

 \submitto{JCAP}

\section{Introduction}

Precise measurements of the temperature anisotropies of the Cosmic
Microwave Background (CMB), particularly those from the Wilkinson
Microwave Anisotropy Probe (WMAP) \cite{WMAP1,WMAP2}, form the
sturdiest foundations of the current (``concordance'') cosmological
model describing a universe dominated by cold dark matter (CDM) and
a cosmological constant \cite{Coles}, and therefore known as
$\Lambda$CDM for short. It is an essential component of this model
that the primordial metric perturbations that gave rise to the
galaxies and large-scale structure we observe around us today should
be Gaussian and statistically homogeneous (i.e. stationary)
\cite{Guth1982,Starobinskij1982,Bardeen1983} and this means that the
temperature fluctuations observed in the CMB should be Gaussian and
statistically isotropic.

Detailed analysis of the WMAP data has shown that any departures
from the standard framework are small and of uncertain statistical
significance. Although some anomalous behaviour has been reported
{\cite{Yadav2008}, there remains no clear evidence of primordial
non-Gaussianity, but there are several indications of departures
from statistical anistropy across the CMB sky. Among the interesting
phenomena revealed by detailed analyses of the pattern of CMB
temperature fluctuations are an extremely cold spot
\cite{Vielva,Cruz1,Cruz2,Cruz3,Cruz4,Cayon,Cspot}, unusual
alignments between large-scale harmonic modes of the temperature
pattern \cite{Schwarz2004, Copi2004,Katz2004,Land2005a,Land2005b,
Land2005c, Land2005d, Land2005e,Land2007, Copi2006, Copi2007}
(sometimes dubbed ``The Axis of Evil'') and a global hemispherical
power asymmetry
\cite{Eriksen2004a,Park2004,Eriksen2007,Hoftuft2009,Hansen2009}.

One must not get carried away with the these features because they
are - almost without exception - based on {\em a posteriori}
evaluations of statistical significance. If the occurrence of an
anomalous feature $A$ in a given model $M$ is  $P(A|M)\simeq 0.001$
(say) then that does not necessarily mean that the probability of
the model given that the anomaly is observed $P(M|A)$ is similarly
low. One has to take into account all the other data that are {\em
not} anomalous before deciding on a true measure of the significance
of a departure. Attempts to do this rigorously have generally
confirmed that current observations are not sufficiently compelling
to suggest that the standard model needs to be abandoned
\cite{HL09,GAWE,HL10,zb10}. Moreover, other analyses suggest the
further possibility that the WMAP temperature fluctuations may be
affected by foregrounds or other systematic problems
\cite{ccno07,cnc07,FP09,sc10}. Even slight effects of this type
could seriously hamper our attempts to uncover evidence of physics
beyond the standard model.

But, although the evidence for the examples of global asymmetry
discussed above is by no means overwhelming, taken together they do
at least suggest the possibility that we may live in universe which
is described by a background cosmology that is globally anisotropic,
i.e. one not described by a Friedmann-Robertson-Walker (FRW) model
and it is therefore incumbent upon us to consider alternatives to
the standard model in order to learn best how to use the data to
confirm or rule out variations on the standard cosmology.

The approach we follow here is to study the Bianchi models, i.e.
cosmological models based on exact solutions to Einstein's General
Theory of Relativity involving homogeneous but not necessarily
isotropic spatial sections. The Bianchi classification groups  all
possible spatially homogeneous but anisotropic relativistic
cosmological models into types depending on the symmetry properties
of their spatial hypersurfaces \cite{Gris,Ellis1}; we discuss these
models in more detail later in the paper.

The Bianchi models are not particularly strongly motivated from the
point of view of fundamental physics, but do nevertheless represent
a promising and potentially profitable first step away from the
standard cosmological framework. For example, it has been known for
some time that interesting localized features in the CMB temperature
pattern can occur in Bianchi models with negative spatial curvature
\cite{Collins1,Dautcourt1,Tolman1,Tolman2,Tolman3,Barrow1}.  The
physical origin of such features lies in the focussing effect of
space on the geodesics that squeezes the pattern of the anisotropic
radiation field into a small region of the sky. The observed lack of
large-scale asymmetries in the CMB temperature has in the past been
used to place constraints on the global rotation and shear allowed
in Bianchi models \cite{Barrow1,BFS96,khb97}. More recently,
however, attention has shifted to the possibility of using the
additional parameters available in such models to reproduce a cold
spot such as that claimed to exist in the WMAP data. Since we now
know that our Universe is {\em close} to isotropic, attention has
focussed on the subset of the  Bianchi types that contain the FRW
model as a limiting case.The model which appears to best able to
reproduce the anomalous cold spot is the Bianchi VII$_h$ case
\cite{Jaffe1,Jaffe2,McEwan1,McEwan2,Bridge}, although Bianch V also
has negatively curved spatial sections and can therefore, in
principle, also produce localised features \cite{sung1}.

However, as well as forming distinctive features in the temperature
pattern, anisotropic cosmological models also generate
characteristic signatures in the polarized component of the
background radiation. Thomson scattering generates polarization as
long as there is a quadrupole anisotropy in the temperature field of
the radiation incident upon the scattering particle. In the
concordance cosmology the temperature and polarization patterns are
(correlated) stochastic fields arising from their common source in
scalar and tensor perturbations arising from inflation. In a Bianchi
cosmology, however, the patterns are coherent and have a
deterministic relationship to one another owing to their common
geometric origin. It has recently been shown
\cite{sung1,Pontzen1,pontzen} that only in special cases are the
properties of the polarization field produced in Bianchi VII$_h$
consistent with the latest available WMAP polarization data
\cite{WMAPPol} because such models generally involve a large
odd-parity (B-mode) \cite{Kam,Hu1} contribution that exceeds the
experimental upper limit. For a discussion of CMB polarization in
the context of the so-called Axis of Evil, see ref. \cite{FE10}.

But their ability to produce localized features is not the only
reason to be interested in the temperature and polarization patterns
produced in Bianchi models. For example, history provides a host of
connections between these models and the understanding of the
interaction between electromagnetic radiation and  gravitational
waves \cite{rees,nanos,NS80,BP80,Pol85,fpc94}.

Our aim in this paper is to give a relatively gentle introduction to
the Bianchi models and show how to calculate the temperature and
polarization properties of the radiation field expected to arise
within them, using methods outlined in a previous short paper
\cite{sung1}. Our aim in doing this is neither to provide an
exhaustive set of alternatives to the standard cosmology nor to
perform a detailed statistical analysis of the patterns we
calculate. Instead we plan to elucidate some of the general
properties of the radiation field in anisotropic cosmological models
belonging to the different Bianchi types.

The outline is as follows. In the next section we describe the
Bianchi classification in general terms for the benefit of the
non-expert. In Section 3 we introduce the specific formalism of
Bianchi cosmological models, i.e. exact solutions of the Einstein
equations involving space-times with symmetries described by the
various Bianchi types. In Section 4 we explain how we compute the
radiation patterns in these models and then, in Section 5,  show
examples of the results obtained. We summarize and present our
conclusions in Section 6.

\section{Bianchi Spaces}

In the Friedman models on which the standard Big Bang cosmology is
based, hypersurfaces of constant time are defined to be those on
which the matter density is the same throughout space. We can
construct a more general definition of homogeneity by requiring that
all comoving observers see essentially the same version of cosmic
history. In mathematical terms this means that there must be some
symmetry that relates what the Universe looks like as seen by
observer A to what is seen in a coordinate system centred on any
other observer B. The possibly space-times consistent with this
requirement possess symmetries can be classified into the various
Bianchi types, which we now discuss. For more details, see
\cite{bgen,cl2}.

The Bianchi classification is based on the construction of
space-like hypersurfaces upon which it is possible to define at
least three independent vector fields, $\xi_\alpha$, that satisfy
the Killing's equation
\begin{equation} \xi_{i; j} + \xi_{j; i}=0. \end{equation}
The vectors that satisfy this are called Killing vectors; the
semicolons denote covariant derivatives. Suppose the Killing vectors
are denoted $\xi_{\alpha}$, where Greek indices can run from $0$ to
$3$. The commutators of the ${\bf \xi}_{\alpha}$ are defined by
\begin{equation} [\xi_{\alpha}, \xi_{\beta}] \equiv
\xi_{\alpha}\xi_{\beta}-\xi_{\beta}\xi_{\alpha}=C_{\alpha\beta}^{\delta}
\xi_{\delta}, \end{equation} where the $C_{\alpha\beta}^{\delta}$
are called structure constants. These are antisymmetric, in the
sense that, \begin{equation}
C_{\alpha\beta}^{\delta}=-C_{\beta\alpha}^{\delta}.
\end{equation}

One can understand how the structure constants arise by considering
symmetry transformations. In three dimensions, spatial homogeneity
is relates to the existence of three independent sets of curves with
tangent vectors $\xi_\alpha$. An infinitesimal symmetry
transformation takes an arbitrary point P (with coordinates $x_i$)
to the point P$^{\prime}$ with coordinates $x_i+\delta x_i$, where
\begin{equation}
\delta {\bf x} = {\bf \xi}_{\rm P} \delta t
\end{equation}
for $\xi_{\rm P}$ some linear combination of the $\xi_{\alpha}$
defined at P. The same symmetry transformation takes the tip of an
arbitrary infinitesimal vector  ${\bf \zeta}_{\rm P} \delta s$ at P
to a new position ${\bf \xi}_{{\rm P} + {\bf \zeta}_{\rm P} \delta
s}\delta t$ at $P^{\prime}$. This means that \begin{equation} {\bf
\zeta}_{\rm {P}} \delta s \rightarrow {\bf
\zeta}^{\prime}_{P^{\prime}} \delta s = {\bf \zeta}_{\rm P} \delta s
+ \xi_{{\rm P} + {\bf \zeta}_{\rm P} \delta s} \delta t -{\bf
\xi}_{\rm P} \delta t.
\end{equation} It is now possible to compare the transformed vector
${\bf \zeta}^{\prime}_{{\rm P}^{\prime}}$ with ${\bf \zeta}_{\rm
P^{\prime}}$, the `actual' vector field defined at P$^{\prime}$. The
difference between the two vectors is the Lie derivative of ${\bf
\zeta}$ taken along ${\bf \xi}$:
\begin{equation}
\Lied_{\bf \xi} {\bf \zeta} \equiv \lim_{\delta t \rightarrow 0}
\frac{{\bf \zeta}^{\prime}_{{\rm P}^{\prime}}-{\bf \zeta}_{\rm
P}}{\delta t} = {\bf \xi}\cdot {\bf \nabla} {\bf \zeta} - {\bf \zeta
}\cdot {\bf \nabla} {\bf \xi}.
\end{equation}
This tells us, roughly speaking, by how much we must turn a vector
after it is carried by a symmetry transformation from P to
P$^{\prime}$ in order for it to point in the same direction as it
did before the transformation.

Instead of choosing an arbitrary vector we can now take one of the
$\xi_{\alpha}$, and instead of an arbitrary direction we transform
it along another of the $\xi_{\alpha}$. The type of space is
specified by the Lie derivatives obtained for this type of
operation:
\begin{equation}
\Lied_{{\bf \xi}_\beta} {\bf \xi}_{\alpha} =
C_{\alpha\beta}^{\delta} {\bf \xi}_{\delta};
\end{equation}
since it is defined as the difference between two vectors the Lie
derivative must itself be a vector and so it can be expressed as a
linear combination of the three independent basis vectors. This
gives us the structure constants we described above.

The set of $n$ Killing vectors will have some $n$-dimensional group
structure, say $G_n$, that depends on the equivalence classes of the
structure constants $C_{\alpha\beta}^{\delta}$. This is used to
classify all spatially homogeneous cosmological models. The most
useful form of this classification proceeds as follows. On any
particular spacelike hypersurface, the Killing vector basis can be
chosen so that the structure constants can be decomposed as
\begin{equation}
C_{\alpha\beta}^{\eta}=\epsilon_{\alpha\beta\gamma}n^{\gamma\eta}+\delta_{\beta}^{\eta}
a_{\alpha} - \delta_{\alpha}^{\eta} a_{\beta}, \end{equation} where
$\epsilon_{\alpha\beta\gamma}$ is the total antisymmetric tensor and
$\delta_{\alpha}^{\beta}$ is the Kronecker delta. The tensor
$n^{\alpha\beta}$ is diagonal with entries, say, $n_1$, $n_2$, and
$n_3$. The vector $a_{\alpha}=(a,0,0)$ for some constant $a$. All
the parameters $a$ and $n_{\alpha}$ can be normalized to be $\pm 1$
or zero. If $an_2n_3=0$ then $n_2$ and $n_3$ can be set to $\pm 1$
and $a$ is then conventionally taken to be $\sqrt{|h|}$ where $h$ is
a parameter used in the classification. The possible combinations of
$n_1$ and $a$ then fix the different Bianchi types according to the
table.
\begin{table}
\begin{center}
\begin{tabular}{|l|c|c|c|c|c|}
\hline
Bianchi Type & Class & $a$ & $n_1$ & $n_2$ & $n_3$\\
\hline
 I & A & 0 & 0 & 0 & 0\\
 II & A & 0 & + & 0 & 0\\
 VI$_0$ & A & 0 & 0& + & -\\
VII$_0$ & A & 0 & 0 & + & + \\
VIII & A & 0 & - & + & + \\
IX &  A& 0 & + & + & + \\
IV & B & +  & 0 & 0 & +\\
V & B& +  & 0 & 0 & 0\\
VI$_{h}$ & B & + & 0 & + & -\\
VII$_h$ &B & + & 0 & + &+ \\
\hline
\end{tabular}
\end{center}
\caption{The Bianchi types shown in terms of whether the various
parameters used to construct the classification are zero, positive
or negative; the designation of Class A or Class B depends on
whether $a=0$, or not, respectively. The parameter $h$ is defined by
$h=a^2/n_1n_2$. The spaces of particular interest in this paper are
I, V, VII$_0$, VII$_h$ and IX because these contain the isotropic
FRW spaces as limiting cases; the others are shown just for
completeness.}
\end{table}

It is interesting also to think about the {\em generality} of the
different types. This can be expressed in terms of the dimension $p$
of the Bianchi group, which gives the dimension of the orbit of
$C_{\alpha\beta}^{\delta}$ as a subset of all 9 of the distinct
components. The Killing vectors must satisfy the  Jacobi identities,
so
\begin{equation}
\epsilon^{\alpha\beta\gamma} \left[ [\xi_{\alpha},\xi_{\beta}],
\xi_{\gamma}\right]=0.
\end{equation}
This implies that
\begin{equation}
n^{\alpha\beta}a_{\beta}=0,
\end{equation}
so that the orbits of any particular group type are at most
six-dimensional. The (isotropic) spaces that feature in the Friedman
models have $G_6$ symmetry groups with $G_3$ subgroups, so that the
zero curvature ($K=0$) FRW model can be thought of as a special case
of Bianchi Types I or VII$_0$. Likewise the open ($K<0$) FRW model
is a special case of types V or VII$_h$. The closed FRW case ($K>0$)
is a special case of Bianchi type IX.

We are interested in cosmological models that are close to the
completely isotropic case described by the FRW metric, but not all
the Bianchi types contain this as a special limiting case
\cite{Ellis1,Gris}. Those that do are types I, V, VII$_{0}$,
VII$_{h}$ and IX; we do not discuss the other cases any further in
this paper. Bianchi I and Bianchi VII$_{0}$ are spatially flat,
Bianchi IX is positively curved and Bianchi types V and VII$_{h}$
have negative spatial curvature. The ``open'' (i.e. negatively
curved) cases are of particular interest as they permit the
focussing of anisotropic patterns into small regions of the sky. The
scalar curvature $R$ of the spatial sections which is given in terms
of the Bianchi parameters as the following convenient form
\begin{equation}\label{curvature}
R = \frac{1}{2}(2n_1n_2+2n_1n_3+2n_2n_3-n^2_1-n^2_2-n^2_3)-6a^2
\end{equation}
For Bianchi V we have $n_1=n_2=n_3=0$ so that $R=-6a^2$. In Bianchi
VII$_h$ we have $n_1=0$ but $n_2\neq 0$ and $n_3\neq 0$; the
parameter $h$ is defined by $h=a^2/n_2n_3$, it is related to the
parameter $x$ which defines the 'spiralness' of the temperature
patterns, \begin{equation} x=\sqrt{\frac{h}{1-\Omega_0}}
\end{equation}
For reference, the scalar curvatures of all the models we discuss
hereafter are shown in Table 2.
\begin{table}
\begin{center}
\begin{tabular}{|l|l|ll|}
  \hline
  Type   &   $R$                                &     K                &               \\
  \hline
 $I$     &   0                                  &     0                &      Flat    \\
 $V$     &  $-6a^2$                       &     $<0$             &      Open    \\
 $VII_0$ &  $-\frac{1}{2}(n_2-n_3)^2 $          &     $0 $             &      Flat    \\
 $VII_h$ & $-6a^2-\frac{1}{2}(n_2-n_3)^2$ & $<0 $     &      Open    \\
 $IX$    & $n_1n_2+n_1n_3+n_2n_3-\frac{1}{2}(n_1^2+n_2^2+n_3^2)$ &$>0 $&      Closed   \\
 \hline
\end{tabular}
\end{center}
\caption{Dependence of the scalar curvature $R$ on the parameters
$a$ and $n_i$ involved in the construction of the Bianchi
classification for those models with an FRW limit.}
\end{table}

\section{Bianchi Cosmologies}
\label{sec:Bianchi}

\subsection{Basics}

The models we consider are based on Einstein's general theory of
relativity and we use the field equations in the form
\begin{equation}
G_{ab}\equiv R_{ab}-\frac{1}{2} R g_{ab} = T_{ab}-\Lambda g_{ab},
\end{equation}
with $R_{ab}$ being the Ricci Tensor, $R$ the Ricci scalar, $T_{ab}$
the energy-momentum tensor and $\Lambda$ the cosmological constant.
Indices $a$ and $b$ run from $0$ to $3$ (c.f. $i$, $j$  and
$\alpha$, $\beta$ which run from $1$ to $3$). We use units where
$8\pi G=c=1$. In terms of a coordinate system $x^{a}$, the metric
$g_{ab}$ is written
\begin{equation}
ds^2=g_{ab} dx^{a}dx^{b}=(h_{ab}-u_{a}u_{b}) dx^{a}dx^{b},
\end{equation}
where $u^{a}$ is the fluid velocity; the signature of $g_{ab}$ is
$(-+++)$.  As we have already explained, the components of the
metric $g_{ab}$ describing a Bianchi space are invariant under the
isometry generated by infinitesimal translations of the Killing
vector fields. In other words, the time-dependence of the metric is
the same at all points. The Einstein equations relate the
energy-momentum tensor $T_{ab}$ to the derivatives of $g_{ab}$ so if
the metric is invariant under a given set of operations then so are
the physical properties encoded in $T_{ab}$.

Before proceeding further, let us comment further on the degree of
{\em generality} of the various Bianchi models which we touched on
in the previous section. An alternative way to quantify this, rather
than looking at the group structure, is to work out the number of
arbitrary constants needed to specify the solutions. This seems more
interesting from a physical point of view, as we are interested in
the solutions to the field equations rather than the groups
themselves. The number of arbitrary constants depends on the form of
the energy-momentum tensor. In Table 3 we give the results for
vacuum and perfect fluid equations of state. The appearance of
$h=-1/9$ as a special case in this table relates to the fact that
two of the Einstein constraint equations become null identities for
this particular choice of $h$. From the Table it emerges that the
``most general'' vacuum solutions are types VII$_h$, VI$_{h}$, VIII,
VI$_{h=-1/9}$ and IX, all of which have four arbitrary parameters.
The least general is the Bianchi Type I vacuum solution, which has
only one.

\begin{table}
\begin{center}
\begin{tabular}{|l|c|c|c|}
\hline
Bianchi Type & Group Dimension & Vacuum & Fluid\\
 & $p$ & $r$ & $s$\\
\hline
 I & 0 & 1 & 2\\
 II & 3 & 2 & 5\\
 VI$_0$ & 5 & 3 & 7\\
VII$_0$ & 6 & 4 & 8\\ VIII & 6 & 4 & 8\\ IX & 6 & 4 & 8\\ IV & 5 & 3
& 7\\ V & 3 & 1 & 5\\ VI$_{h}$ & 6 & 4 & 8\\ VII$_h$ & 6 & 4 & 8\\
VI$_{h=-1/9}$ & 6 & 4 & 7\\ \hline
\end{tabular}
\end{center}
\caption{The Bianchi types shown in terms of the group dimension $p$
and number of arbitrary constants needed to specify the model on a
given constant time surface, in vacuum ($r$) and with a perfect
fluid equation of state ($s$).}
\end{table}

For cases describing perfect fluids the situation is a little more
subtle. One would expect to have four additional parameters to
specify these compared to the vacuum solutions, but the table shows
that $s=r+4$ is not always the case. This is the case because the
Einstein equations place additional restrictions on the form of
$T_{ab}$  allowed in Types I and II. For example, if a perfect fluid
is added to the vacuum Type I solution then the form of the metric
requires all the time-space components of the Ricci tensor to be
identically zero. This means that the energy-momentum tensor
$T_{ab}$ must have
\begin{equation}
T_{0\alpha}=0
\end{equation}
for $\alpha=1$, $2$ or $3$. This in turn means that the matter must
be comoving, i.e. its velocity is $u_{a} = \delta_{a}^{0}$. Only one
free parameter is therefore needed to specify the solution, the
energy density $\mu$. The perfect fluid case of Bianchi
VI$_{h=-1/9}$ is also peculiar, in that it is not as general as
Bianchi VI$_{h}$, VII$_{h}$, VIII or IX because the degeneracy
described above only appears in vacuum.

\subsection{Example: The Kasner Solution}

General solutions in closed form of the Einstein equations are only
known for some special cases of the Bianchi types, which
demonstrates the difficulty of finding meaningful exact solutions in
situations of restricted symmetry. There is, however, one very well
known example - the Kasner solution \cite{Kasner}- which is a useful
illustration of the sort of behaviour one can obtain and which
therefore provides a useful pedagogical route into a more general
treatment of Bianchi cosmologies.

The Kasner  metric, which describes a space belonging to Bianchi
Type I, has the form
\begin{equation} ds^2= dt^2 - X_1^2(t)dx_1^2 - X_2^2 dx_2^2 - X_3^2
dx_3^2. \end{equation} Substituting this metric into the Einstein
equations (with $\Lambda=0$ and a perfect fluid with pressure $p$
and density $\mu$) yields \begin{equation}\frac{\ddot{X_i}}{X_i} -
\left(\frac{\dot{X_i}}{X_i}\right)^2 +3
\left(\frac{\dot{X_i}}{X_i}\right) \left(\frac{\dot{a}}{a}\right) =
\frac{1}{2}\left(\mu-p\right),\end{equation}  in which
$a^3=X_1X_2X_3$. Note that this  emerges from the diagonal part of
the Einstein equations so the summation convention does not apply.
One also obtains \begin{equation} \frac{\dot{X_1}\dot{X_2}}{X_1X_2}
+ \frac{\dot{X_2}\dot{X_3}}{X_2X_3} +
\frac{\dot{X_3}\dot{X_1}}{X_3X_1} = \mu.
\end{equation} This is easy to interpret: the spatial sections expand at a rate
$\dot{X_i}/X_i$ in each direction. The mean rate of expansion is
just\begin{equation} \frac{\dot{a}}{a}=\frac{1}{3}\left(
\frac{\dot{X_1}}{X_1}+\frac{\dot{X_2}}{X_2}+\frac{\dot{X_3}}{X_3}\right).
\end{equation}
In the neighbourhood of an observer at the centre of a coordinate
system $x_i$, fluid particles will move with some velocity $u_i$. In
general, \begin{equation} \frac{\partial u_i}{\partial
x_j}=\frac{1}{2}\left(\frac{\partial u_i}{\partial
x_j}-\frac{\partial u_j}{\partial x_i}\right)+
\frac{1}{2}\left(\frac{\partial u_i}{\partial x_j}+\frac{\partial
u_j}{\partial x_i}\right)= \omega_{ij}+\theta_{ij},\end{equation}
where $\omega_{ij}$ is the rate of rotation: in more familiar
language, the vorticity vector $\omega_i= \epsilon_{ijk}\omega_{jk}$
which is just the curl of $u_i$. The tensor $\theta_{ij}$ can be
decomposed into a diagonal part and a trace-free part according to
\begin{equation} \theta_{ij}=\frac{1}{3}\delta_{ij} \theta+\sigma_{ij},
\end{equation} where $\sigma_{ii}=0$. In this description $\theta$,
$\sigma_{ij}$ and $\omega_{ij}$ respectively represent the
expansion, shear and rotation of a fluid element. In the Kasner
model we have
\begin{equation} \theta=3(\dot{a}/a) \end{equation} and \begin{equation} \omega_{ij}=0.
\end{equation} As we shall see below, more complicated Bianchi models have non-zero rotation. We can
further write evolution equations for \begin{equation} \sigma_i =
\frac{\dot{X_i}}{X_i}-\frac{\dot{a}}{a}. \end{equation} In
particular we get \begin{equation} \dot{\sigma_i}+\theta\sigma_i=0
\end{equation} which can be immediately integrated to give
\begin{equation} \sigma_i=\frac{\Sigma_i}{a^3}, \end{equation} where the
$\Sigma_i$ are constants such that $\Sigma_1+\Sigma_2+ \Sigma_3=0$.
The Kasner solution itself is for a vacuum $p=\mu=0$, which has a
particularly simple behaviour described by $X_i=A_it^p$ where
$p_1+p_2+p_3=p_1^2+p_2^2+p_3^2=1$. Notice that in general these
models possess a shear that decreases with time. They therefore tend
to behave more like an FRW model as time goes on. Their behaviour as
$t\rightarrow 0$ is, however, quite complicated and interesting.

\subsection{Tetrad Frame}

It is convenient to follow \cite{Ellis1},  introducing a tetrad
basis constructed From a local coordinate system $x^{i}$ by
\begin{equation}  {\bf e}_a= e_{a}^{i} \frac{\partial}{\partial
x^{i}}\label{tet1}
\end{equation}
such that \begin{equation} g_{ab}=e_{a}^i e_{b}^j
g_{ij}=e_{a}^{i}e_{bi} = {\rm diag} (-1, +1, +1, +1)\end{equation}
meaning that the tetrad basis ${\bf e}_a$ is orthonormal. The Ricci
rotation coefficients, \begin{equation}
\Gamma_{abc}=e_{a}^{i}e_{ci;j} e_b^{j}, \end{equation} are the
tetrad components of the Christoffel symbols; semicolons denote
covariant derivatives. In general, the operators defined by equation
(\ref{tet1}) do not commute: they generate a set of relations of the
form
\begin{equation} [{\bf e}_a, {\bf e}_b]=\gamma_{ab}^{c} {\bf
e}_{c}\label{comm}. \end{equation} These Ricci rotation coefficients
are just \begin{equation} \Gamma_{abc}= \frac{1}{2}
(\gamma_{abc}+\gamma_{cab}-\gamma_{bca}).
\end{equation} The matter flow is described in terms of the expansion
$\vartheta_{ab}$ and shear $\sigma_{ab}$:
\begin{eqnarray}
u_{a;b}& =& \omega_{ab}+\vartheta_{ab}-\dot{u}_au_{b}\nonumber\\
\sigma_{ab}& = & \vartheta_{ab}-\frac{1}{3} \vartheta h_{ab}\equiv
\vartheta_{ab}-Hh_{ab},
\end{eqnarray}
where $\vartheta=\Tr(\vartheta_{ab})=\vartheta_{aa}$ and the
magnitude of the shear is $\sigma^2=\sigma^{ab}\sigma_{ab}/2$. We
now take the time-like vector in our basis to be the fluid flow
velocity so that $u^a=\delta_{0}^a$ and $u_a=-\delta_{a}^0$. The
remaining space-like vectors  form an orthonormal triad, with a set
of commutation relations like that shown in equation (\ref{comm})
but with an explicit time dependence in the ``structure constants''
describing the spatial sections:
\begin{equation}
[{\bf e}_i, {\bf e}_j]=\gamma_{ij}^{k}(t) {\bf e}_{k}.
\end{equation}
Without loss of generality we can write \begin{equation}
\gamma_{ij}^{k}=\epsilon_{ijl}n^{lk}+\delta_{j}^{k}a_{i}-\delta_i^{k}a_{j},
\end{equation}
for some tensor $n_{ij}$ and some vector $a_i$. The Jacobi
identities require that $n_{ij}a^{j}=0$ so we choose $a^{j}=(a,0,0)$
and $n_{ij}={\rm diag}(n_1,n_2,n_3)$. The four remaining free
parameters are used to construct the Bianchi classification
described briefly above, and more in detail elsewhere
\cite{Gris,Ellis1,Ellis2,Ellis3,tilt,Ellis4}.

\section{Radiation Transport in Anisotropic Cosmologies}

Having established some general results about Bianchi models, we now
turn to the problem of calculating the temperature and polarization
patterns they produce. Our general method is to generalize the
Liouville equation so that it can be applied to a complete
description of photons travelling through a curved space-time. This
requires that we set up radiation distribution functions that
incorporate all the Stokes parameters needed to specific polarized
radiation. We also need to include a source term that describes the
effects of Thomson scattering by free electrons for the entire
history from decoupling to the observed epoch. In doing this we
follow closely the methods of \cite{Dautcourt1} and also
\cite{Tolman1}. Note, however, that they use a different definition
of the Ricci rotation coefficients.
\subsection{Radiation Description}
Our expansion of the distribution functions into multipoles is based
on the usual Stokes parameters, and on spin-weighted spherical
harmonics. The transfer equation for polarized radiation propagating
through space-time can be described by a (complex) photon
distribution comprising components with spin-weights 0 and 2, i.e.
\begin{equation}\label{photon_distribution}
\hat{N} \equiv {N^0 \choose N^2} = \frac{1}{ch^4\nu^3} {I+iV \choose
Q-iU}
\end{equation}
Here $I$, $Q$, $U$ and $V$ are the usual Stokes parameters that
describe polarized radiation. As we shall see, however, $V$ (which
measures circular polarization) does not arise in this context. The
{\em degree of (linear) polarization} is defined by
\begin{equation}\label{Polarization_degree}
P=\frac{\sqrt{Q^2+U^2}}{I}
\end{equation}
The polarization orientation is described by an angle $\chi$, where
\begin{equation}\label{polarization_angle}
\chi = \frac{1}{2}\arctan \frac{U}{Q}.
\end{equation}

The relativistic form of the Liouville equation is
\begin{equation}
p^a\frac{\partial}{\partial
x^a}-\Gamma^a_{bc}p^bp^c\frac{\partial}{\partial p^a}=0.
\end{equation}
The unpolarized part of the radiation distribution is described by
 spin-zero quantities, $N$ (i.e. quantities invariant with
respect to rotations around the ray direction $k^i$).  In terms of
an affine parameter $\lambda$ along the photon path one calculates the
total change of $N$ as
\begin{equation}\label{Liouville1}
    \frac{dN}{d\lambda}=l^{\mu}_{(a)}p^a\frac{\partial N}{\partial
    x^{\mu}}+\frac{\partial N}{\partial p^a}\frac{dp^a}{d\lambda}
\end{equation}
and the photon path $p^a(\lambda)$ is determined by the geodesic equation
expressed in the  tetrad notation we introduced above as
\begin{equation}\label{geo0}
 \frac{dp^c}{d\lambda}=-\Gamma^c_{\phantom{0}ab}p^ap^b=-\varepsilon^2\gamma^c
 \end{equation}
  so that
\begin{equation}\label{Liouville2}
    \frac{dN}{d\lambda}= p^a l^{\mu}_{(a)}\frac{\partial N}{\partial
    x^{\mu}}-\Gamma^c_{\phantom{0}ab}p^ap^b\frac{\partial N}{\partial
    p^c}
\end{equation}
in which  $p^a\equiv(\varepsilon,\varepsilon k^i)$. $l^a$ are space-time unit vectors and $\varepsilon$ is energy of photon. Where the $\gamma^a$ is defined as
\begin{equation}
\gamma^a=\Gamma^a_{\phantom{0}00}+\Gamma^a_{\phantom{0}0i}k^i
         +\Gamma^a_{\phantom{0}i0}k^i+\Gamma^a_{\phantom{0}ik}k^ik^k.
\end{equation}
If the change in
$N$ along a photon path arises from collisions only, one obtains the
following (Boltzmann) equation
\begin{equation}\label{Boltzmann1}
   \frac{1}{\varepsilon}\frac{dN}{d\lambda}=  l^{\rho}_{(0)}\frac{\partial N}{\partial
    x^{\rho}}+k^i l^{\rho}_{(i)}\frac{\partial N}{\partial
    x^{\rho}}-\varepsilon\gamma^0\frac{\partial N}{\partial\varepsilon}
    +\frac{\gamma^i}{\sqrt{2}}(m^i\bar{\eth}+\bar{m}^i\eth)N
\end{equation}
In the case of polarized radiation we need to extend the description
of the radiation field to include both spin-0 and spin-2 components.
This requires us to generalize $N$  which can be decomposed into
parts $N^0$ and $N^2$. In the transfer equation for $N^2$ these give
rise to additional terms related to the change of angles $\theta$
and $\phi$, and and extra rotation $\psi$ of polarization (see the
Appendix in \cite{sung0}).
\begin{eqnarray}
 \frac{\gamma^i}{\sqrt{2}}(m^i\bar{\eth}_2+\bar{m}\eth_2)N^2
               &=&2i\cos\theta \frac{d\phi}{d\lambda}N^2+ \frac{\gamma^i}{\sqrt{2}}(m^i\bar{\eth}+\bar{m}\eth)N^2
\end{eqnarray}
and
\begin{equation}\label{rotation_angle}
-2iN^2\frac{d\psi}{d\lambda}=-2i\cot\theta \frac{d\phi}{d\lambda}N^2-2N^2m^i\bar{m}^k\varepsilon(\Gamma^k_{\phantom{0}0i}+k^l\Gamma^k_{\phantom{0}li}
).
\end{equation}
(see  \cite{Tolman1} for details). The first two extra terms cancel
out in the Liouville equation so we obtain the following simplified
form
\begin{eqnarray}\label{Liouville5}
 \mathcal{D}_AN^A &\equiv &  l^{\rho}_{(0)}\frac{\partial N^A}{\partial
    x^{\rho}}+k^i l^{\rho}_{(i)}\frac{\partial N^A}{\partial
    x^{\rho}}-\varepsilon\gamma^0\frac{\partial N^A}{\partial\varepsilon}
    +\vartheta N^A \nonumber\\
    & & + \delta^2_AiN^2(\Gamma^k_{\phantom{0}0i}\epsilon^{ikl}k^l
    +\Gamma^k_{\phantom{0}li}k^lk^m\epsilon^{ikm} ).
\end{eqnarray}
We used the antisymmetry of $a$ and $c$ in $\Gamma^a_{bc}$ and the
relation,
$\frac{1}{2}(m^l\bar{m}^j-\bar{m}^lm^j)=-\frac{i}{2}\epsilon^{ljk}k^k$
to obtain the imaginary terms. The operator $\mathcal{D}_A$ on the
left hand side preserves spin weight ($A=0,2$); the corresponding
angular operator is
$\vartheta=\gamma^i/\sqrt{2}(m^i\bar{\eth}+\bar{m}^i\eth)$.

\subsection{Scattering}

Photon scattering can be described by the addition of a source term
to the right hand side of the Liouville equation so that it can be
written
\begin{equation}\label{scattering1}
    \mathcal{D}_AN^A=\tau(-N^A+J_A),
\end{equation}
in which $J_A$ takes the form
\begin{equation}\label{scatteringEmission1}
J_A=\int [p_{AB}(\theta,\phi,\theta',\phi' )N^B(\theta',\phi')+
 \hat{p}_{AB}(\theta,\phi,\theta',\phi' )\bar{N}^B(\theta',\phi')]\frac{d\Omega'}{4\pi}
\end{equation}
in terms of the scattering matrices $p_{AB}$ and $\hat{p}_{AB}$:
\begin{eqnarray}\label{scatteringMatrics}
p_{00}& = &
\frac{1}{2}+\frac{3}{8}k^{ik}k^{ik'}+\frac{1}{2}k^ik^{i'}\nonumber\\
p_{02} & = &
\bar{\hat{p}}_{02}=-\frac{3}{4}k^{ik}\bar{m}^{ik'}\nonumber\\
p_{22} & =& \frac{3}{2}m^{ik}\bar{m}^{ik'}\nonumber\\
\hat{p}_{00}& = &
\frac{1}{2}+\frac{3}{8}k^{ik}k^{ik'}-\frac{1}{2}k^ik^{i'}\nonumber\\
p_{20} & = & \hat{p}_{20}=-\frac{3}{4}m^{ik}k^{ik'}\nonumber\\
\hat{p}_{22}& = & \frac{3}{2}m^{ik}m^{ik'}.
\end{eqnarray}
The emission term $J_A$ contains only harmonics up to $l=2$, since
all other terms vanish in virtue of the orthogonality relations for
spherical harmonics
\begin{eqnarray}\label{scatteringEmission2}
J_0 &=&\texttt{Re}N^0_0+\frac{i}{3}k^{i}\texttt{Im}N^0_i
  +\frac{1}{10}k^{ik}\texttt{Re}N^0_{ik}-\frac{3}{10}k^{ik}\texttt{Re}N^2_{ik}\\
J_2&=&-\frac{1}{5}m^{ik}\texttt{Re}N^0_{ik}+\frac{3}{5}m^{ik}\texttt{Re}N^2_{ik}
\end{eqnarray}
The radiation modes with $l\leq2$ are damped as well as re-radiated
by Thomson scattering, while higher-order modes $l>2$ are only
damped.
\subsection{Transfer Equation}
We now expand the distribution function in our Boltzmann equation in
terms of multipole components:
\begin{eqnarray}\label{transfer_eq}
  N^0 &=& N^0_0+N^0_ik^i+N^0_{ij}k^{ij}+ \dots ,\nonumber\\
  N^2 &=& N^2_{ij}m^{ij} + \dots
\end{eqnarray}
where  $k^i$ expresses the three-dimensional ray direction, i.e.
\begin{eqnarray}\label{direction}
  k^i &=& (\cos\theta,\sin\theta \cos\phi, \sin\theta \sin\phi ) \nonumber \\
  k^{ik} &=& k^ik^k-\frac{1}{3}\delta^{ik} \nonumber \\
  m^i &=& \frac{1}{\sqrt{2}}(\frac{\partial k^i}{\partial \theta}
        +\frac{i}{\sin \theta}\cdot\frac{\partial k^i}{\partial
             \phi}) \nonumber \\
 m^{ij} &=& m^im^k
\end{eqnarray}
in which $m^ik_i=0$, $m^i\bar{m}_i=k^ik_i=1$ and the bar indicates
complex conjugate. The number of indices of the $k$ and $m$
polynomials characterizes the multipole order of the corresponding
contributions to anisotropy and polarization. This leads to the
following equations for the moments of these quantities
\begin{eqnarray}
\int k^ik^k\frac{d\Omega}{4\pi} &=& \frac{1}{3}\delta^{ik}\nonumber\\
\int k^{ij}k^{kl}\frac{d\Omega}{4\pi} &=& \frac{1}{15}\Big( \delta^{ik}\delta^{jm}+\delta^{im}\delta^{jk} -\frac{2}{3}\delta^{ij}\delta^{kl} \Big)\nonumber\\
\int m^{ij}\bar{m}^{km}\frac{d\Omega}{4\pi} &=&\frac{1}{10}(\delta^{ik}\delta^{jm}+\delta^{im}\delta^{jk})-\frac{1}{15}\delta^{ij}\delta^{km}\nonumber\\
\int m^{ij}\bar{m}^{km}k^r\frac{d\Omega}{4\pi}
&=&\frac{i}{30}(\delta^{ik}\varepsilon^{mjr}+\delta^{jk}\varepsilon^{mir}
  +\delta^{jm}\varepsilon^{kir}+\delta^{im}\varepsilon^{kjr} ),
\end{eqnarray}
where the integration is taken over the unit sphere.

We are now in a position to write down equations for the evolution
of the components of the distribution function, as follows:
\begin{eqnarray}\label{transfer1}
&& \partial_t N^0_0
    -\frac{1}{3}\Gamma^0_{\phantom{0}kk}\partial_{\varepsilon}N^0_0
    -\frac{2}{15}\Gamma^0_{\phantom{0}kl}\partial_{\varepsilon}N^0_{kl}
    -\frac{1}{3}\Gamma^k_{\phantom{0}ll}N^0_k
    -\frac{2}{5}\Gamma^0_{\phantom{0}kl}N^0_{kl}
      =  -\frac{1}{L}i\texttt{Im}N^0_0\nonumber\\
&&\partial_t N^0_i
   +(\hat{A}^k_i\partial_{\varepsilon}+\hat{B}^k_i)N^0_k+\hat{C}^{kl}_iN^0_{kl}
      =  -\frac{1}{L}(\texttt{Re}N_i+\frac{2}{3}i\texttt{Im}N^0_i)  \nonumber \\
&&\partial_t N^0_{ij}
    +\hat{E}_{ij}\partial_{\varepsilon}N^0_0
    +(\hat{D}^k_{ij}\partial_{\varepsilon}+\hat{H}^k_{ij})N^0_k
    +(\hat{F}^{kl}_{ij}\partial_{\varepsilon}+\hat{G}^{kl}_{ij})N^0_{kl}\nonumber\\
&&   \qquad \qquad \qquad \qquad \qquad \qquad
       =  -\frac{1}{L}(\frac{9}{10}\texttt{Re}N^0_{ij}
          +\frac{3}{10}\texttt{Re}N^2_{ij}+i\texttt{Im}N^0_{ij}) \nonumber\\
&&\partial_t N^2_{ij}
    -\frac{1}{3}\Gamma^0_{\phantom{0}kk}\partial_{\varepsilon}N^2_{ij}+\hat{K}^{kl}_{ij}N^2_{kl}
      =  -\frac{1}{L}(\frac{1}{5}\texttt{Re}N^0_{ij}+\frac{2}{5}\texttt{Re}N^2_{ij}
         + i\texttt{Im}N^2_{ij}  ).
\end{eqnarray}
We have introduced the differential operators $\partial_t =d/dt$ and
$\partial_{\varepsilon}=\varepsilon\partial /\partial\varepsilon$.
The latter is used in the definition of $\zeta$:
\begin{equation}
\zeta \equiv \frac{\partial(\ln N^0_0)}{\partial\ln\varepsilon}
      =\varepsilon\frac{\partial(\ln N^0_0)}{\partial\varepsilon}
\end{equation}
The coefficients arising in the transfer equation can be obtained in
the following form
\begin{eqnarray}\label{Coeff_transfer}
  \hat{A}^{k}_{i} & = &  \frac{1}{5}(\Gamma^0_{\phantom{0}ik}
                         -\Gamma^0_{\phantom{0}ki}+\Gamma^0_{\phantom{0}ll}\delta_{ik})\nonumber \\
  \hat{B}^{k}_{i} & = &  -\Gamma^k_{\phantom{0}0i}+\frac{1}{5}(\Gamma^0_{\phantom{0}ki}
                        -4\Gamma^0_{\phantom{0}ik}+\Gamma^l_{\phantom{0}l0}\delta_{ik})             \nonumber  \\
  \hat{C}^{kl}_{i}& = &  -\frac{2}{5}(\Gamma^k_{\phantom{0}li}+\Gamma^k_{\phantom{0}mm}\delta_{il}
                        +3\Gamma^k_{\phantom{0}00}\delta_{li})\nonumber \\
  \hat{D}^{k}_{ij}& = &  \frac{1}{3}\Gamma^0_{\phantom{0}0l}\delta_{ij}
                        -\frac{1}{2}\Gamma^0_{\phantom{0}0i}\delta_{lj}
                        -\frac{1}{2}\Gamma^0_{\phantom{0}0j}\delta_{li}\nonumber \\
  \hat{E}_{ij}    & = &  \frac{1}{3}\Gamma^0_{\phantom{0}ll}\delta_{ij}-\frac{1}{2}\Gamma^0_{\phantom{0}ij}
                        -\frac{1}{2}\Gamma^0_{\phantom{0}ji} \nonumber \\
  \hat{F}^{kl}_{ij} & = &  \frac{2}{21}\Gamma^0_{\phantom{0}kl}\delta_{ij}
                          -\frac{1}{7}(\delta_{ki}\delta_{lj}\Gamma^0_{\phantom{0}mm}
                          +\Gamma^0_{\phantom{0}ki}\delta_{jl}+\Gamma^0_{\phantom{0}kj}\delta_{il})     \nonumber\\
  \hat{G}^{kl}_{ij}& = & \frac{2}{21}\Gamma^0_{\phantom{0}kl}\delta_{ij}
                         +\frac{2}{7}(\Gamma^0_{\phantom{0}ki}\delta_{jl}+\Gamma^0_{\phantom{0}kj}\delta_{il} )
                         -\Gamma^k_{\phantom{0}0i}\delta_{lj}-\Gamma^k_{\phantom{0}0j}\delta_{li}
                         \nonumber\\
                       & &  -\frac{5}{7}(\Gamma^0_{\phantom{0}ik}\delta_{jl}+\Gamma^0_{\phantom{0}jk}\delta_{il})
                         +\frac{2}{7}\Gamma^0_{\phantom{0}mm}\delta_{ik}\delta_{jl}\nonumber  \\
  \hat{H}^{k}_{ij} & = & -\frac{1}{2}\Gamma^k_{\phantom{0}ij}-\frac{1}{2}\Gamma^k_{\phantom{0}ji}
                         +\frac{1}{3}\Gamma^k_{\phantom{0}mm}\delta_{ij}
                         +\frac{1}{2}\Gamma^j_{\phantom{0}00}\delta_{ik}
                         +\frac{1}{2}\Gamma^i_{\phantom{0}00}\delta_{jk}
                         -\frac{1}{3}\Gamma^k_{\phantom{0}00}\delta_{ij}\nonumber \\
  \hat{K}^{kl}_{ij} & = &  -\frac{2}{9}\Gamma^0_{\phantom{0}kl}\delta_{ij}
                     +\frac{1}{3}(\Gamma^k_{\phantom{0}0i}+\Gamma^k_{\phantom{0}i0})\delta_{lj}
                     +\frac{1}{3}(\Gamma^j_{\phantom{0}0k}+\Gamma^j_{\phantom{0}k0})\delta_{li}
                     +\frac{i}{3}\delta_{ki}\delta_{lj}\Gamma^s_{\phantom{0}tr}\epsilon_{rst}\nonumber\\
\end{eqnarray}
Because  $k^{ij}$ and $m^{ij}$ are both symmetric and traceless, it
follows that $N^0_{ij}$ and $N^2_{ij}$ are too. The coefficients
listed above must therefore be made symmetric and traceless on the
index pairs $ij$ and $kl$.

Equations (\ref{transfer1}) describe the behavior of the
lowest-order angular modes of a general radiation field in a general
space-time, with two main assumptions. First, in the orthonormal
frame, we assume that the `'spatial'' derivatives $\partial_iN$
vanish for all quantities $N$ of interest because of homogeneity.
Second, we assume that the radiation field is described by a Planck
distribution at all times. These assumptions reduce the set of
equations needed to a linear system of coupled ordinary differential
equations with '`time''-dependent coefficients. The patterns that
are produced therefore depend both on the background cosmology and
the initial conditions.

Thomson scattering does not affect the component $\texttt{Re}N^0_0$
but it does the term $\texttt{Im}N^0_0$ that describes circular
polarization. The mode describing linear polarization, $N^2_{ij}$,
is coupled to higher-order modes of the radiation field but not
directly through $N^0_0$. Any non-zero term $\hat{C}^{kl}_i$ would
produce a dipole variation of the radiation distribution represented
by $N^0_i$ but no dipole can be produced this way in the particular
case of Bianchi I (nor indeed for the FRW case). The presence of a
dipole component is inevitable in other Bianchi models since, even
with $N^0_i=0$ initially, $N^0_i$ becomes different from zero  if
$\Gamma^0_{\phantom{0}0i}\neq 0$ or if
$N^0_{kl}\Gamma^k_{\phantom{0}li}+N^0_{ik}\Gamma^k_{\phantom{0}ll}$
differs from zero. In a similar manner a quadrupole component can
always be generated from the isotropic mode $N^0_0$, if $E_{ik}$ is
different from zero which means that the fluid flow possesses some
kind of shearing motion.

It is clear from the system of equations (\ref{transfer1}) that a
gravitational field alone is not able to generate polarization.
Initially unpolarized radiation collisionlessly propagating in an
arbitrary gravitational field must remain unpolarized. However, a
quadrupole mode of unpolarized radiation generates a linear
polarization component at $l=2$, if Thomson scattering is present.
This is the standard mechanism by which polarization is generated
from radiation anisotropies in the early Universe. Since a
quadrupole mode of unpolarized radiation is generated from the
isotropic component if $E_{ik} \neq 0$, the cosmological
gravitational field could therefore be indirectly responsible for
generating polarization if it first generates a quadrupole
anisotropy. Equations (\ref{transfer1}) also show that an initial
monopole $N^0_0$ produces a non-zero quadrupole $N^0_{ij}$ via the
shear $\hat{E}_{ij} =-\sigma_{ij} $ and $N^0_i$ is subsequently
coupled with $N^0_{ij}$ by $\hat{C}^{kl}_{i}$. The effect of shear
on the radiation is to generate a quadrupole anisotropy by
redshifting the it anisotropically; in some models, dipole and
higher order multipoles would also arise.

To summarize, then. In order to get interesting higher-order
patterns in the radiation background we must either have non-zero
shear if there is no initial quadrupole or have non-zero initial
quadrupole and monopole terms if there is no shear.

\subsection{Geodesic Equations}
We now follow the follow the convention in ref. \cite{Dautcourt1} to
construct the equations describing geodesics in the models we
consider.
From each components of eq. (\ref{geo0}) time variation of $\varepsilon$ and direction vector $k^i$ have the form such as
\begin{eqnarray}
\frac{dp^0}{d\lambda}&=&\frac{d\varepsilon}{d\lambda} =-\varepsilon^2\gamma^0\\
\frac{dp^i}{d\lambda}&=&\frac{d\varepsilon}{d\lambda}k^i+\varepsilon\frac{dk^i}{d\lambda}  =-\varepsilon^2\gamma^i
\end{eqnarray}
Using the relations
\begin{eqnarray}
    dp^i &=& k^id\varepsilon+\varepsilon dk^i =k^id\varepsilon+\varepsilon(a^id\theta+\sin\theta b^id\phi )
\end{eqnarray}
and orthogonality of $a^i$ , $b^i$ and $k^i$, we obtain the
following equations for the time variation for $\theta$ and $\phi$:
\begin{eqnarray}
\frac{d\theta}{d\lambda}&=&\frac{1}{\varepsilon}a_i\frac{dp^i}{d\lambda}
    =-\varepsilon a_i\gamma^i \\
\frac{d\phi}{d\lambda}&=&\frac{1}{\varepsilon}\frac{b_i}{\sin\theta}\frac{dp^i}{d\lambda}
=-\varepsilon\frac{b_i}{\sin\theta}\gamma^i
\end{eqnarray}
since the $\gamma^a$ are also represented by Ricci Coefficients
$\Gamma^a_{bc}$. Here, we also neglect the effect shear which would
make ($ \Gamma^0_{ij}=0$). Finally we obtain  variation terms:
\begin{eqnarray}
\frac{1}{\varepsilon}\frac{d\theta}{d\lambda}&=& \sin\theta[a+\cos\phi\sin\phi(n_3-n_2)] \\
\frac{1}{\varepsilon}\frac{d\phi}{d\lambda}&=&
\cos\theta[n_1-n_3+(n_3-n_2)\cos^2\phi]
\end{eqnarray}
The change of polarization angle $\psi$ is expressed by
\begin{eqnarray}
\frac{1}{\varepsilon}\frac{d\psi}{d\lambda}
&=&im^i\bar{m}^k\Gamma^k_{li}k^l-\cot\theta b_l\gamma^l
= \frac{1}{6}n^s_s-n^i_kk^{ik}+\cos\theta\frac{1}{\varepsilon}
\frac{d\phi}{d\lambda}.
\end{eqnarray}
Although the $\Gamma^k_{li}$ contain the Bianchi vector $a_i$, only
the  $n_i$ components appear in the change of $\psi$ since
$a_ik^lk^m\epsilon^{ilm}=0$ by symmetry of the $l$ and $m$ indices.
Consiquently this term does not give different results for the
models such as type I and V or VII$_h$ and VII$_0$. Applying this to
each Bianchi types we obtain, for Types I and V,
\begin{equation}
\frac{d\psi}{d\lambda} =0.
\end{equation}
For Types VII$_h$ and VII$_0$ we get
\begin{eqnarray}
\frac{1}{\varepsilon}\frac{d\psi}{d\lambda}
=2n(1-\sin^2\theta)+\frac{1}{\varepsilon}\cos\theta\frac{d\phi}{d\lambda}
=-n\cos^2\theta
\end{eqnarray}
and for Type IX we have
\begin{equation}
\frac{1}{\varepsilon}\frac{d\psi}{d\lambda} =n.
\end{equation}
Note that we consider the FRW limits i.e. Bianchi vectors as
$n_i=n$. The term $\frac{d\psi}{d\lambda}$ is important since it
gives rise to mixing terms between the E and B modes in the
Boltzmann equation.
\section{Temperature and Polarization Patterns}
In this section we will present representative examples of the
temperature and polarization patterns produced in the models we have
discussed, computed by numerically integrating the system of
equations derived in Section 4. The patterns produced depend on the
parameters chosen for the model in question and also, as we have
explained in the previous section, on the choice of initial data. In
the following, primarily pedagogical, discussion we do not attempt
to normalize the models to fit current cosmological observations but
restrict ourselves to phenomenological aspects of the patterns
produced. The overall level of temperature anisotropy depends on the
choice of parameters in the Bianchi models that express the extent
of its departure from the FRW form. Since we do not tune this to
observations, the amplitude is arbitrary, as in the cosmic epoch
attributed to each of the evolutionary stages. Moreover, the overall
degree of polarization depends strongly on the ionization history
through the optical depth $\tau$ which appears in Eq.
(\ref{scattering1}). We shall not attempt to model this in detail in
this paper either. What is important, however, is that the
geometrical relationship between the temperature and polarization
patterns does not depend on these factors; it is fixed by the
geometric structure of the model, not on its normalization.

Of course we compute only the {\em coherent} part of the radiation
field that arises from the geometry of the model. Any realistic
cosmological model (i.e. one that produces galaxies and large-scale
structure) must have density inhomogeneities too. Assuming these are
of stochastic origin they would add incoherent perturbations on top
of the coherent ones produced by the background model.

For illustrative purposes we have chosen cases where the initial
conditions produce a pure quadrupole anisotropy of {\em tesseral}
form, i.e. and $l=2$ spherical harmonic mode with $m=1$ as shown in
Figure 1.
\begin{figure}
\begin{center}
 \includegraphics[scale=0.30,angle=90]{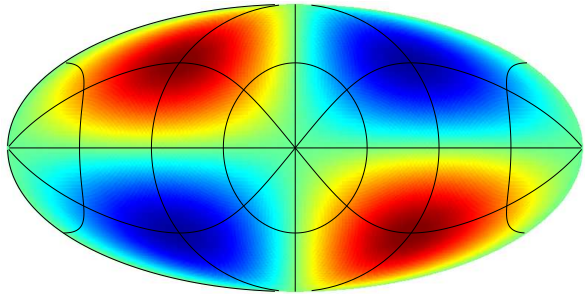}
\end{center}
\caption{Initial configuration chosen for the temperature pattern.}
\end{figure}
We gratefully acknowledge the use of the Healpix software
\cite{healpix} in creating this and all the other all-sky maps shown
hereafter.

Other choices are, of course, possible. A quadrupole with $m=0$
would produce a {\em zonal} pattern, and one with $m=2$ would be a
{\em sectoral mode} \cite{sc10}; these choices are discussed at some
length in ref. \cite{pontzen}. One could also generate more
complicated patterns by having an observer who is not at rest in the
frame we are using, which would introduce an additional dipole
anisotropy. We will not discuss this possibility further in this
paper.

The simplest case is obviously that of Bianchi I, but this is
nevertheless of some interest because a Universe of this type could
in principle account for the presence of a low quadrupole
\cite{BianchiI,BianchiIb}. In this example the temperature pattern
does not evolve at all with time, so one can simply treat the
initial quadrupole as a free parameter. The polarization patterns
arising in this model, which do not evolve with time either, are
shown in Figure 2.
\begin{figure}
\begin{center}
   \includegraphics[scale=0.15,angle=90]{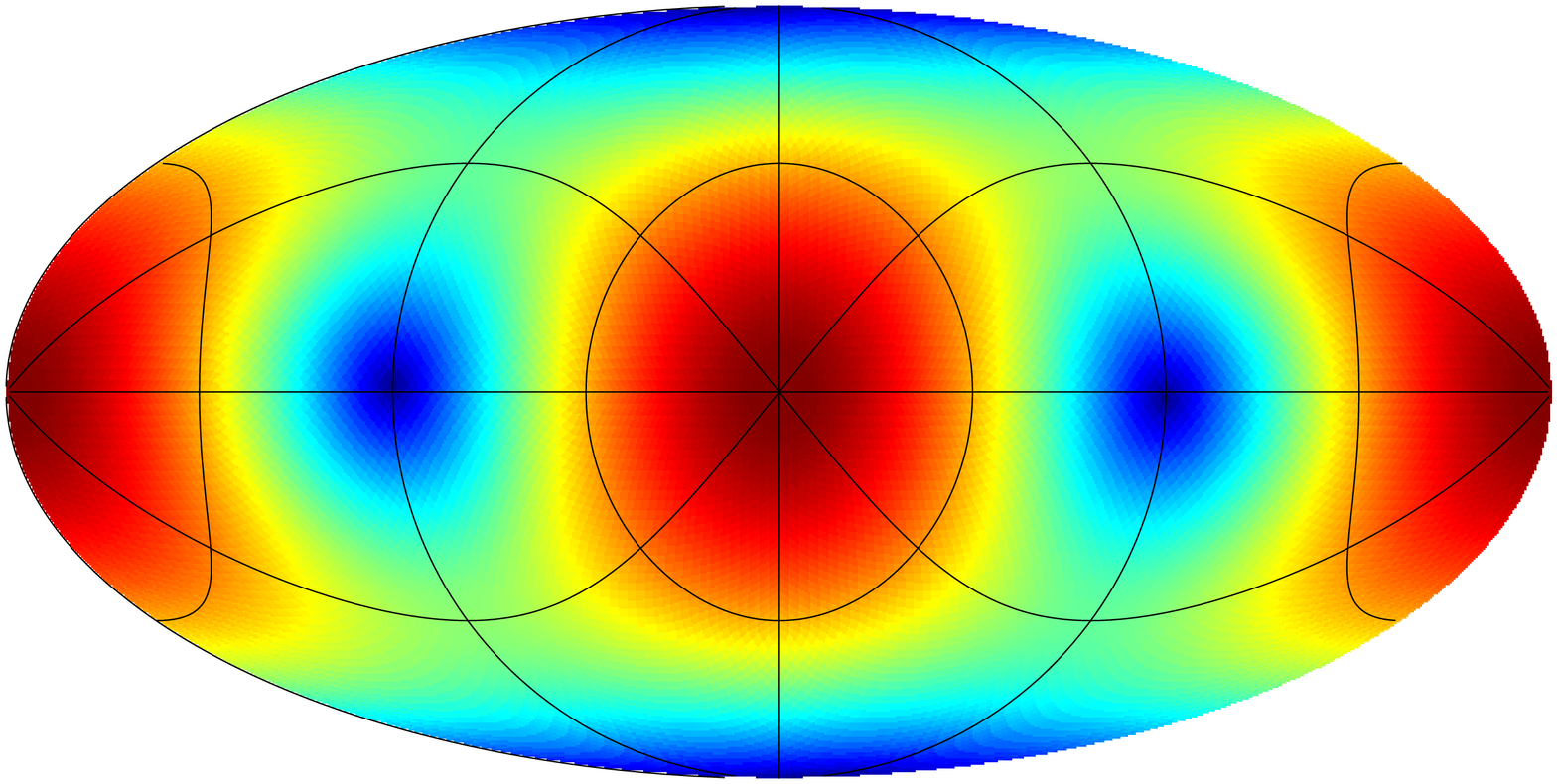}
   \includegraphics[scale=0.15,angle=90]{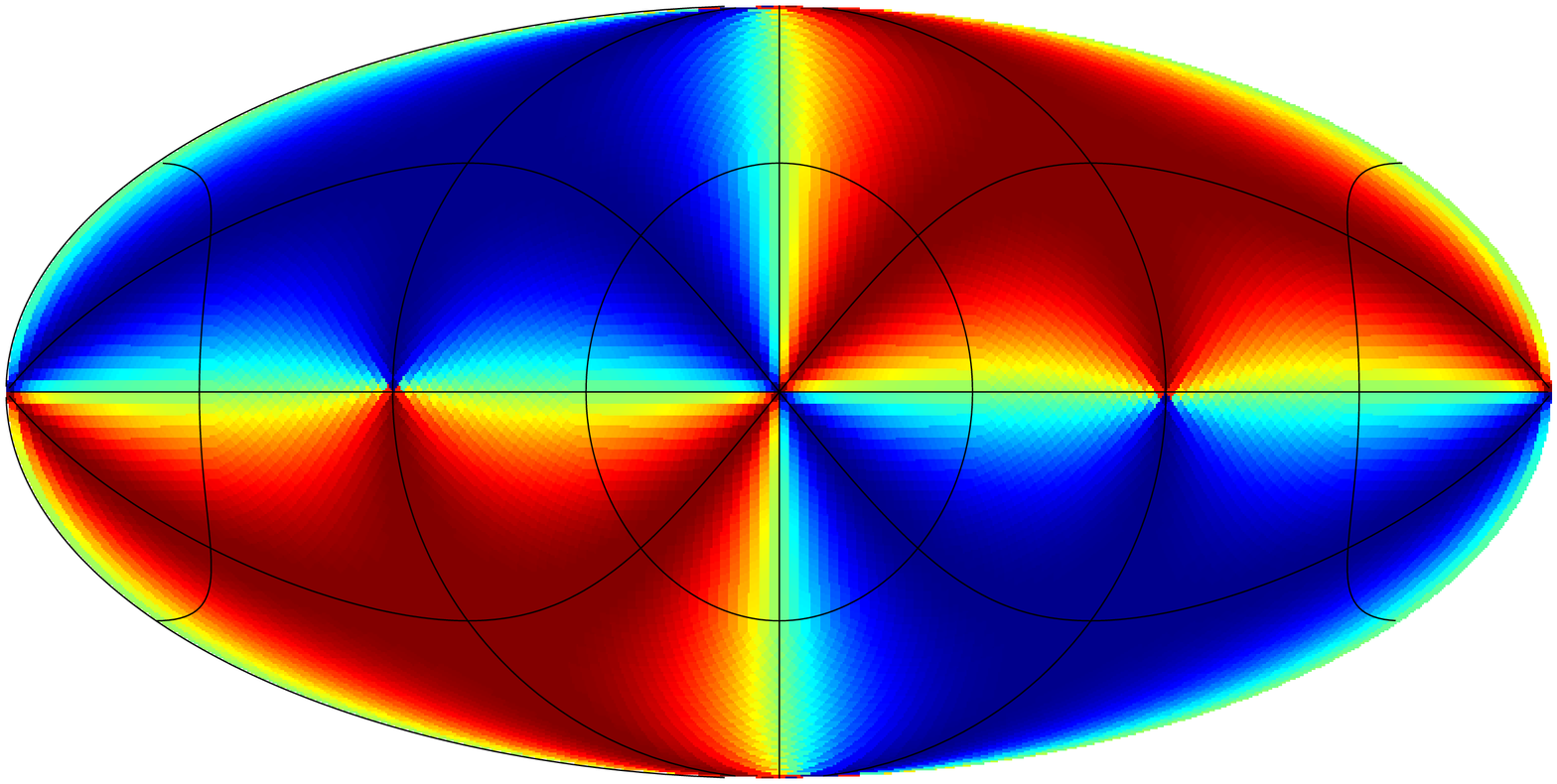}
   \includegraphics[scale=0.15,angle=90]{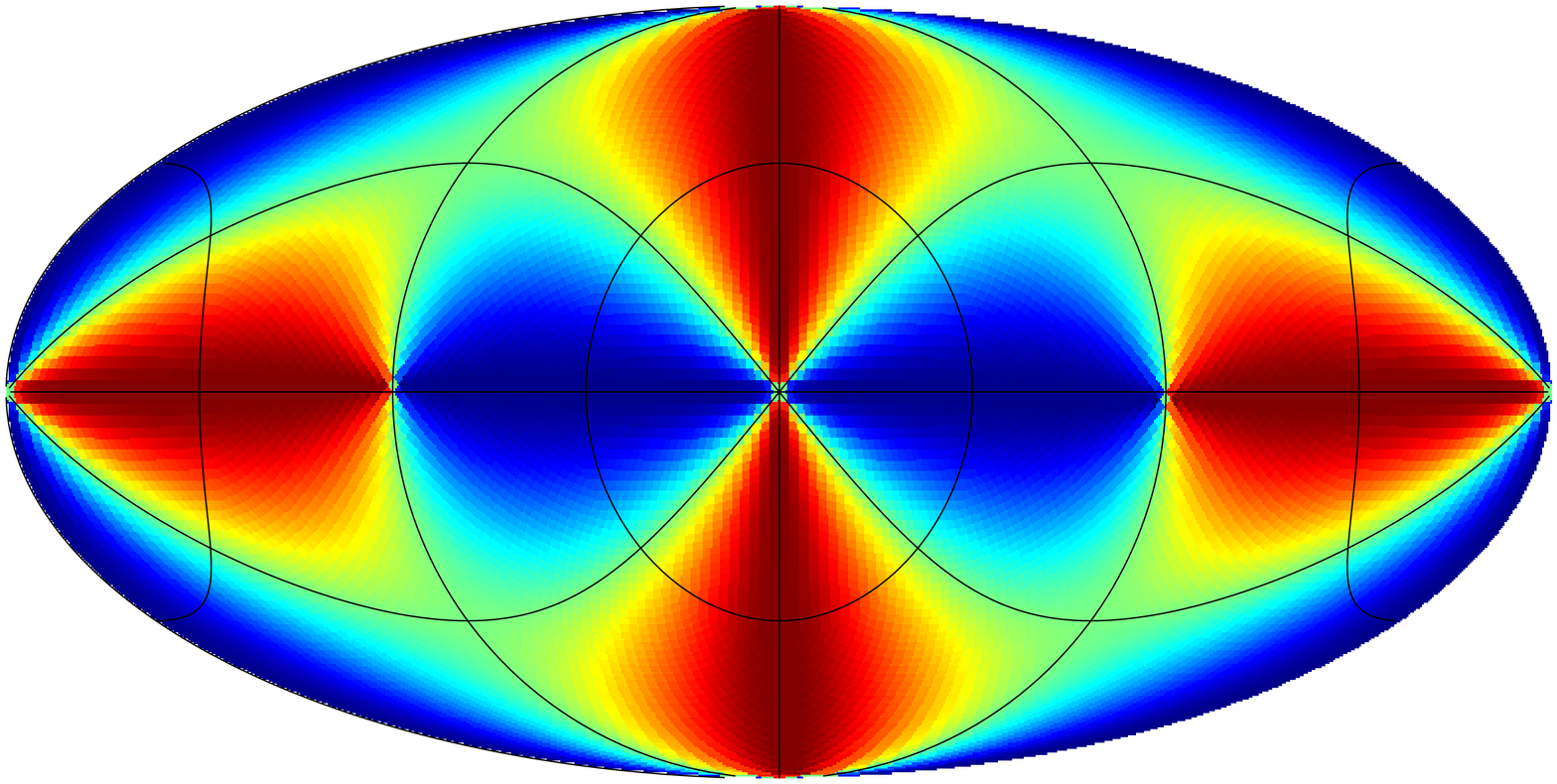}
\end{center}
\caption{Polarization maps, i.e. polarization amplitude (left),
Stokes parameter Q (middle) and U (right) of Bianchi type I.}
\end{figure}

We next turn our attention to Bianchi types V, VII$_h$ and VII$_0$.
These models  have a single preferred axis of symmetry. The
alignment of the shear eigenvectors relative to this preferred axis
determines not only the dynamical evolution of the model through the
field equations, but also the temperature and polarization pattern
which, as we shall see, gets imprinted into the cosmic background
radiation. Figure 3 shows (from bottom to top) the time evolution of
the temperature pattern in these models. Note that, in Type V
(left), the initial quadrupole retains its shape but gets focussed
into a patch of decreasing size as time goes on. This is due to the
effect of negative spatial curvature. In Bianchi Type VII$_0$
(right) the effect of rotation and shear is to twist the initial
quadupole into a spiral shape that winds up increasingly as the
system evolves. In the middle case, Bianchi Type VII$_h$ we have a
combination of the two cases either side: there is both a focussing
and a twist. This case produces the most complicated temperature
pattern.

\begin{figure}
\begin{center}
  \includegraphics[scale=0.15,angle=90]{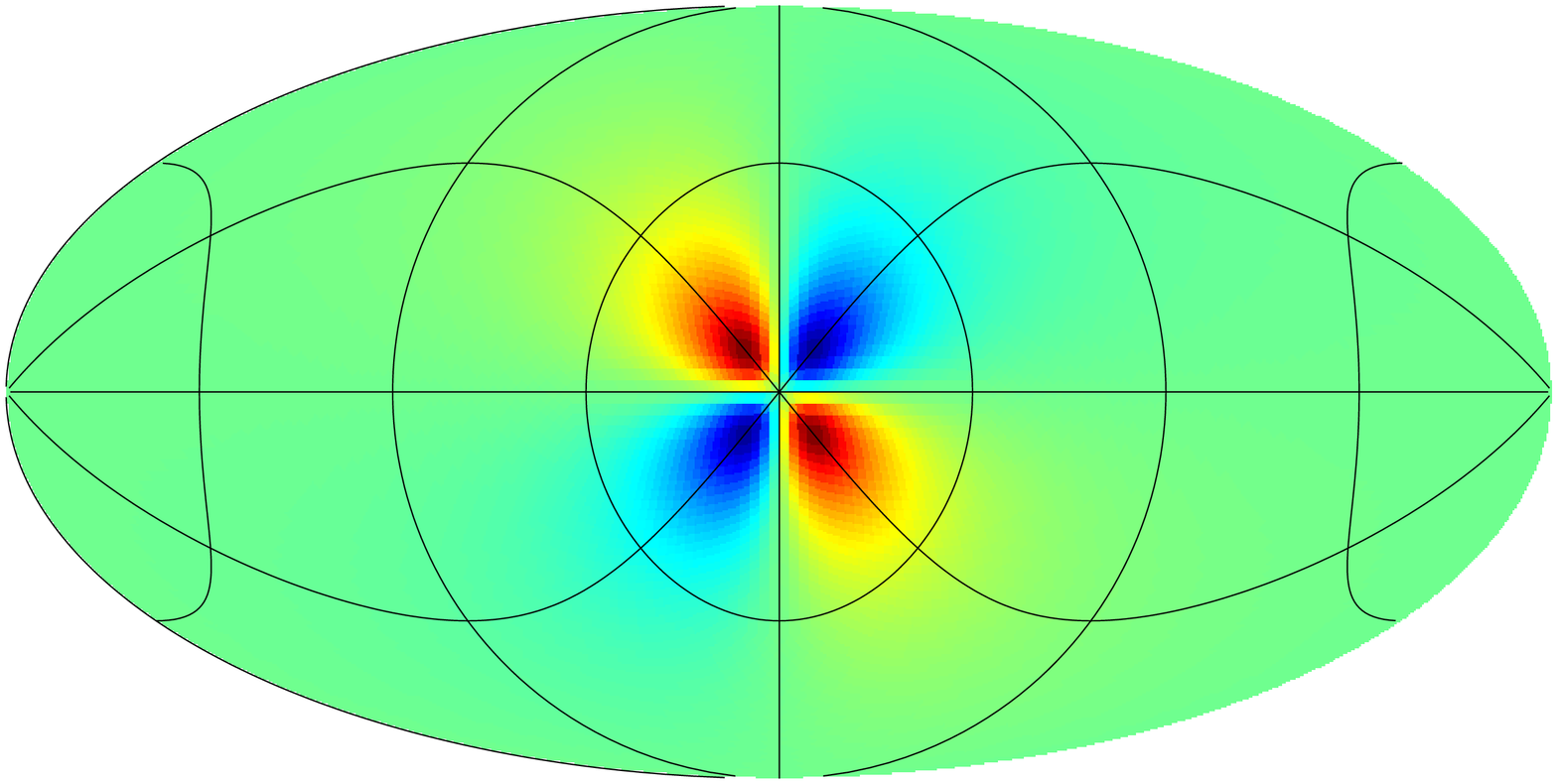}
  \includegraphics[scale=0.15,angle=90]{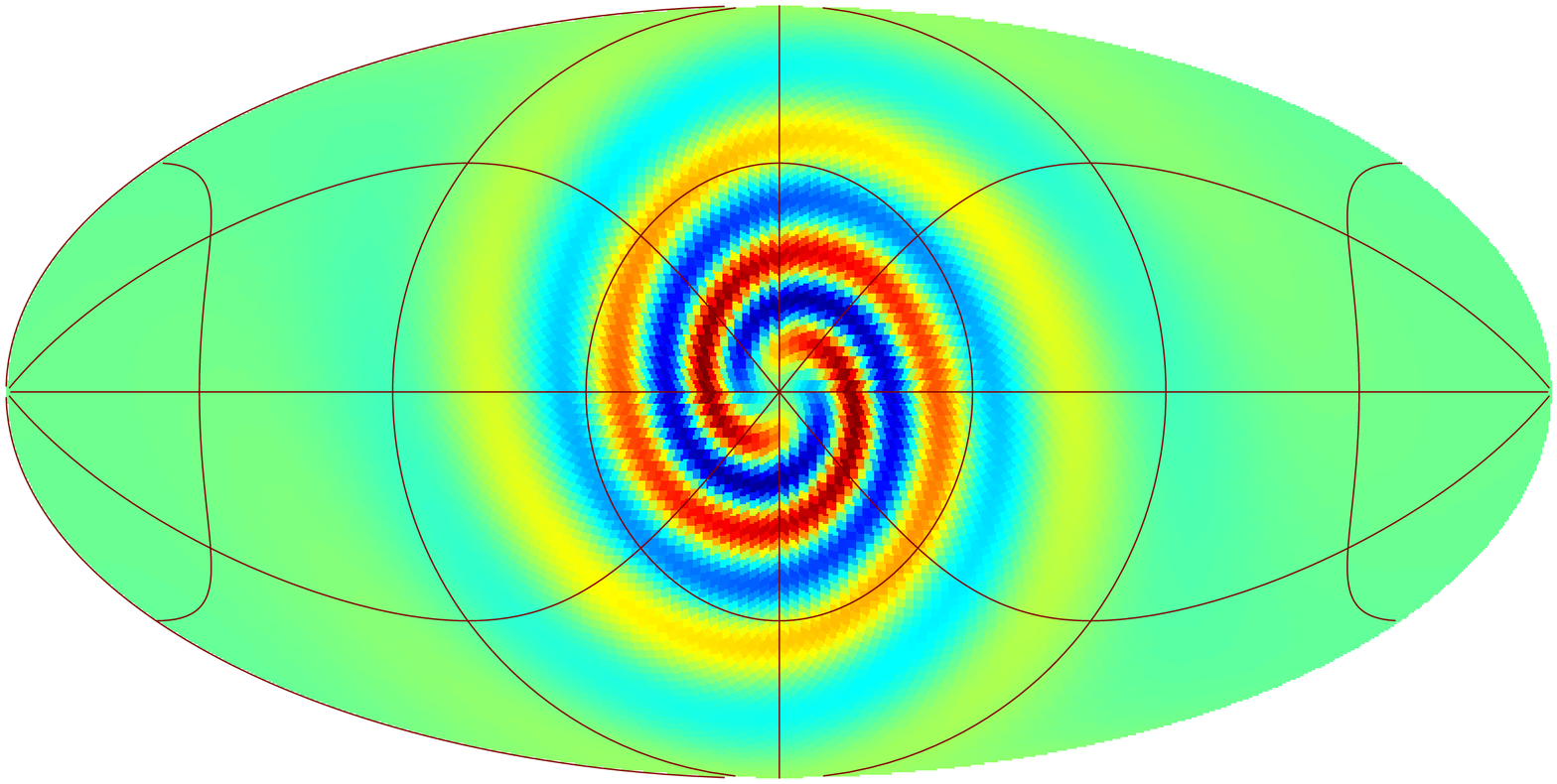}
   \includegraphics[scale=0.15,angle=90]{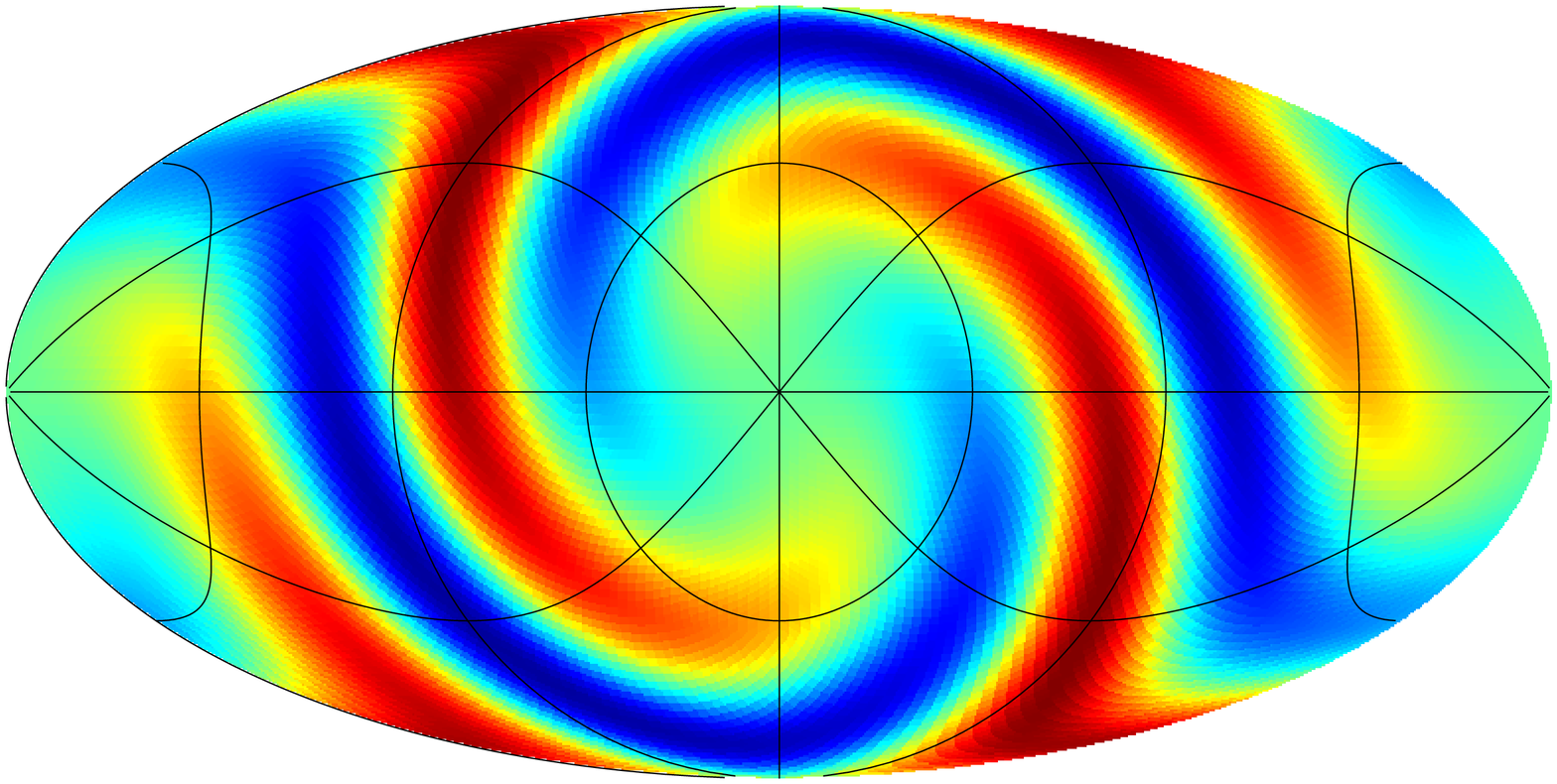}\\
  \includegraphics[scale=0.15,angle=90]{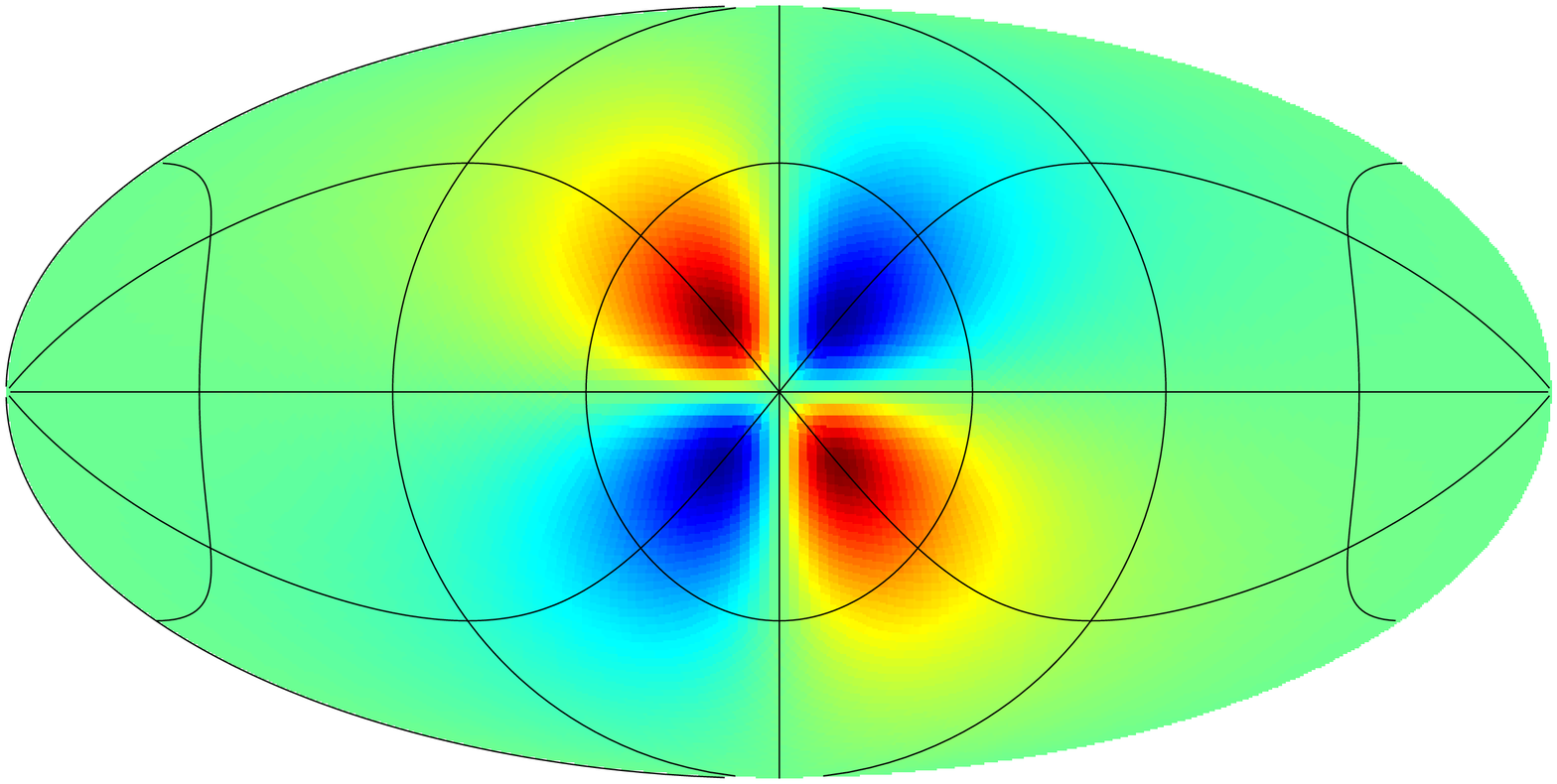}
    \includegraphics[scale=0.15,angle=90]{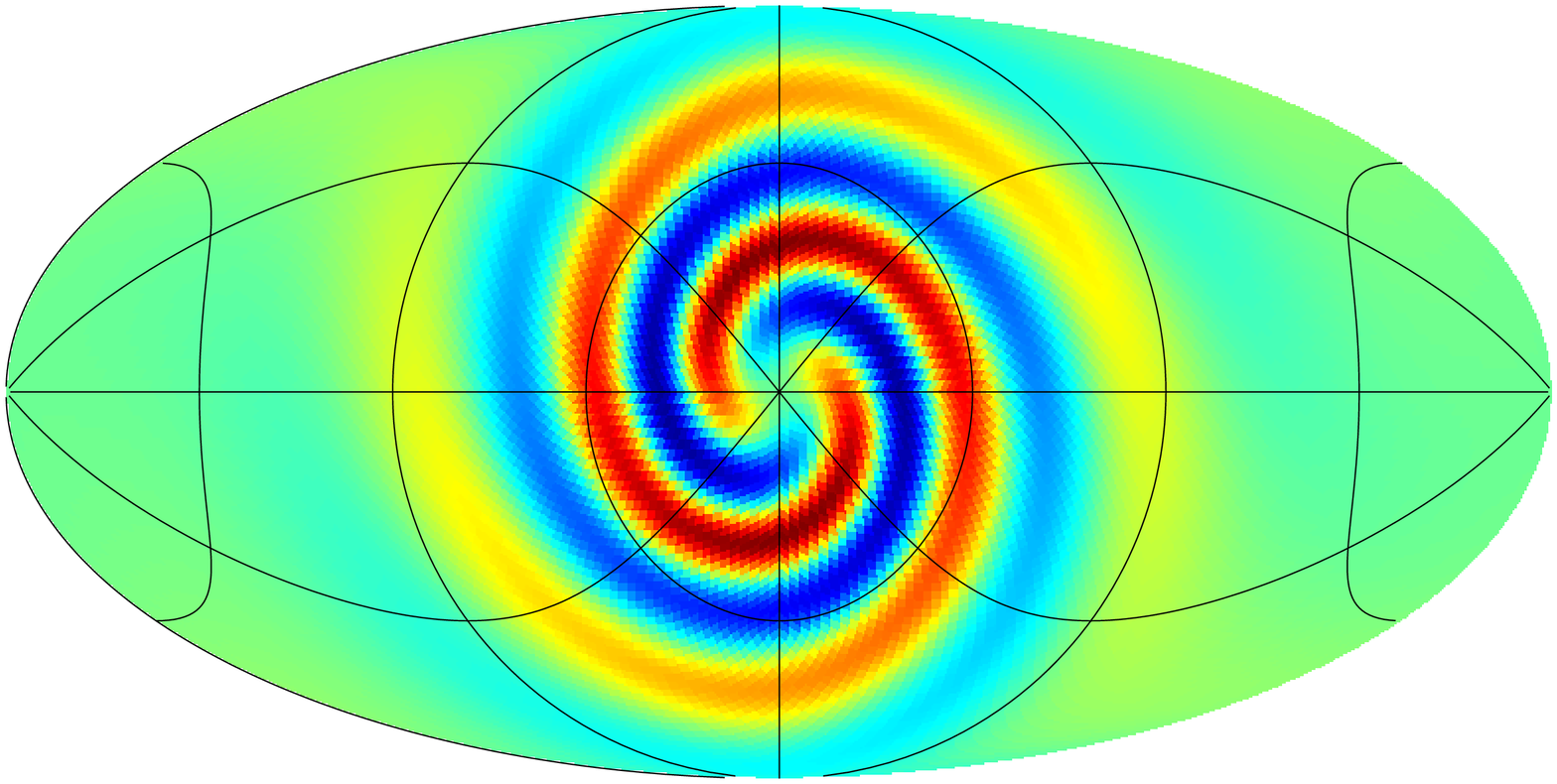}
   \includegraphics[scale=0.15,angle=90]{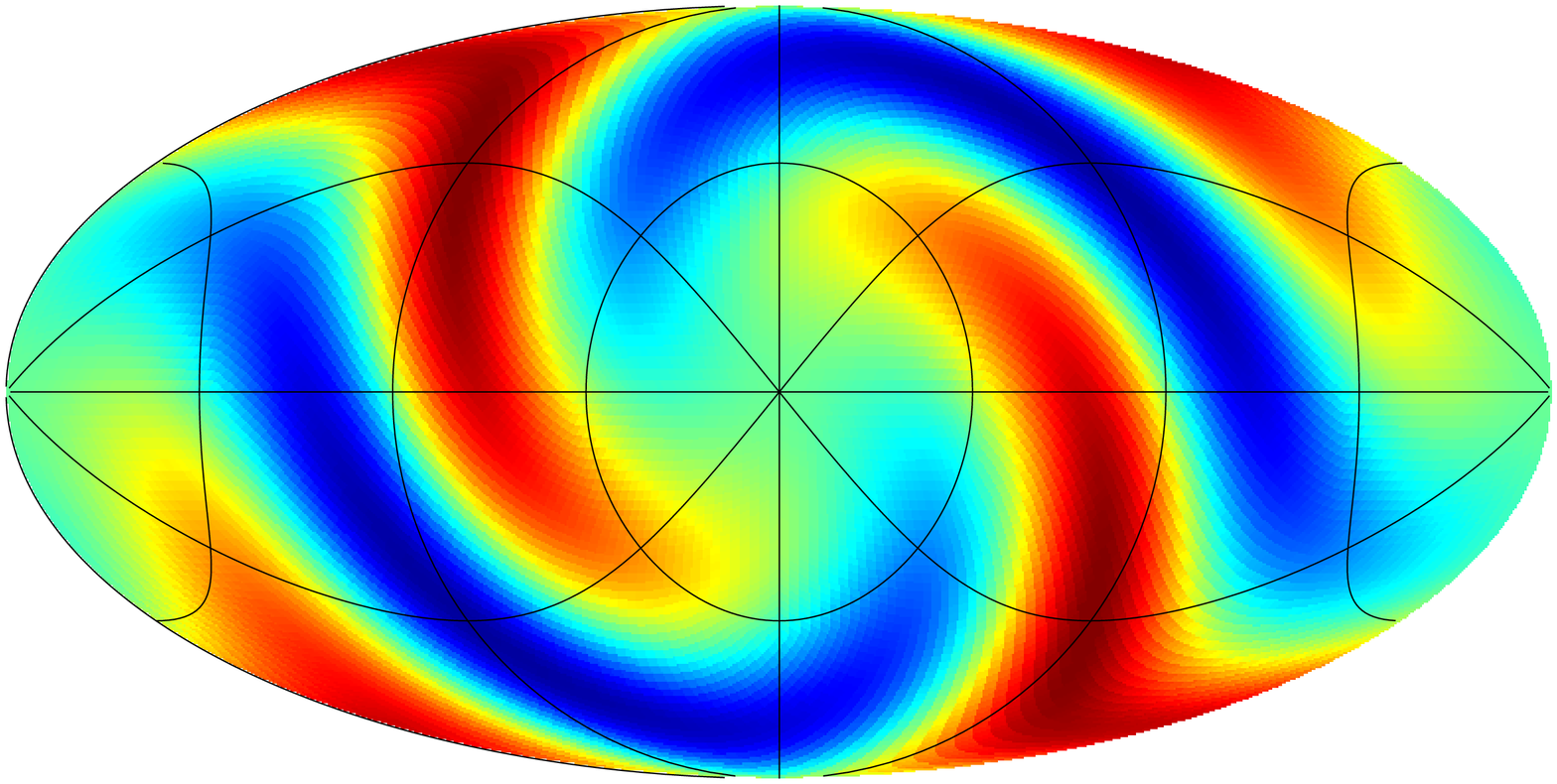}\\
  \includegraphics[scale=0.15,angle=90]{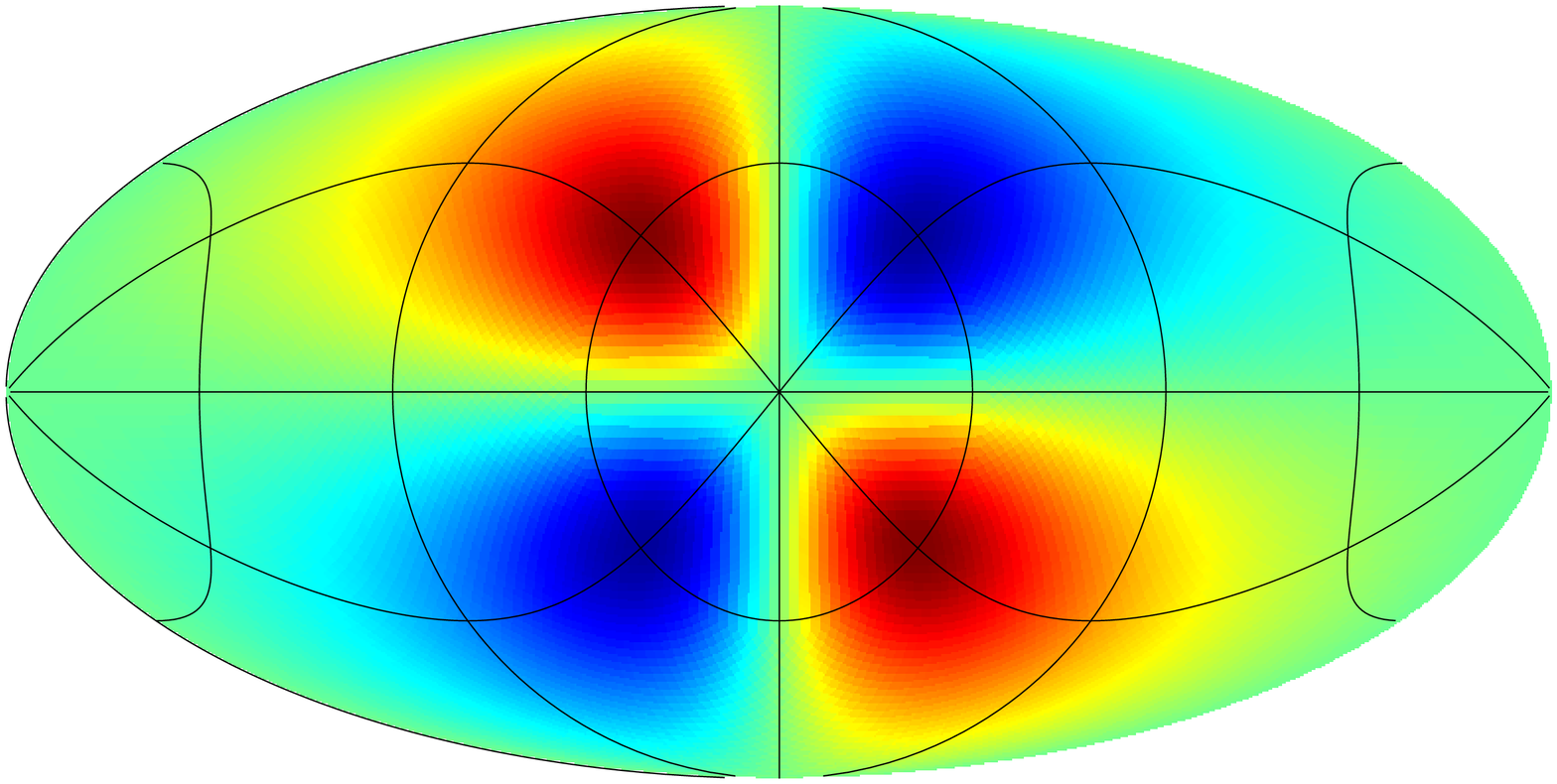}
    \includegraphics[scale=0.15,angle=90]{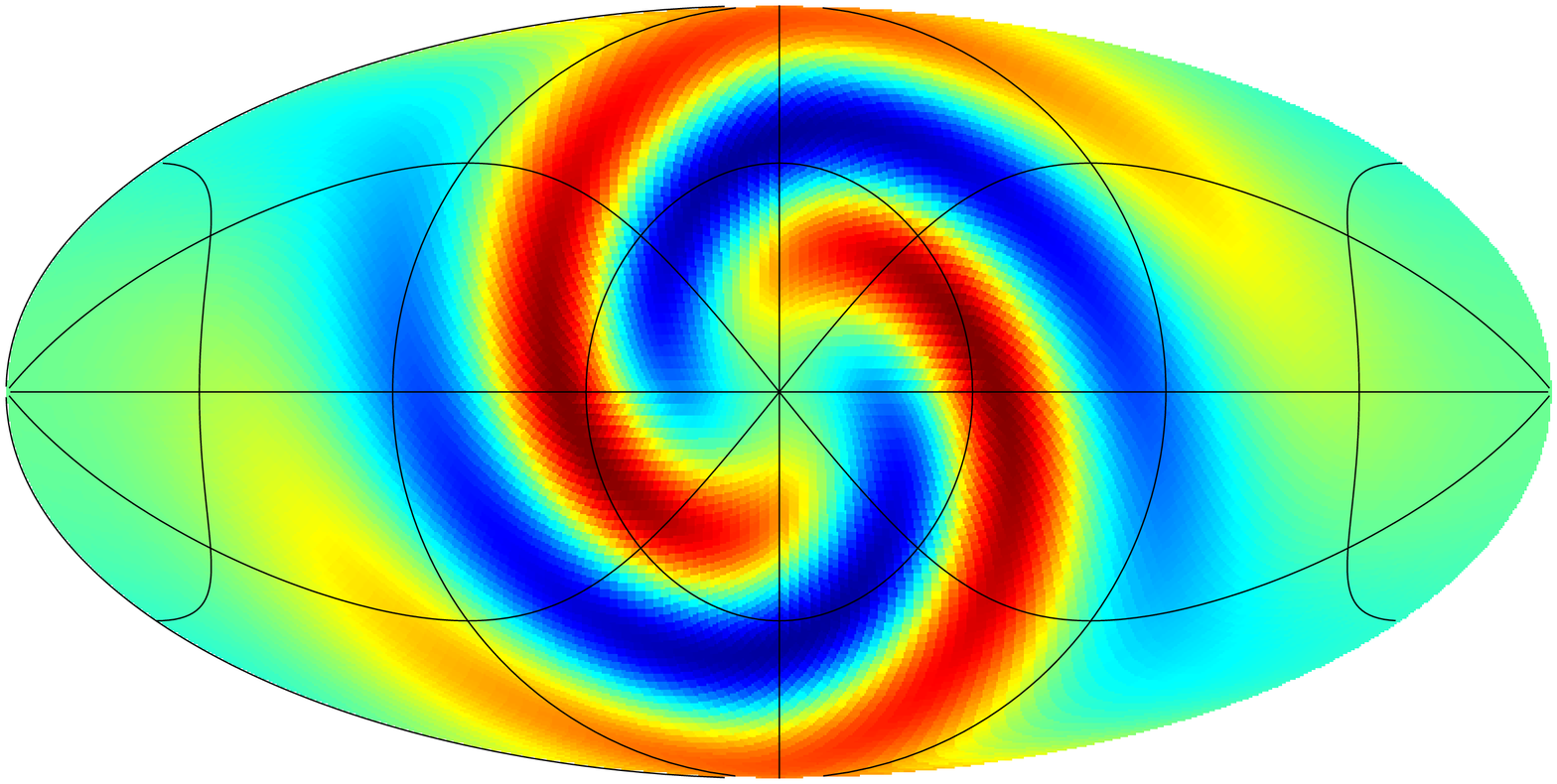}
   \includegraphics[scale=0.15,angle=90]{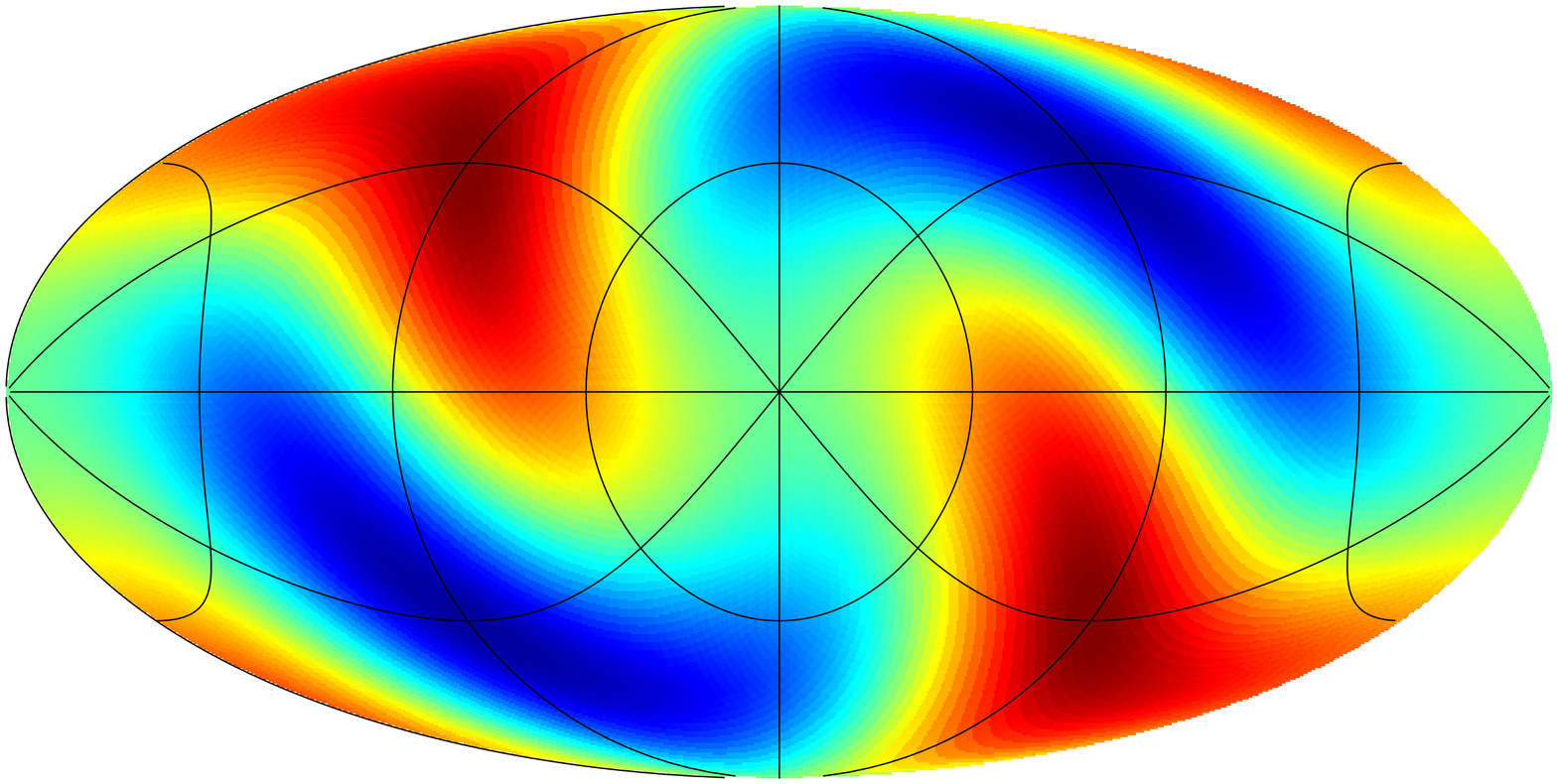}\\
  \includegraphics[scale=0.15,angle=90]{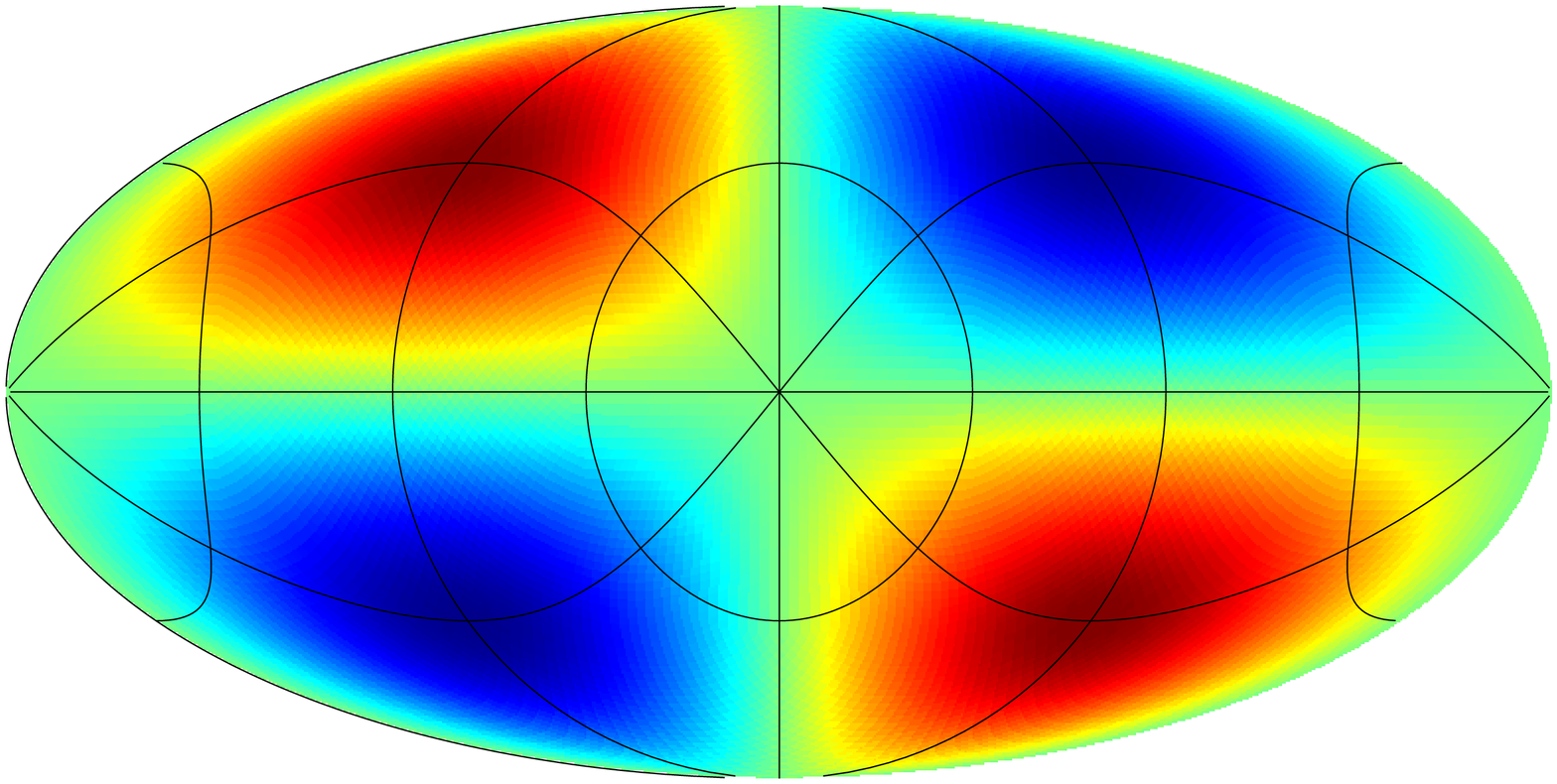}
    \includegraphics[scale=0.15,angle=90]{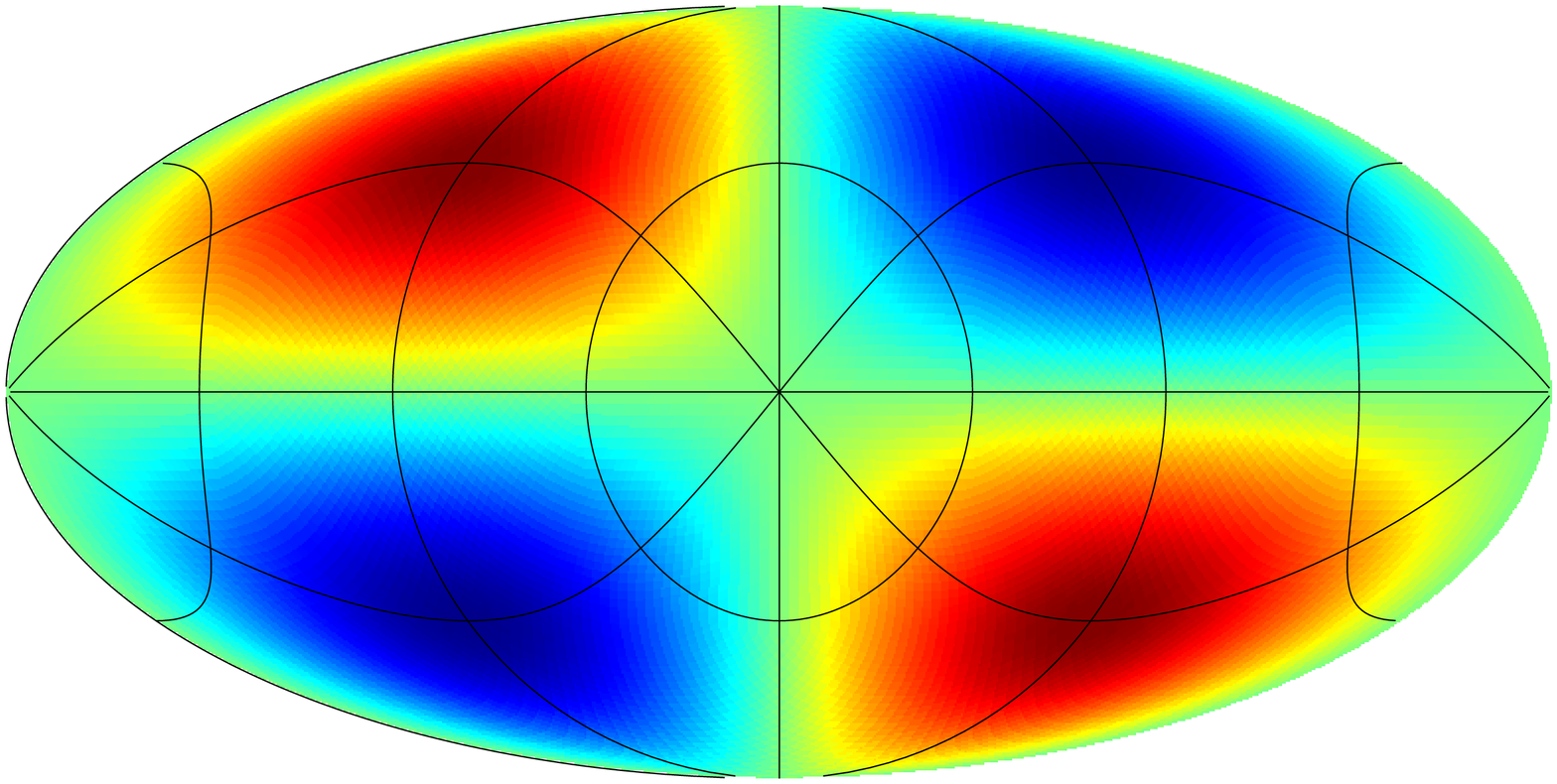}
   \includegraphics[scale=0.15,angle=90]{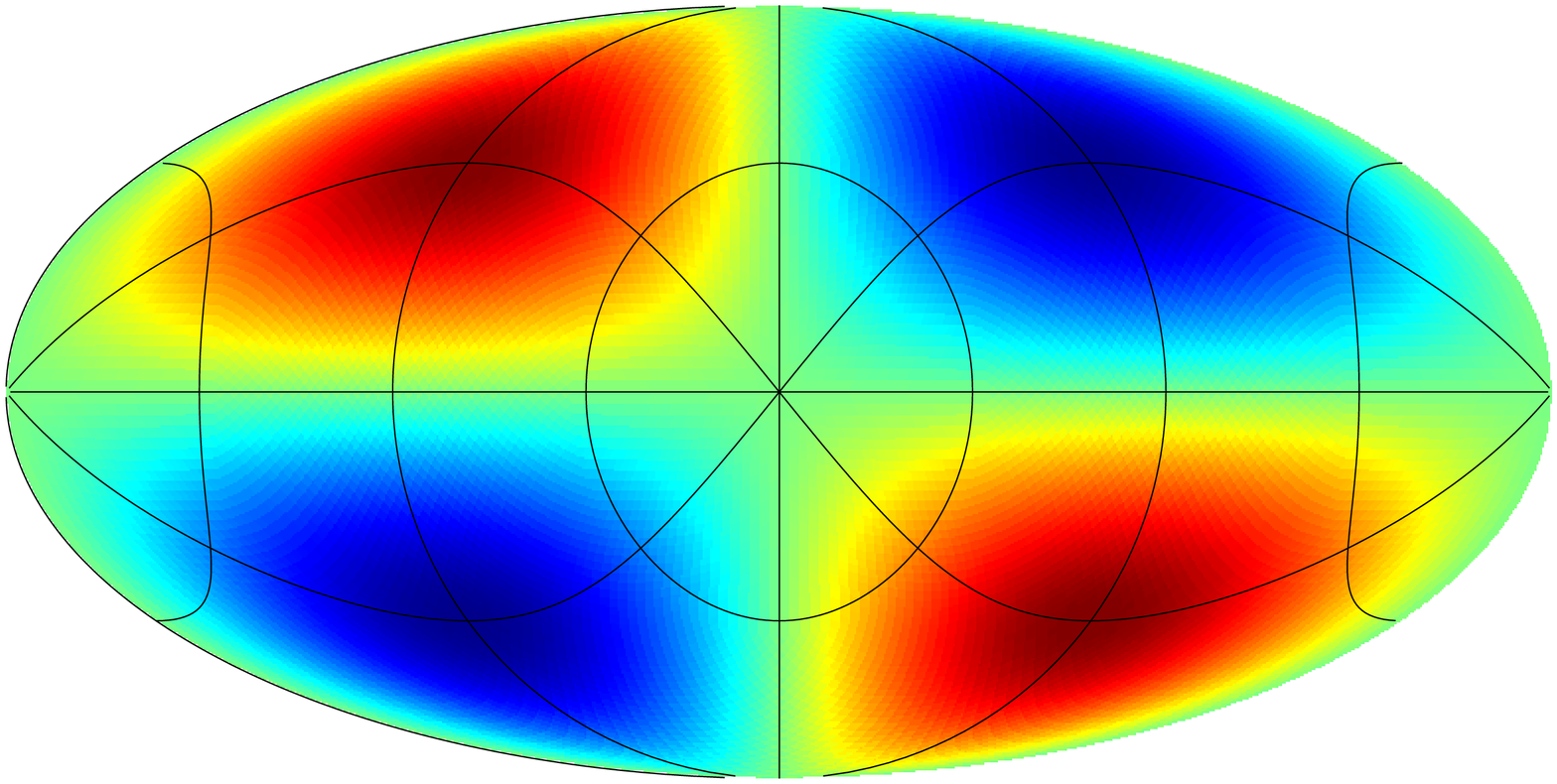}\\
   \end{center}
 \caption{The time evolution of the temperature maps for Bianchi types  V, VII$_h$ and VII$_0$. Time increases from bottom to top.}\label{Ttypesz}
\end{figure}

In the following three figures we examine the polarization pattern
produced in the models shown in Figure 2. First, in Figure 3, we
have Bianchi V. These results show that while the polarization
pattern alters with time in this case, its general orientation on
the sky does not (as is the case with the temperature pattern). The
implications of this for the production of cosmological B-mode
polarization was discussed by ref. \cite{sung1}.

\begin{figure}
\begin{center}
  \includegraphics[scale=0.15,angle=90]{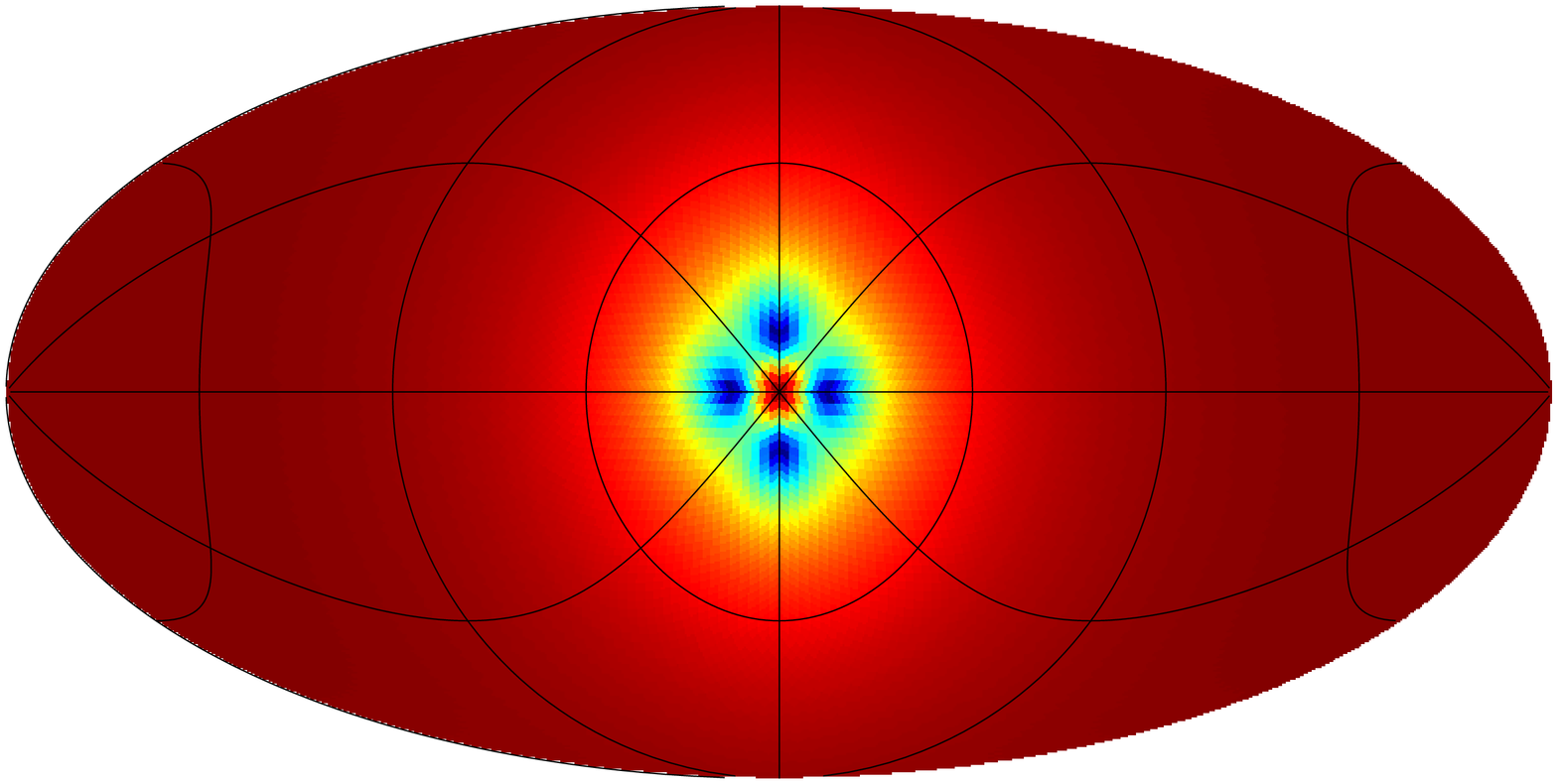}
  \includegraphics[scale=0.15,angle=90]{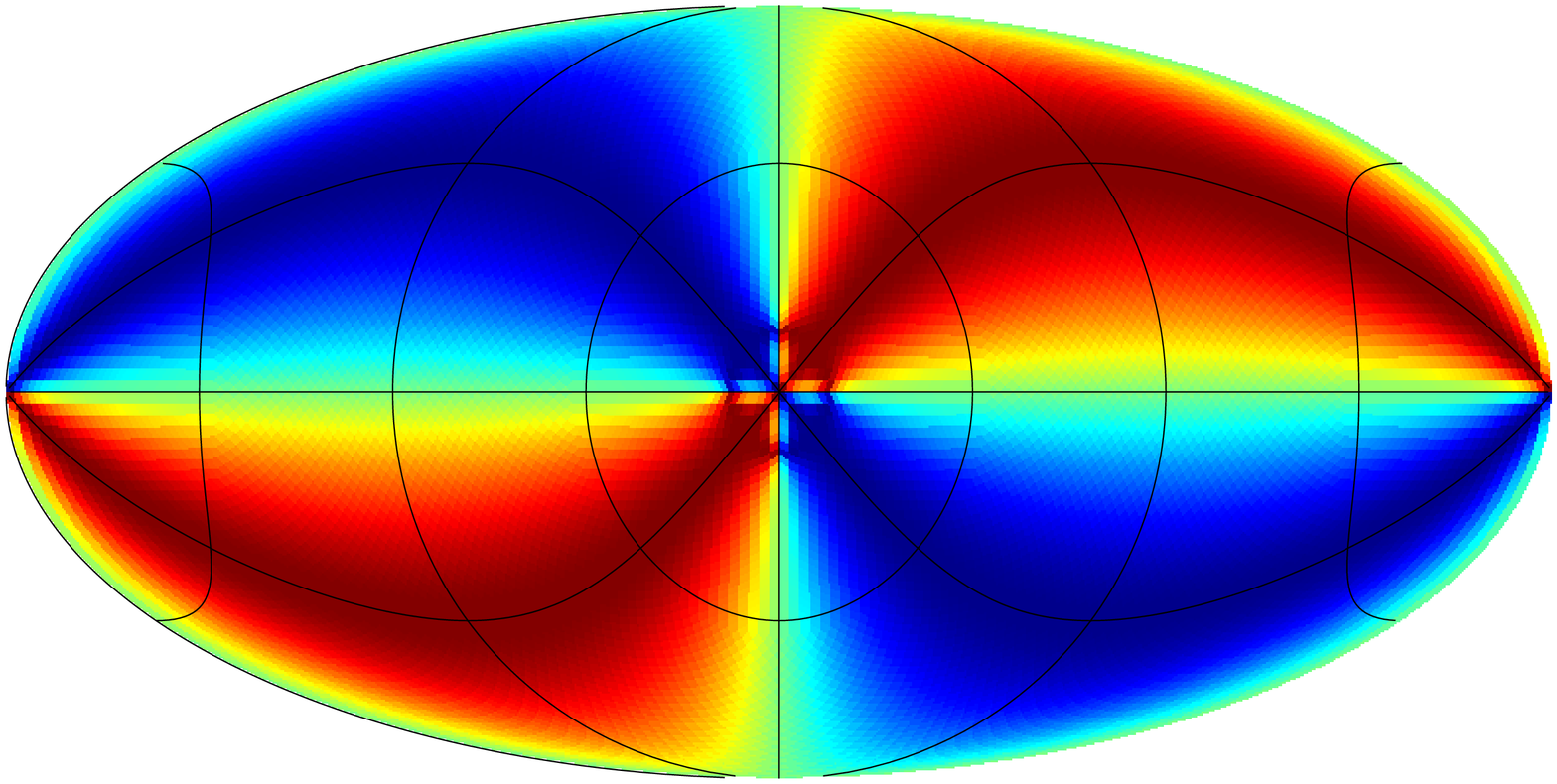}
   \includegraphics[scale=0.15,angle=90]{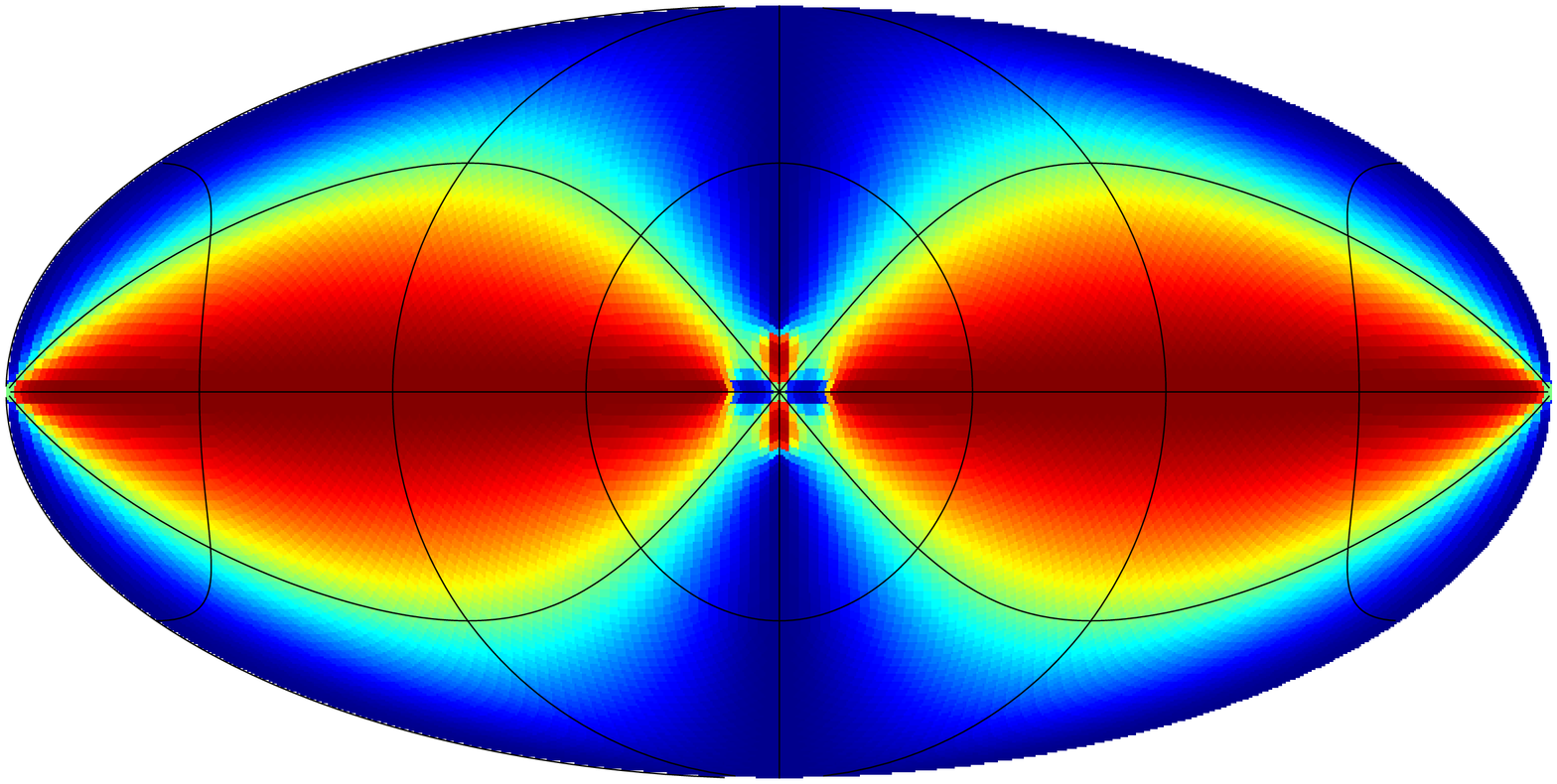}\\
  \includegraphics[scale=0.15,angle=90]{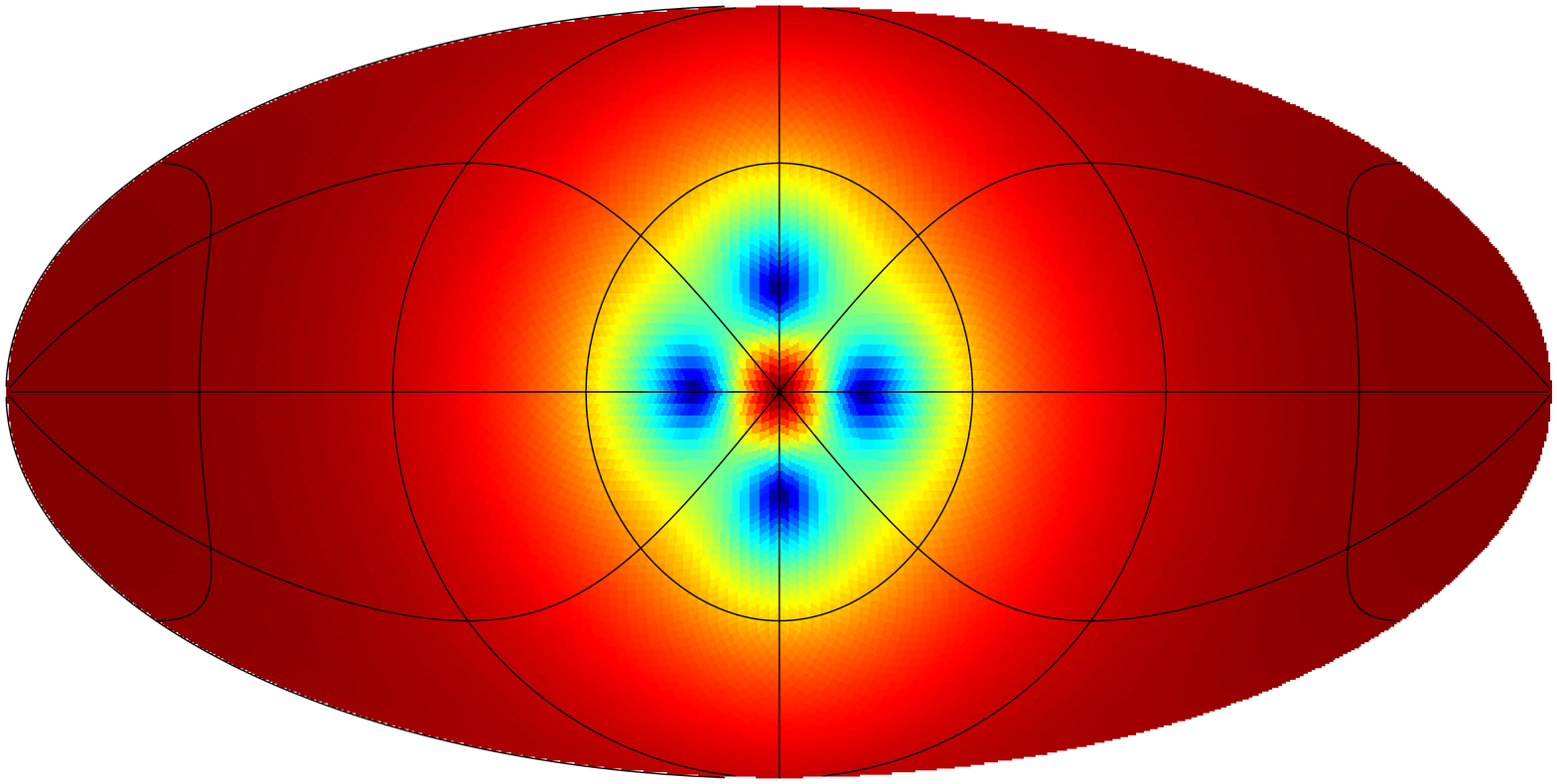}
    \includegraphics[scale=0.15,angle=90]{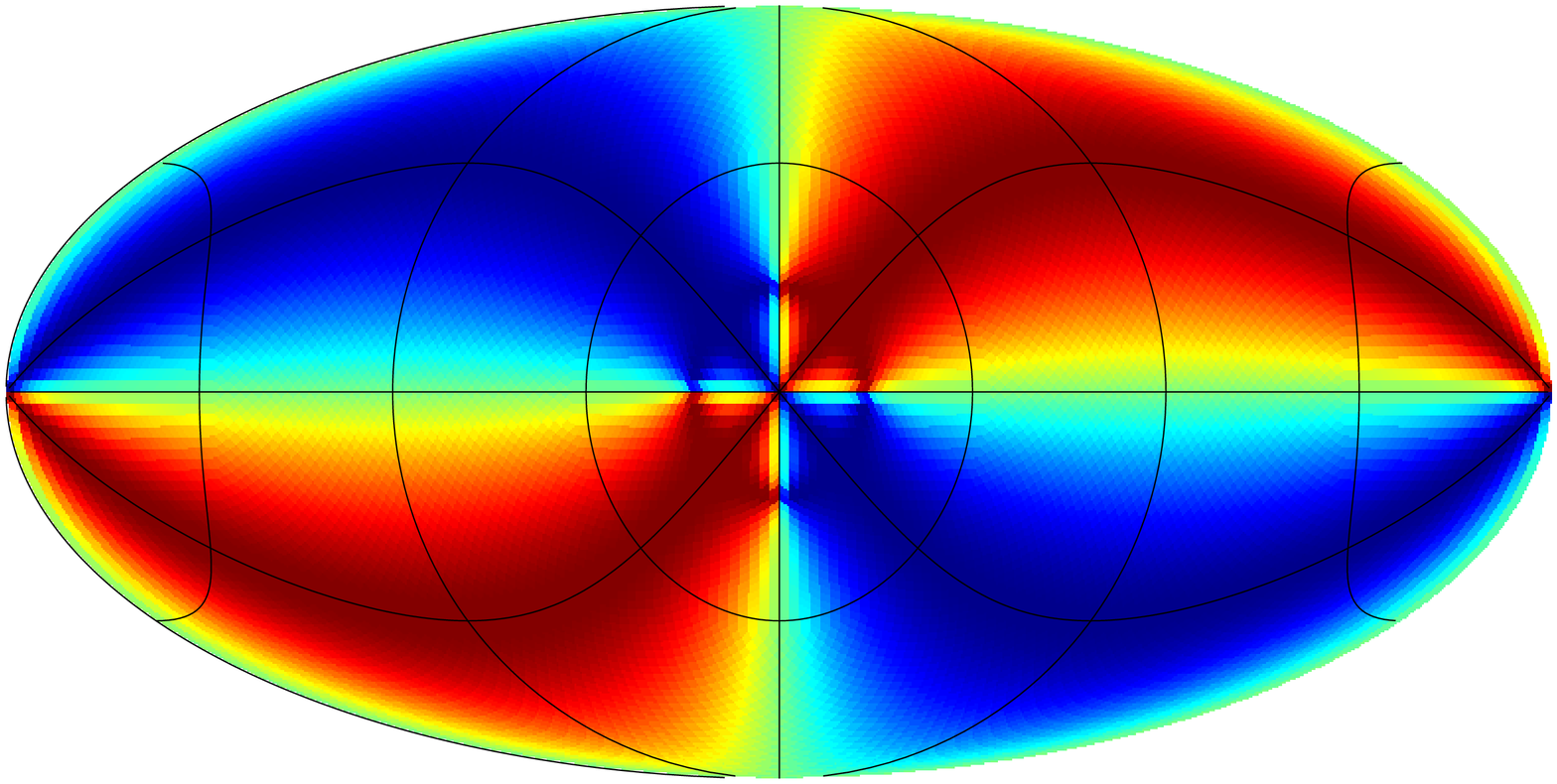}
   \includegraphics[scale=0.15,angle=90]{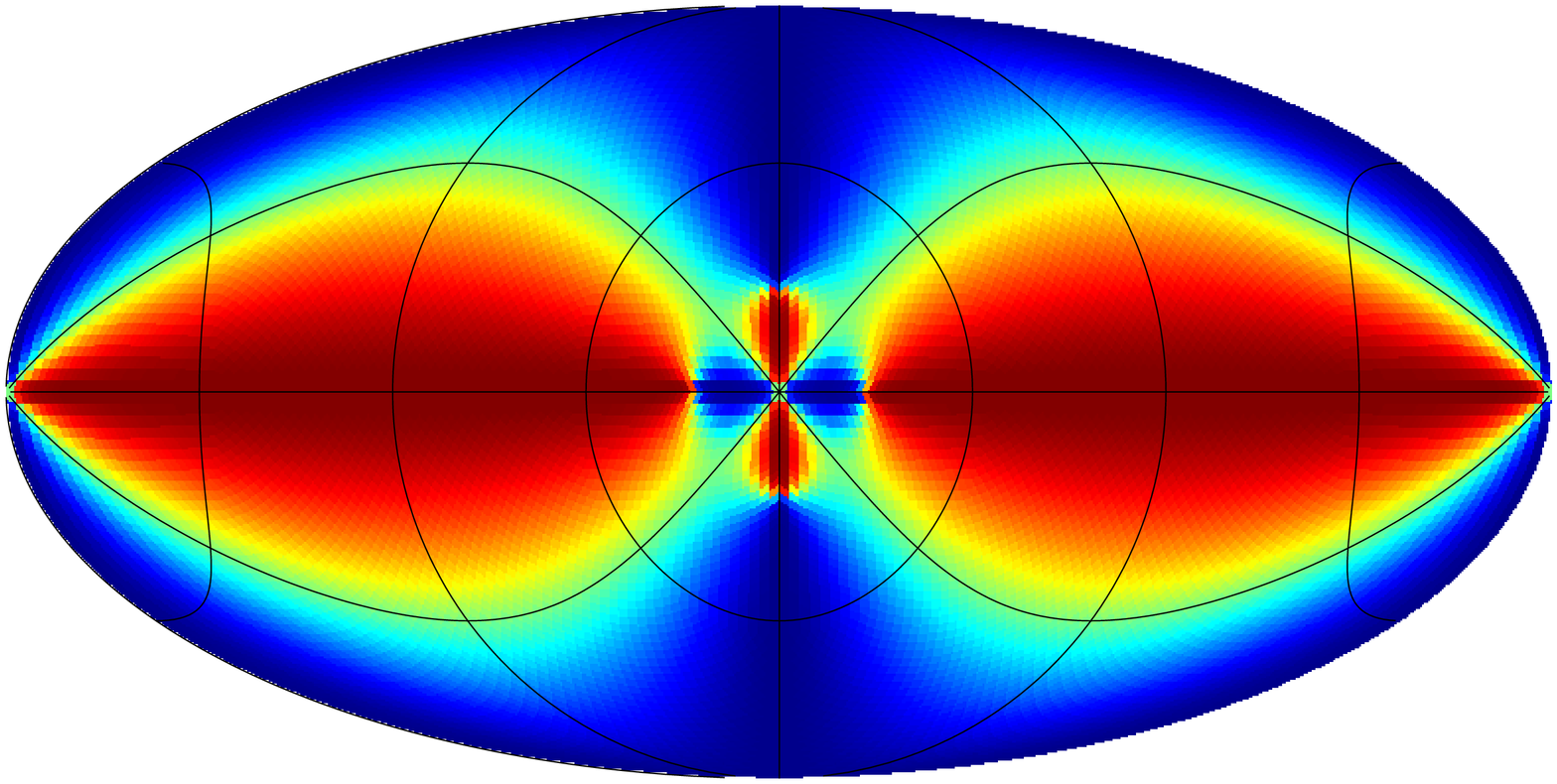}\\
  \includegraphics[scale=0.15,angle=90]{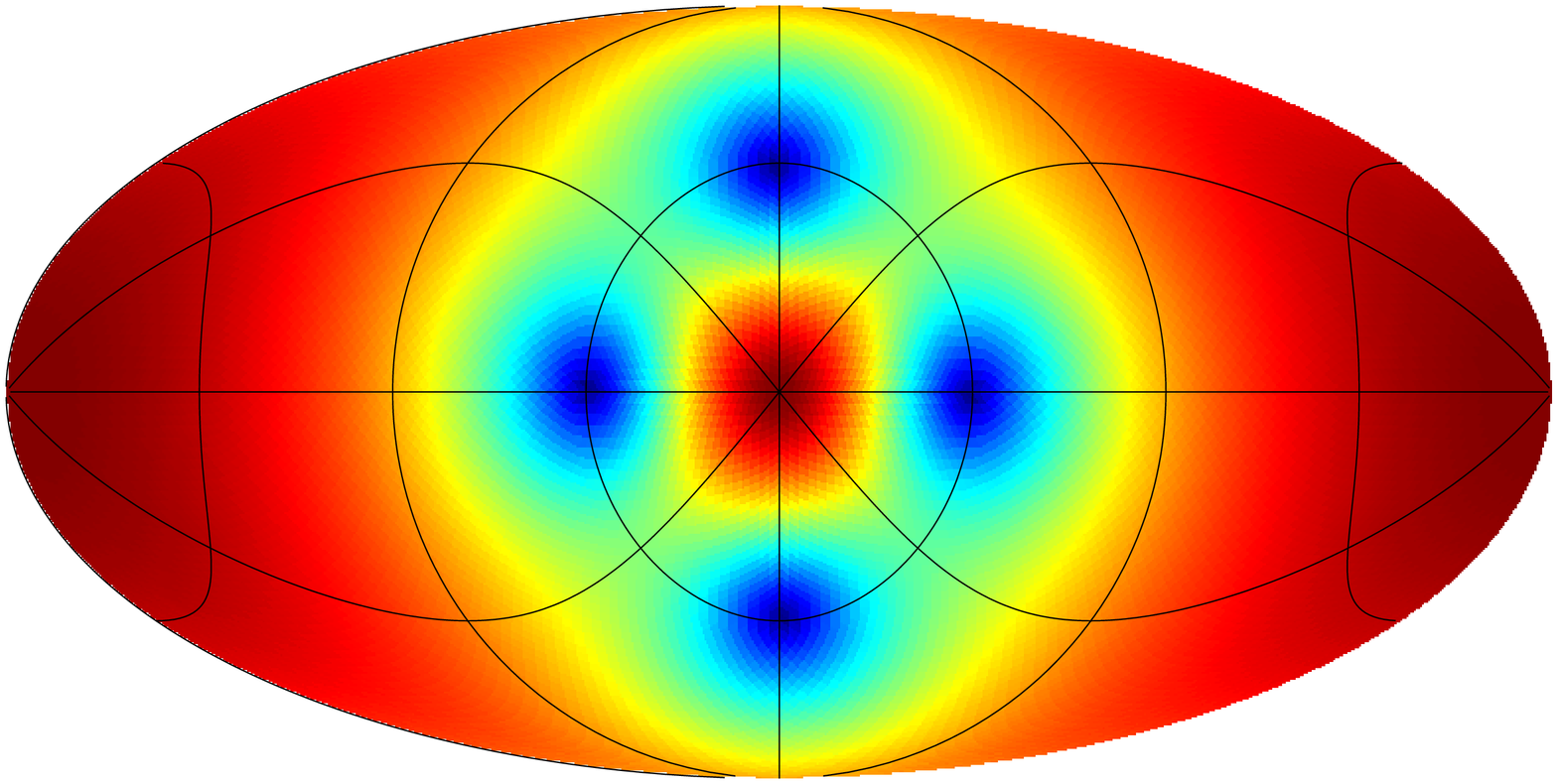}
    \includegraphics[scale=0.15,angle=90]{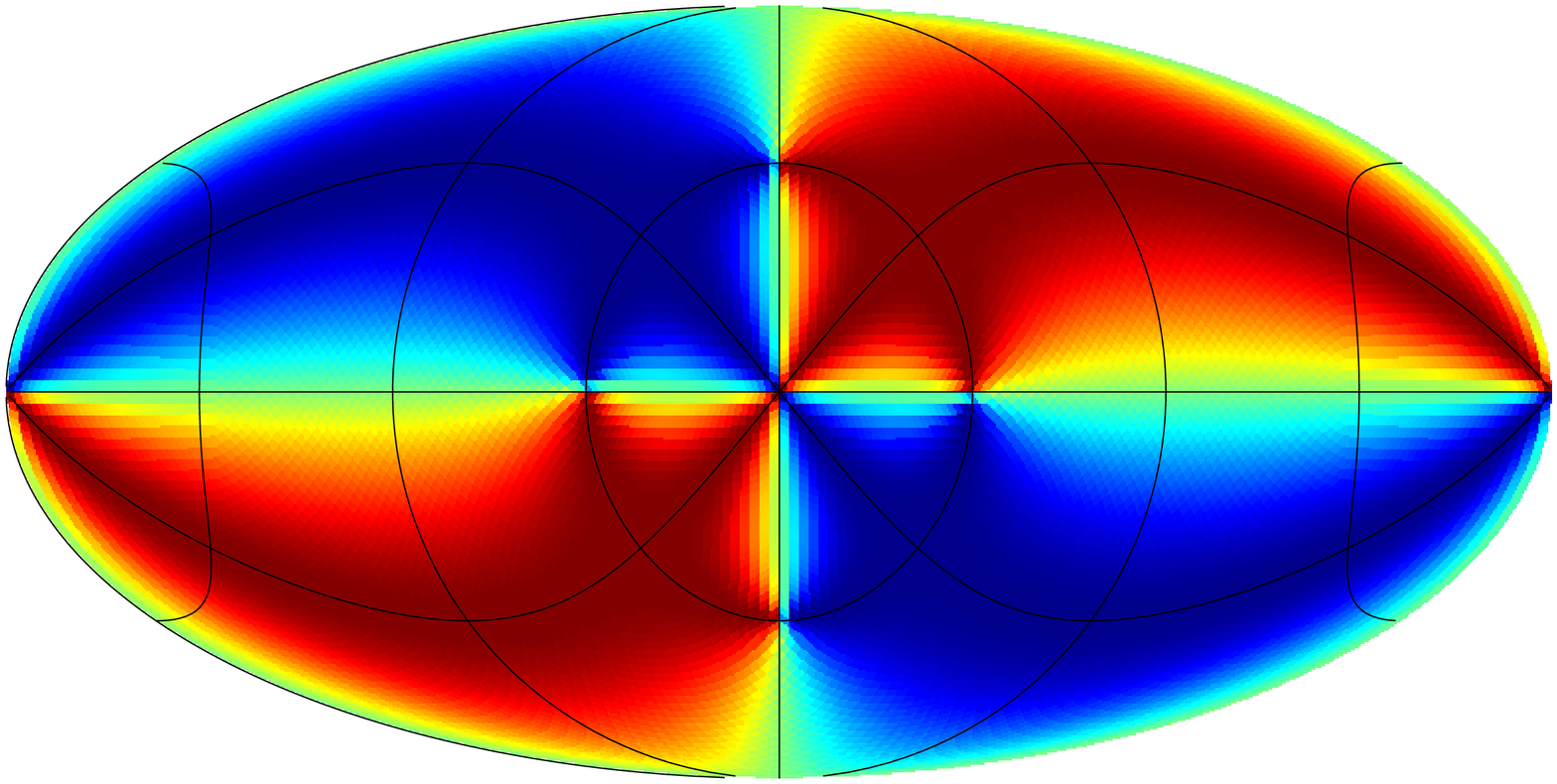}
   \includegraphics[scale=0.15,angle=90]{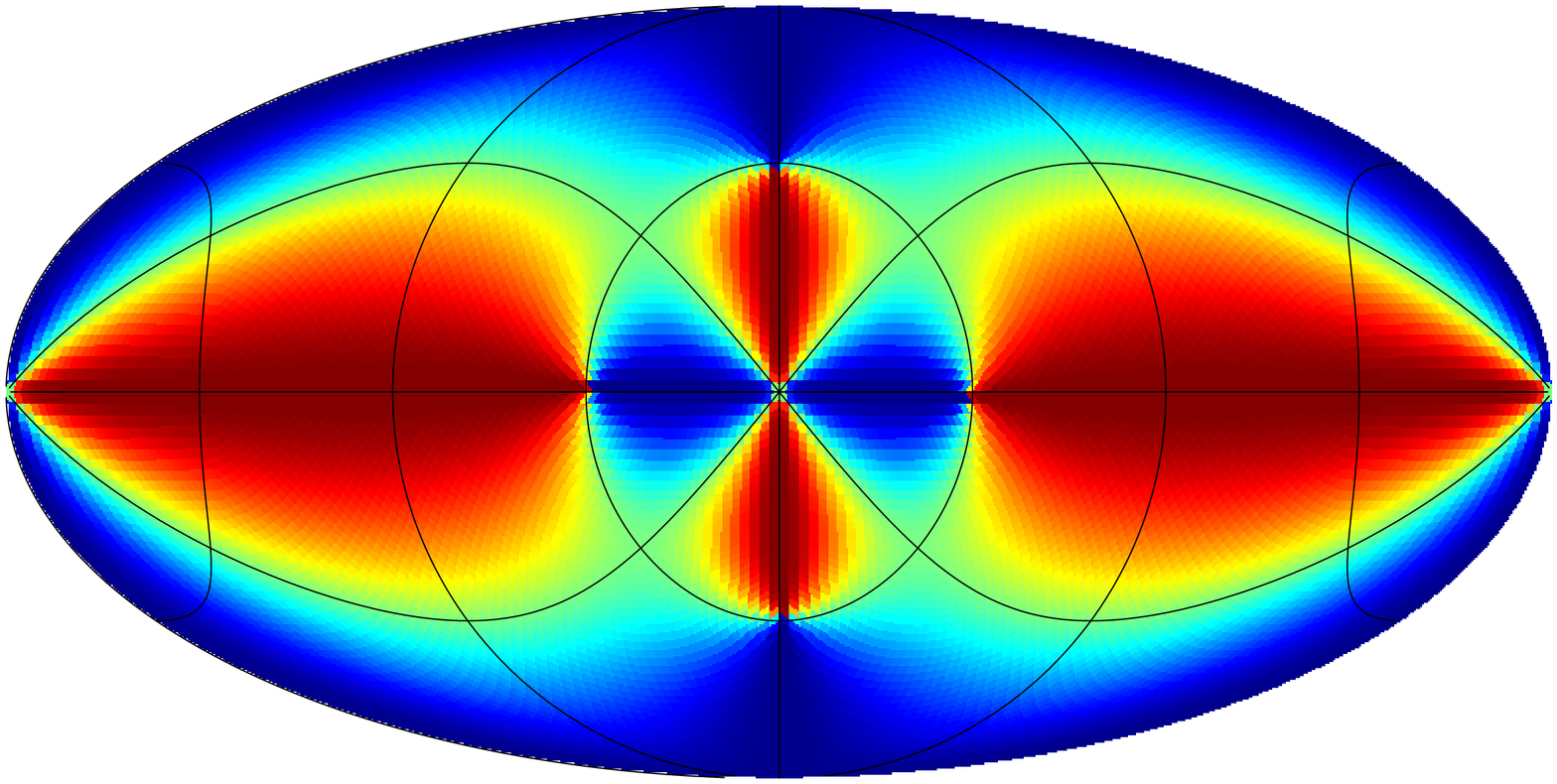}\\
  \includegraphics[scale=0.15,angle=90]{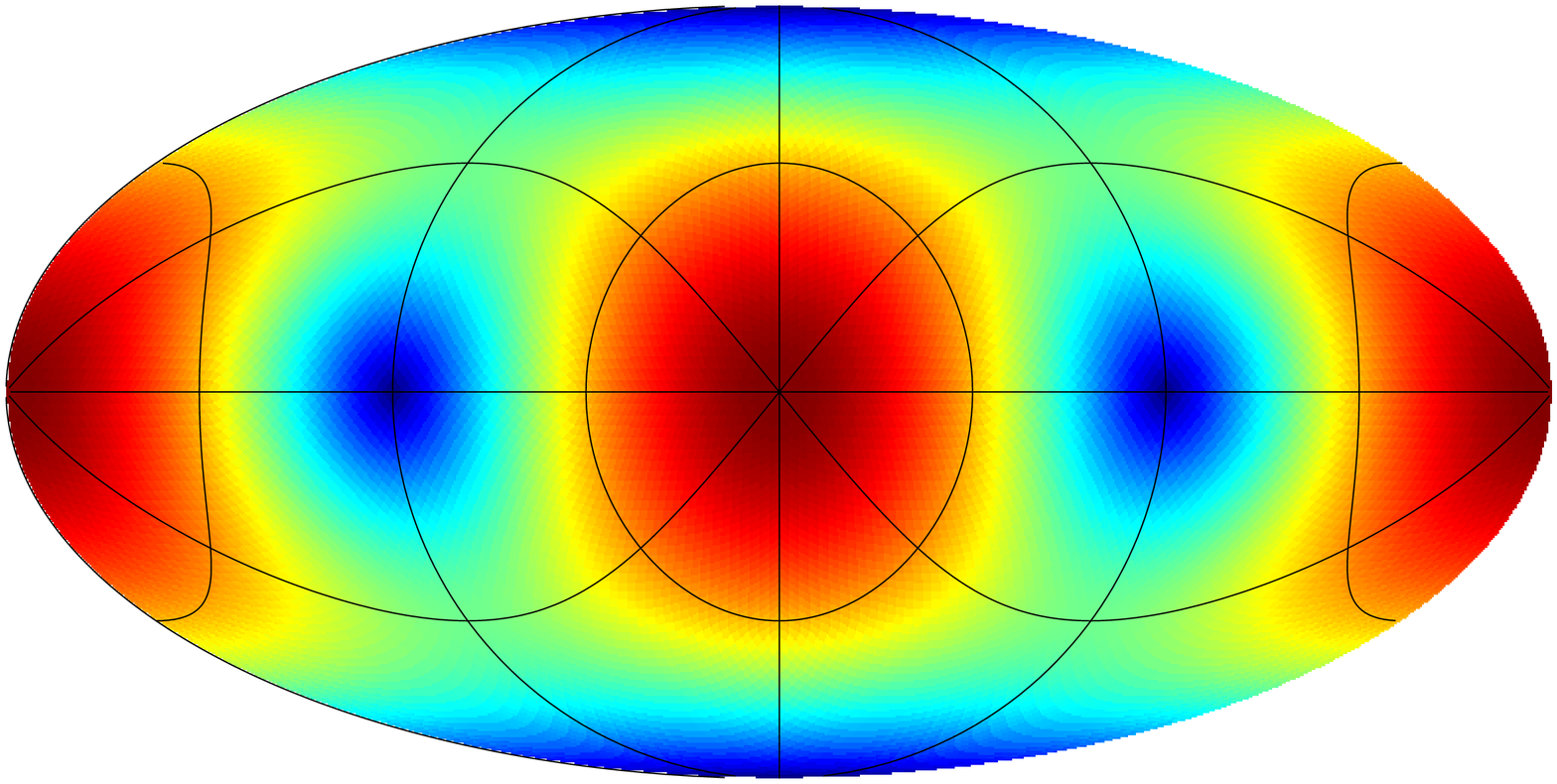}
    \includegraphics[scale=0.15,angle=90]{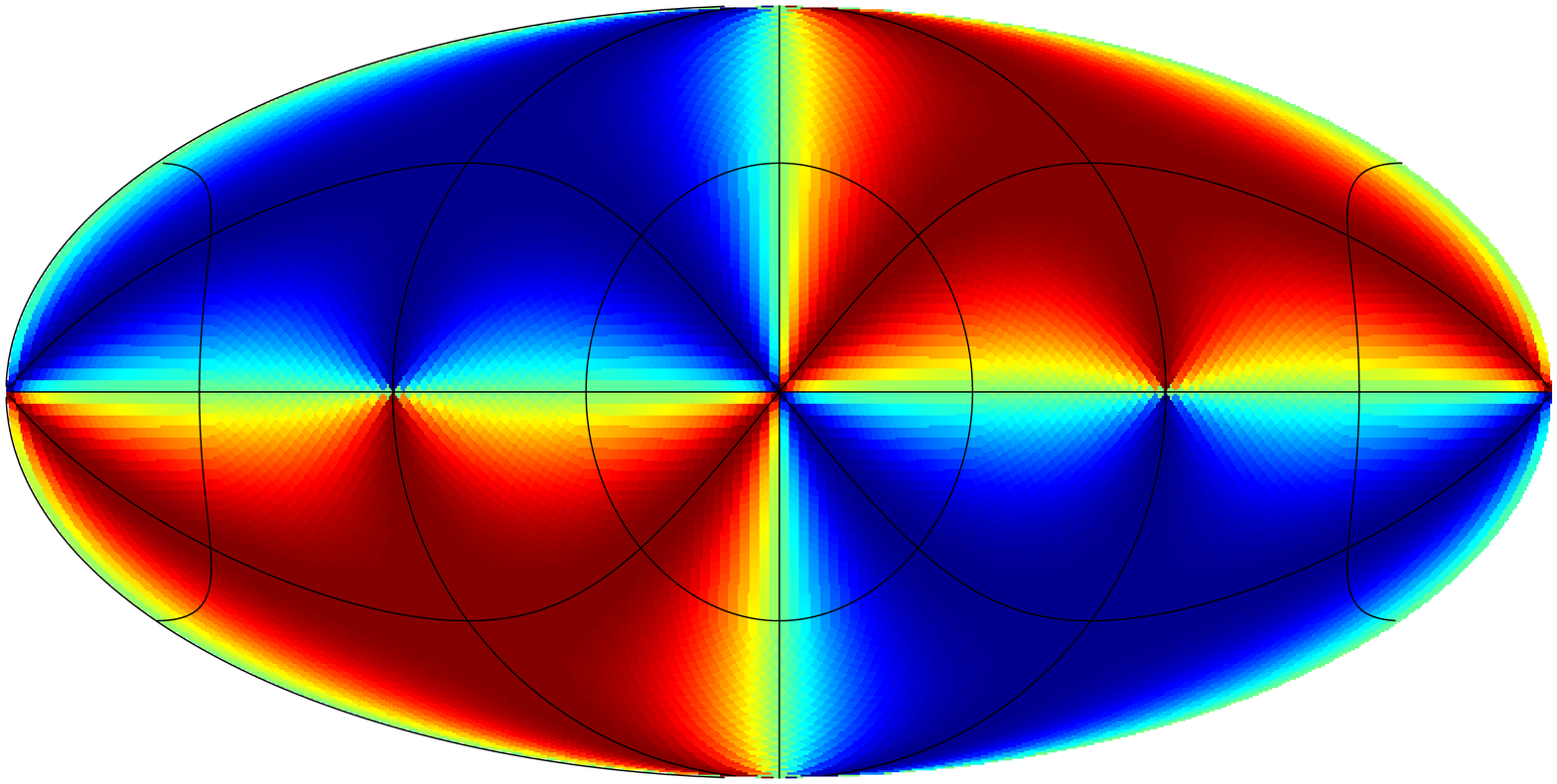}
   \includegraphics[scale=0.15,angle=90]{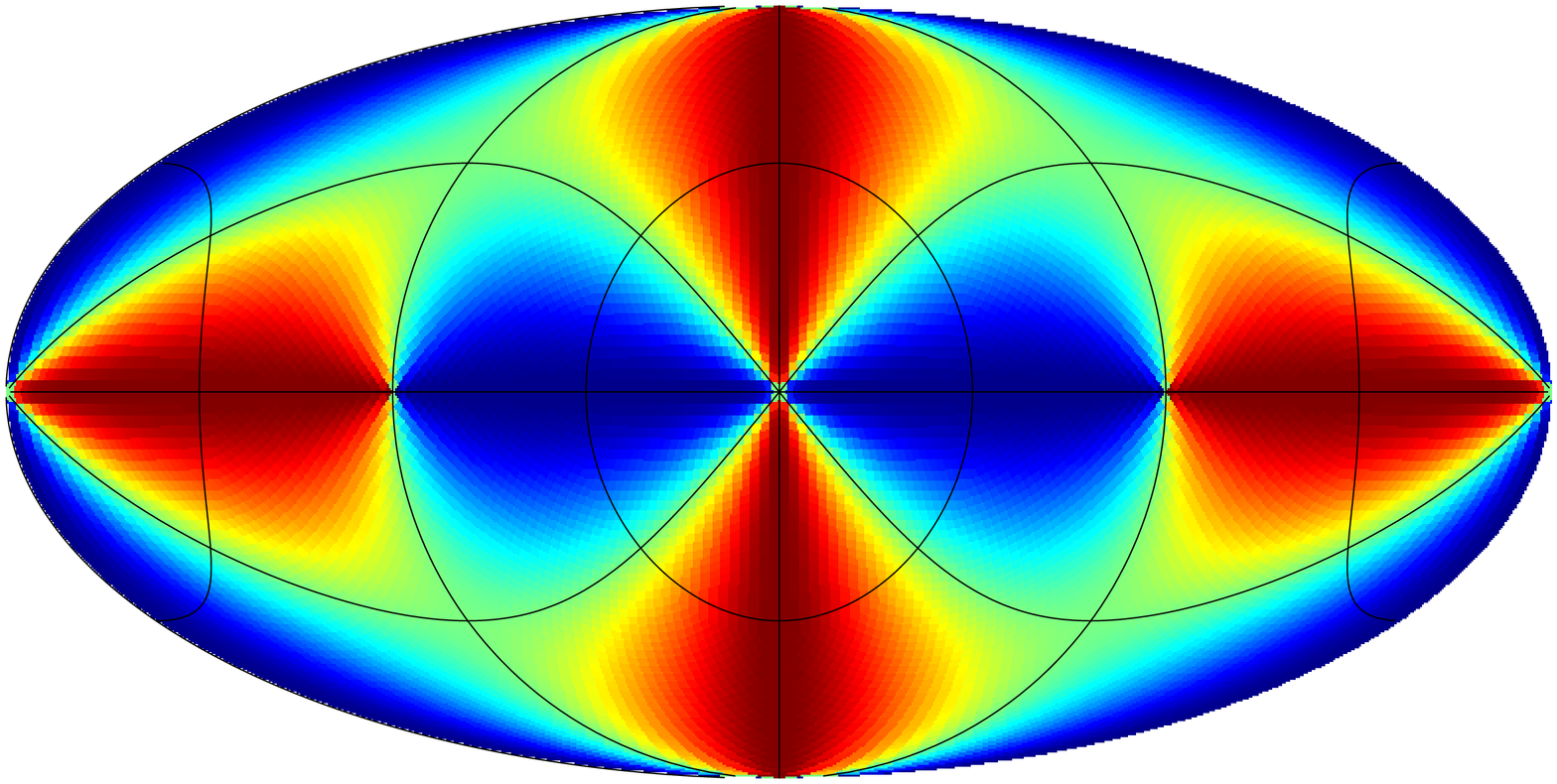}\\
   \end{center}
 \caption{The time evolution of  polarization maps for Bianchi V: degree of linear polarization (left); Stokes parameter Q (middle); and U(right). Time increases from bottom to top.}\label{PVz}
\end{figure}

Figure 5 shows an example of Bianchi Type VII$_h$. Note the
prodigious twisting of the polarized component of the radiation
field, as well as a concentration of the degree of polarization
defined by Equation (\ref{Polarization_degree}). This case
generically produces a large amount of mixing between $Q$ and $U$
during its time evolution.

\begin{figure}
\begin{center}
  \includegraphics[scale=0.15,angle=90]{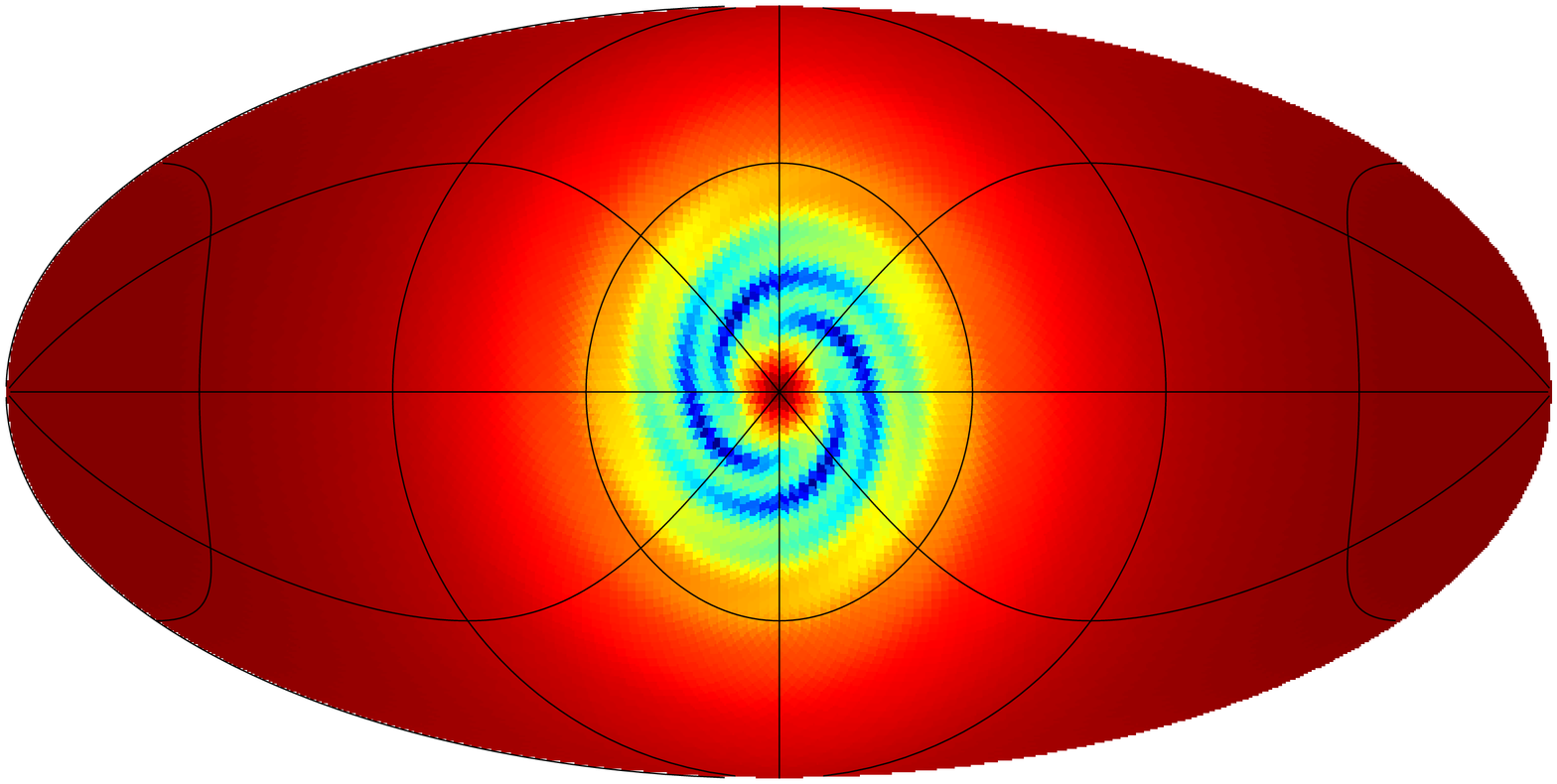}
  \includegraphics[scale=0.15,angle=90]{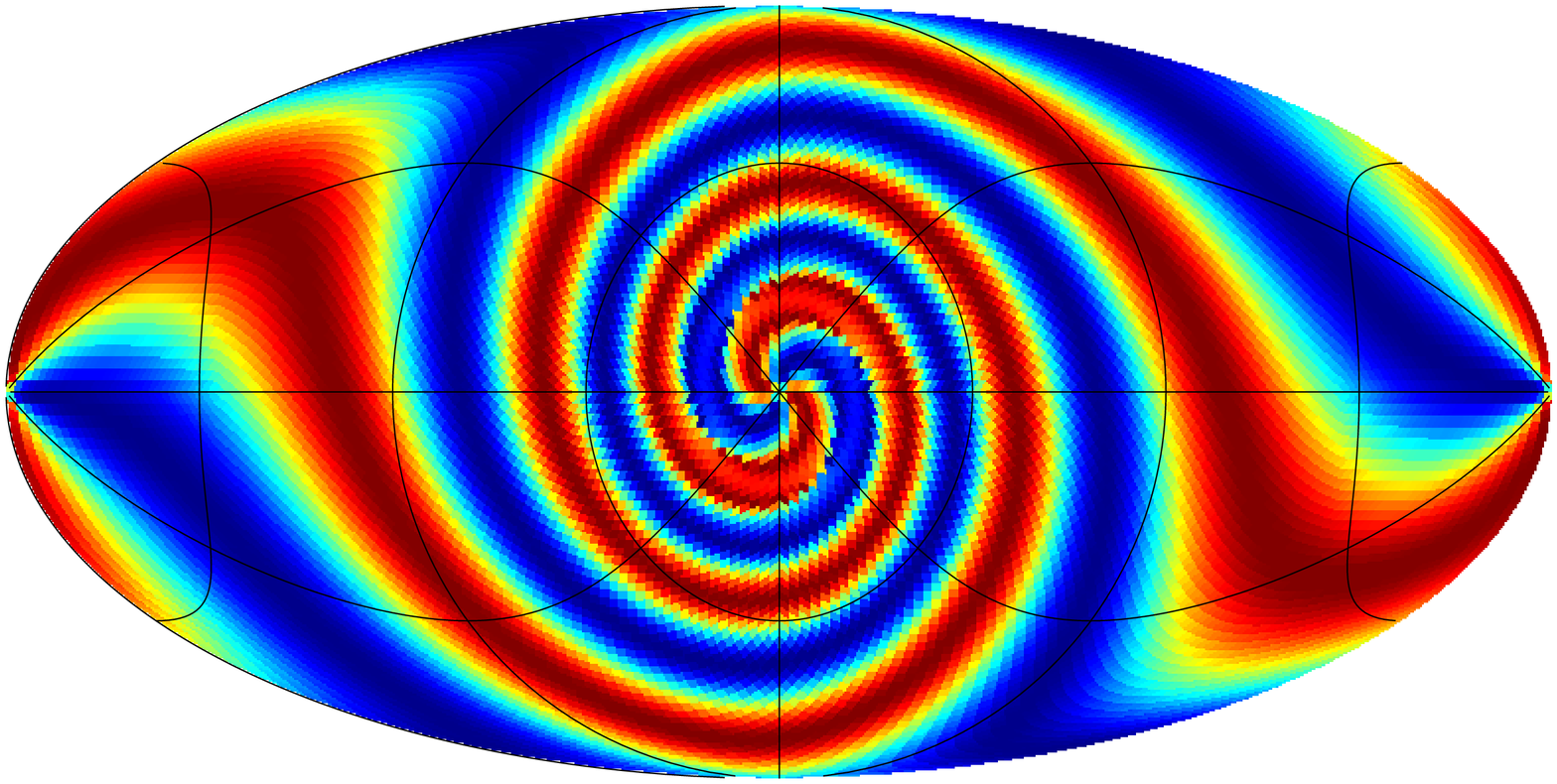}
   \includegraphics[scale=0.15,angle=90]{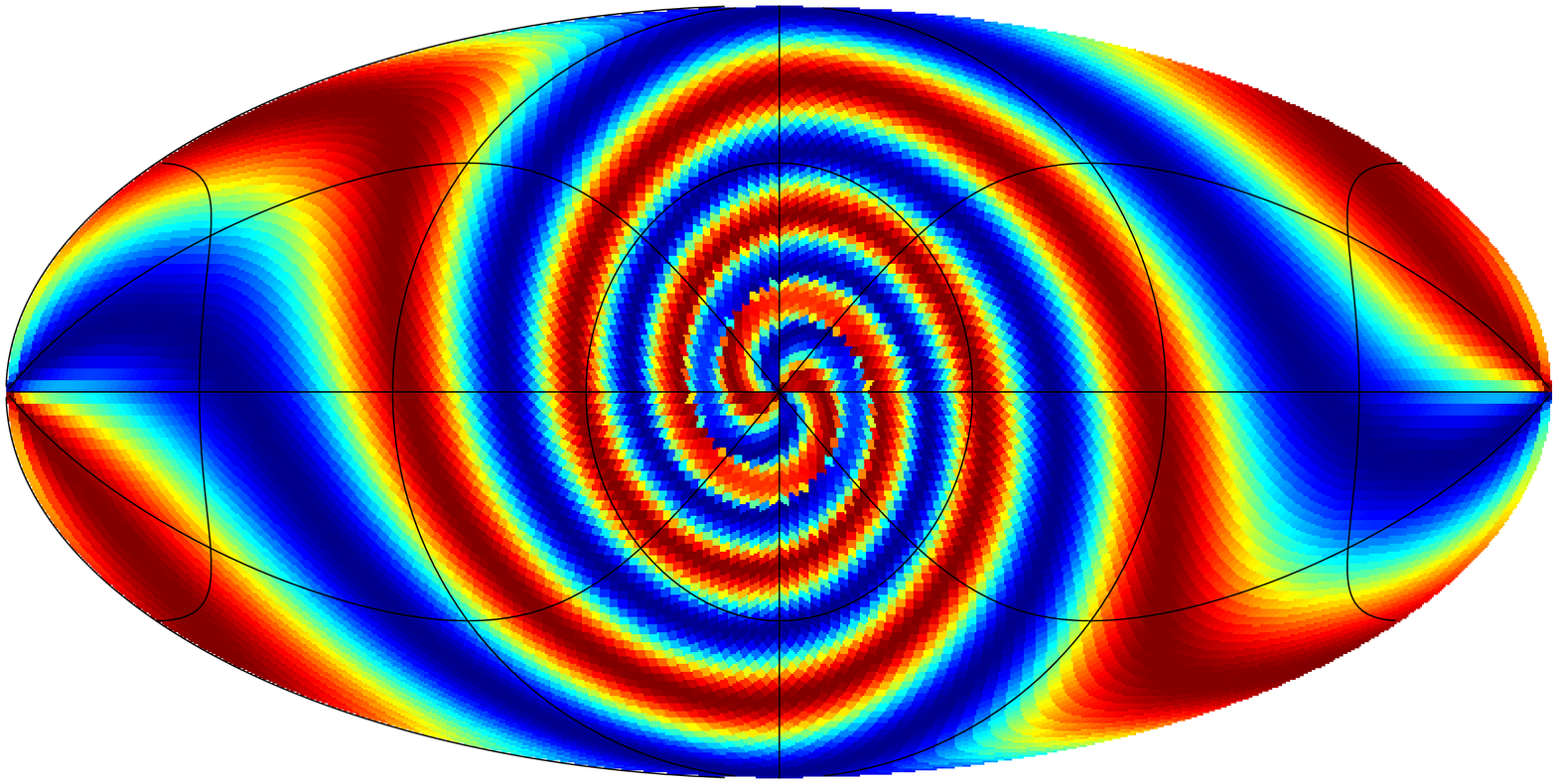}\\
  \includegraphics[scale=0.15,angle=90]{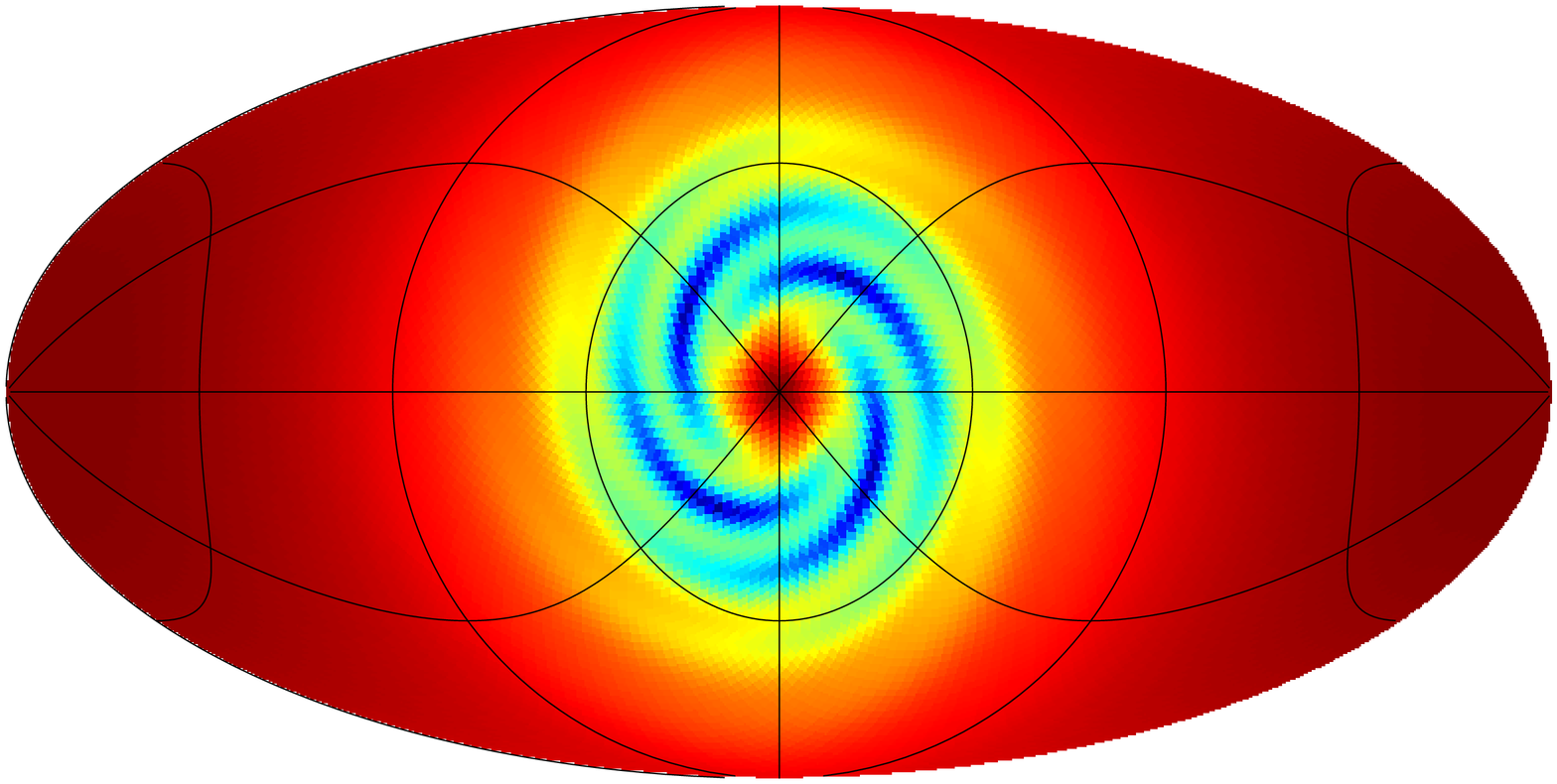}
    \includegraphics[scale=0.15,angle=90]{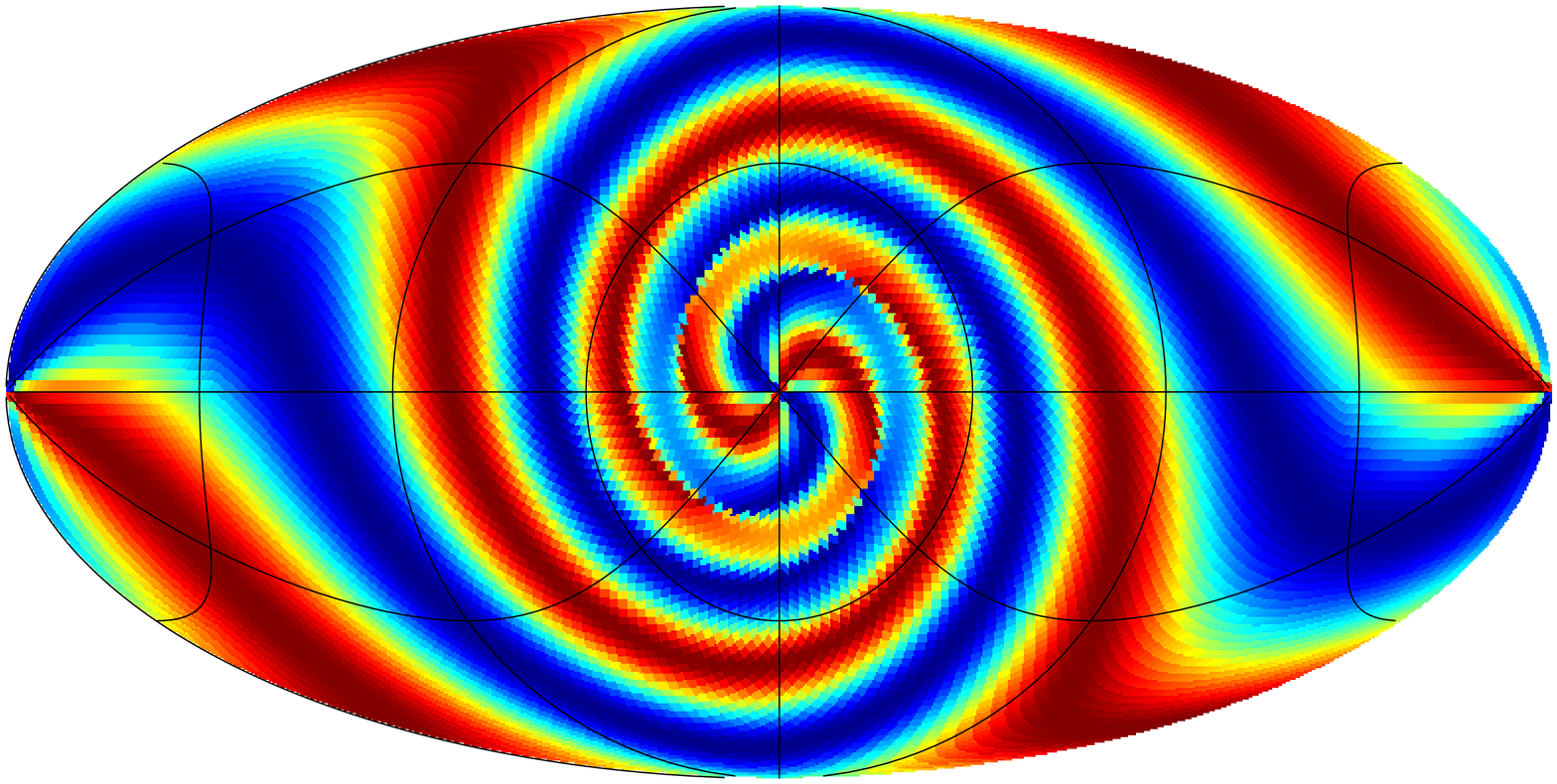}
   \includegraphics[scale=0.15,angle=90]{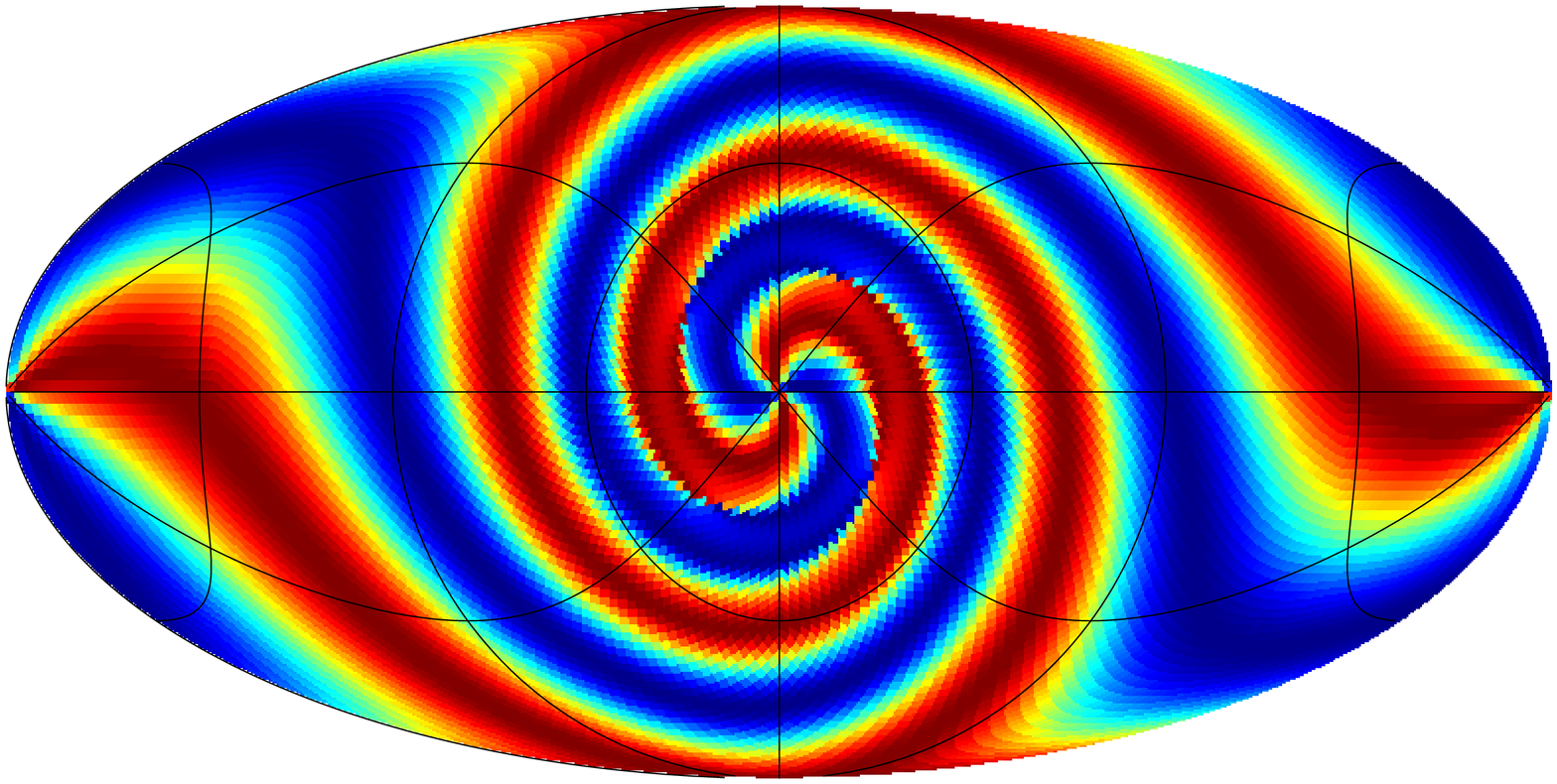}\\
  \includegraphics[scale=0.15,angle=90]{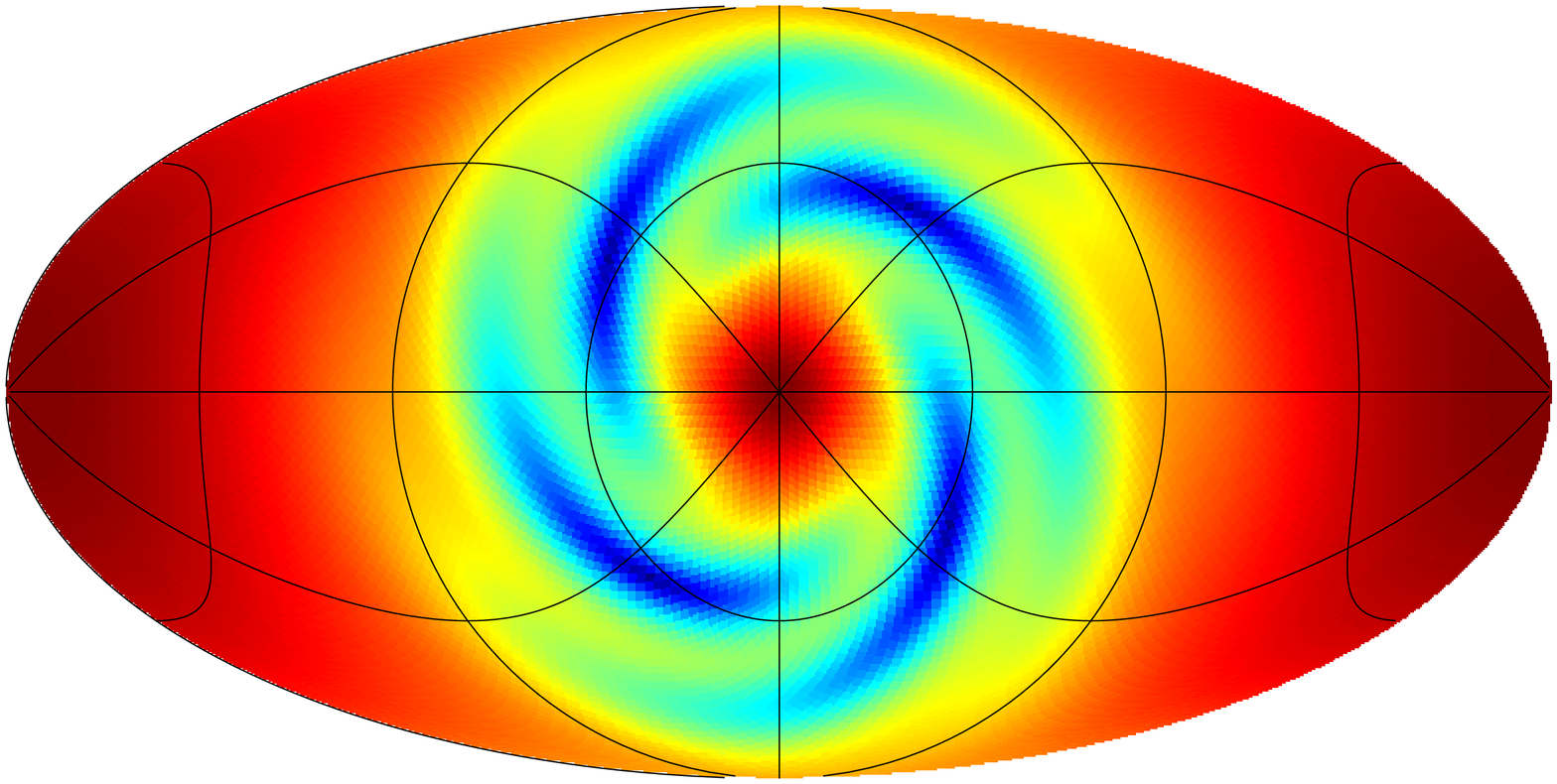}
    \includegraphics[scale=0.15,angle=90]{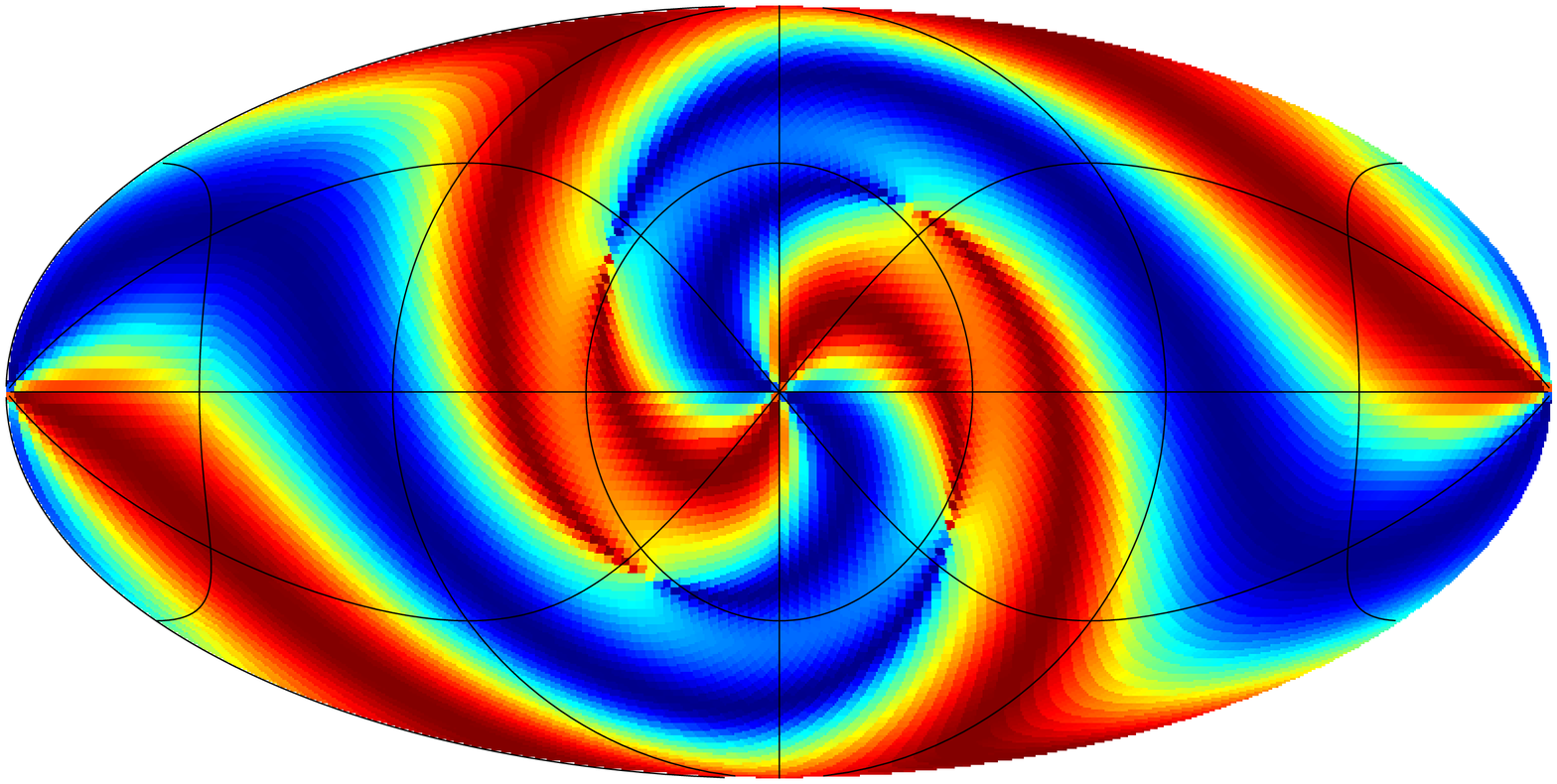}
   \includegraphics[scale=0.15,angle=90]{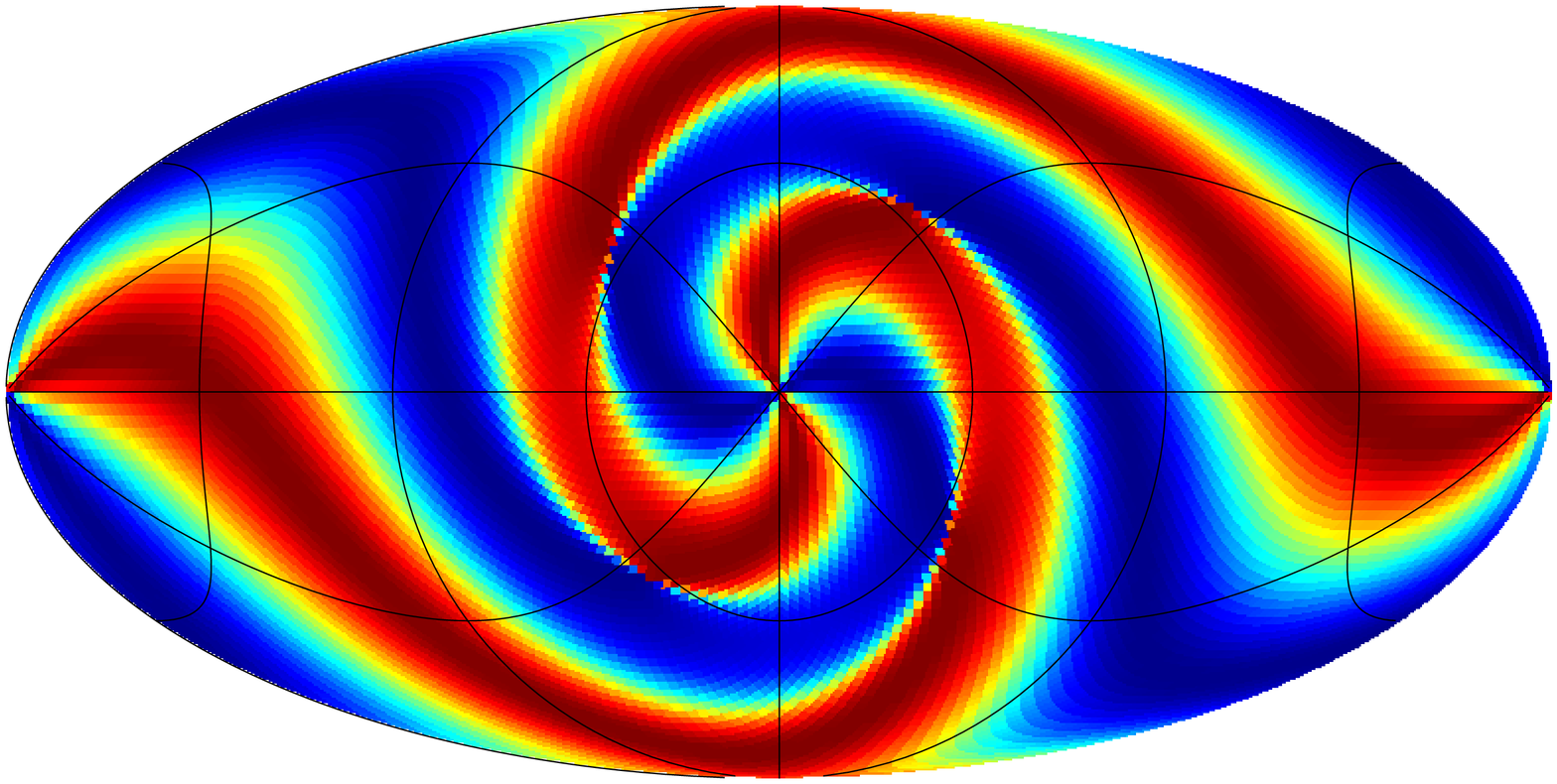}\\
  \includegraphics[scale=0.15,angle=90]{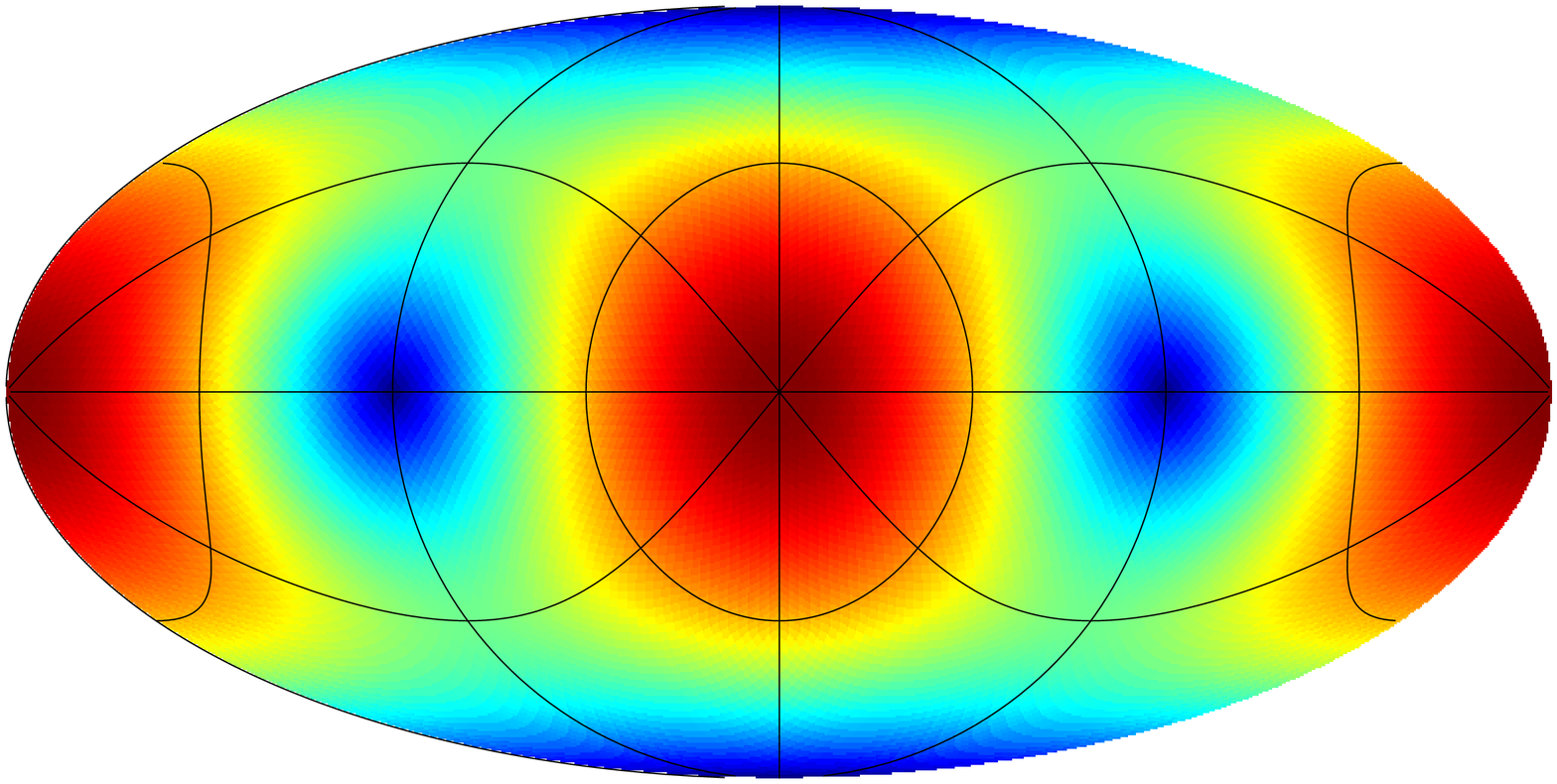}
    \includegraphics[scale=0.15,angle=90]{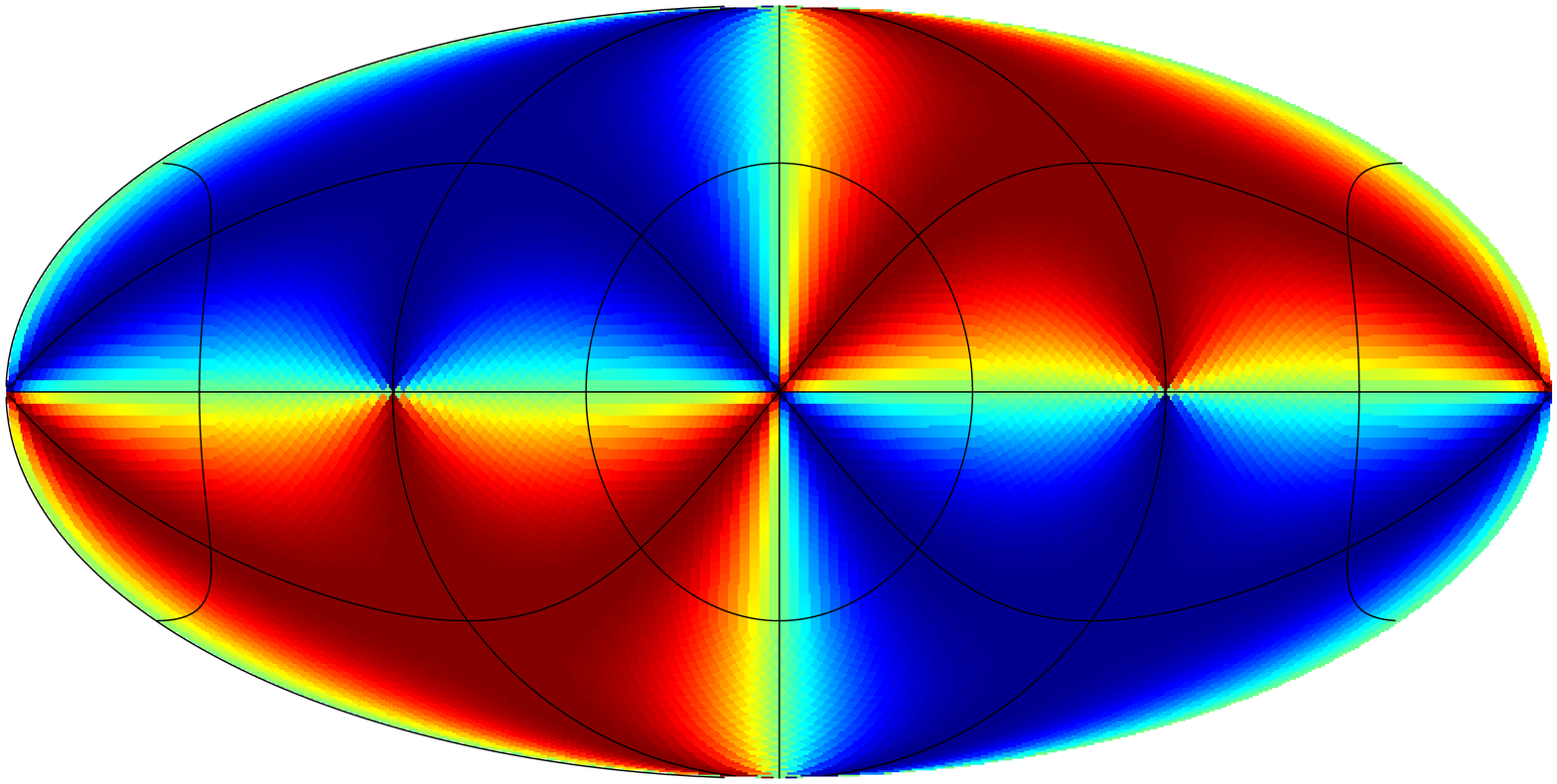}
   \includegraphics[scale=0.15,angle=90]{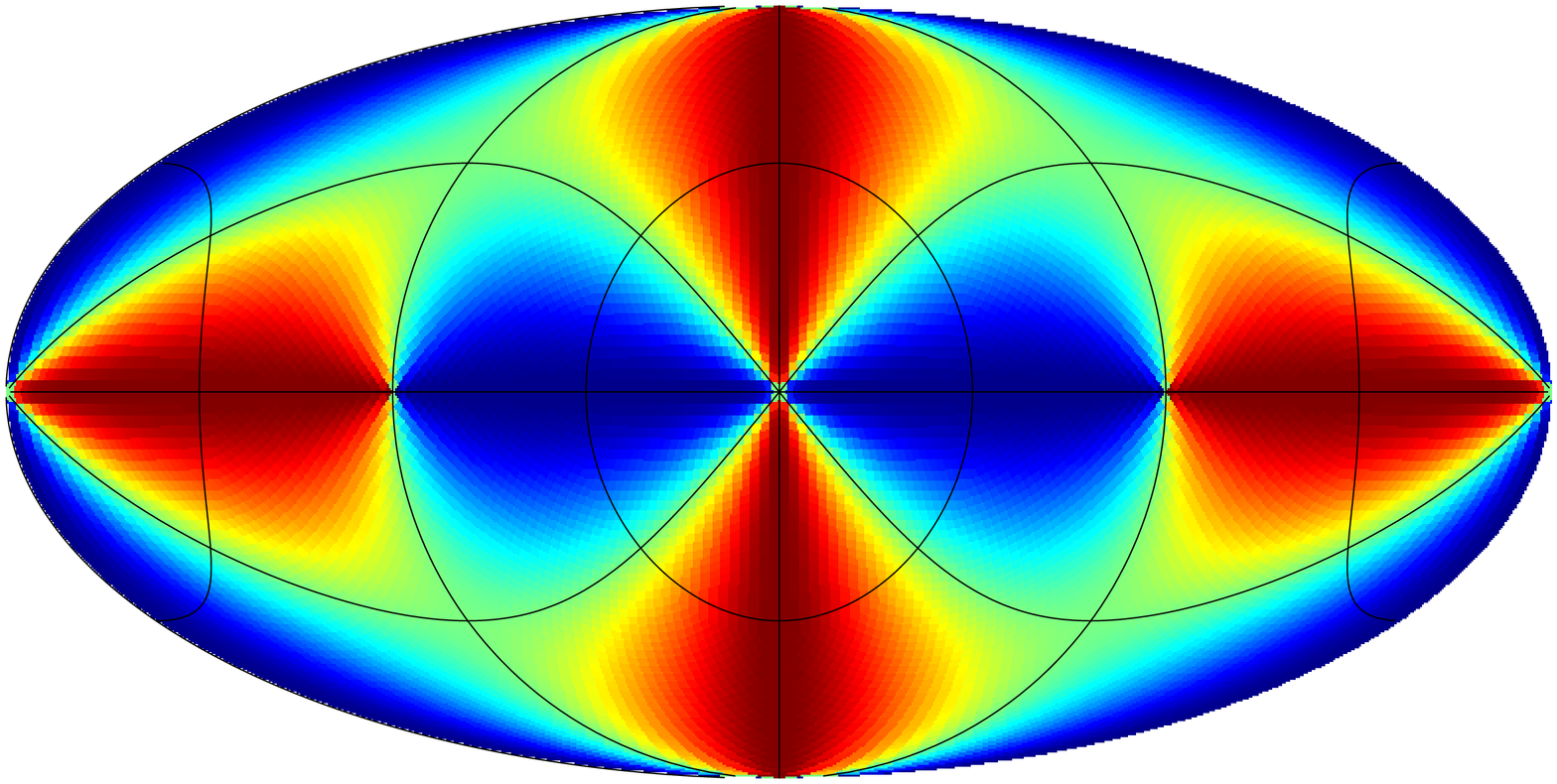}\\
   \end{center}
 \caption{Same as Figure \ref{PVz}, but for Bianchi type VII$_h$.}
\end{figure}

Figure 6 shows, not unexpectedly, that Bianchi Type VII$_0$ produces
a similar interweaving of the $Q$ and $U$ configurations, but
without the focussing effect in the total polarized fraction.
\begin{figure}
 \begin{center}
  \includegraphics[scale=0.15,angle=90]{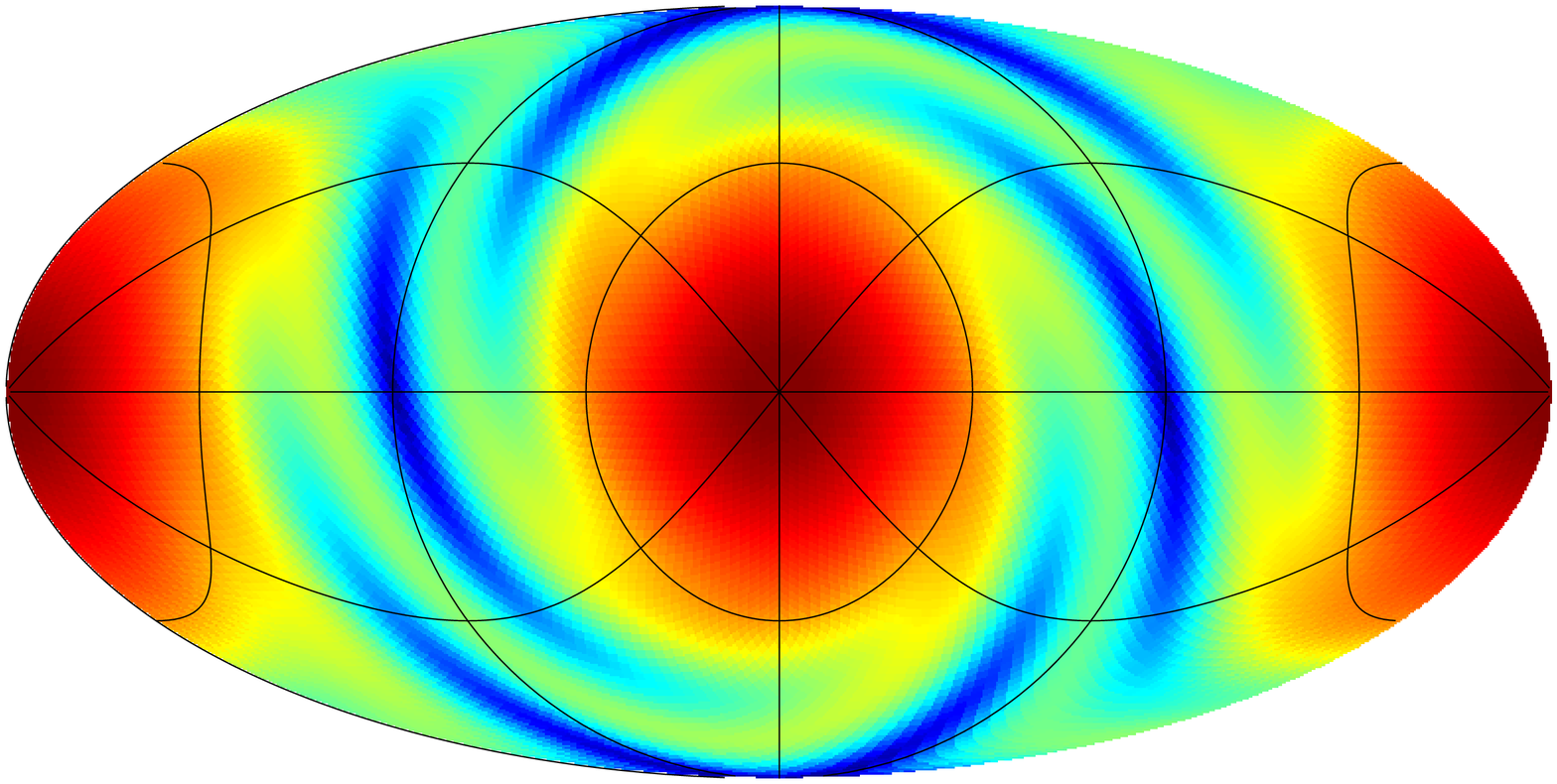}
  \includegraphics[scale=0.15,angle=90]{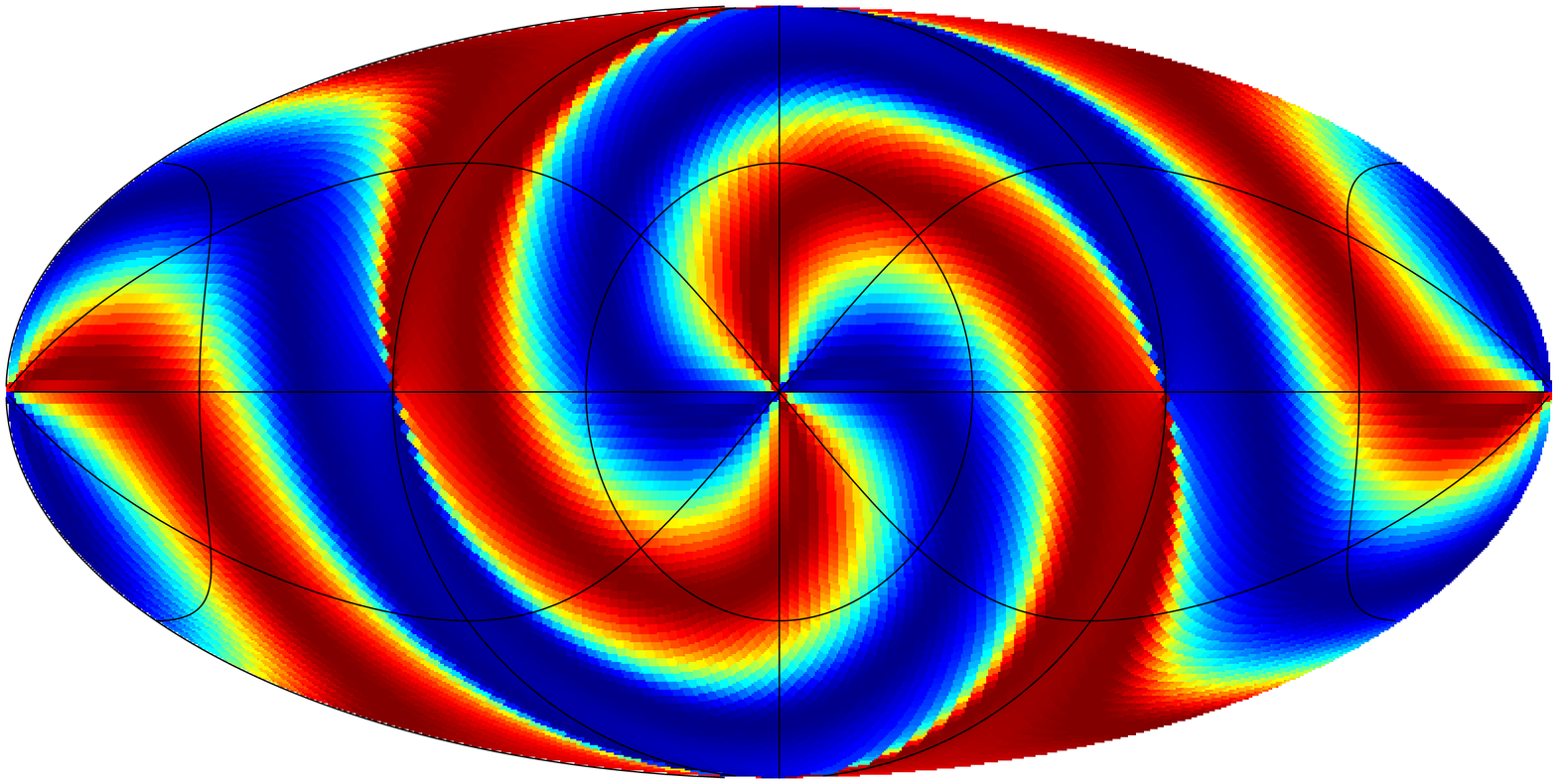}
   \includegraphics[scale=0.15,angle=90]{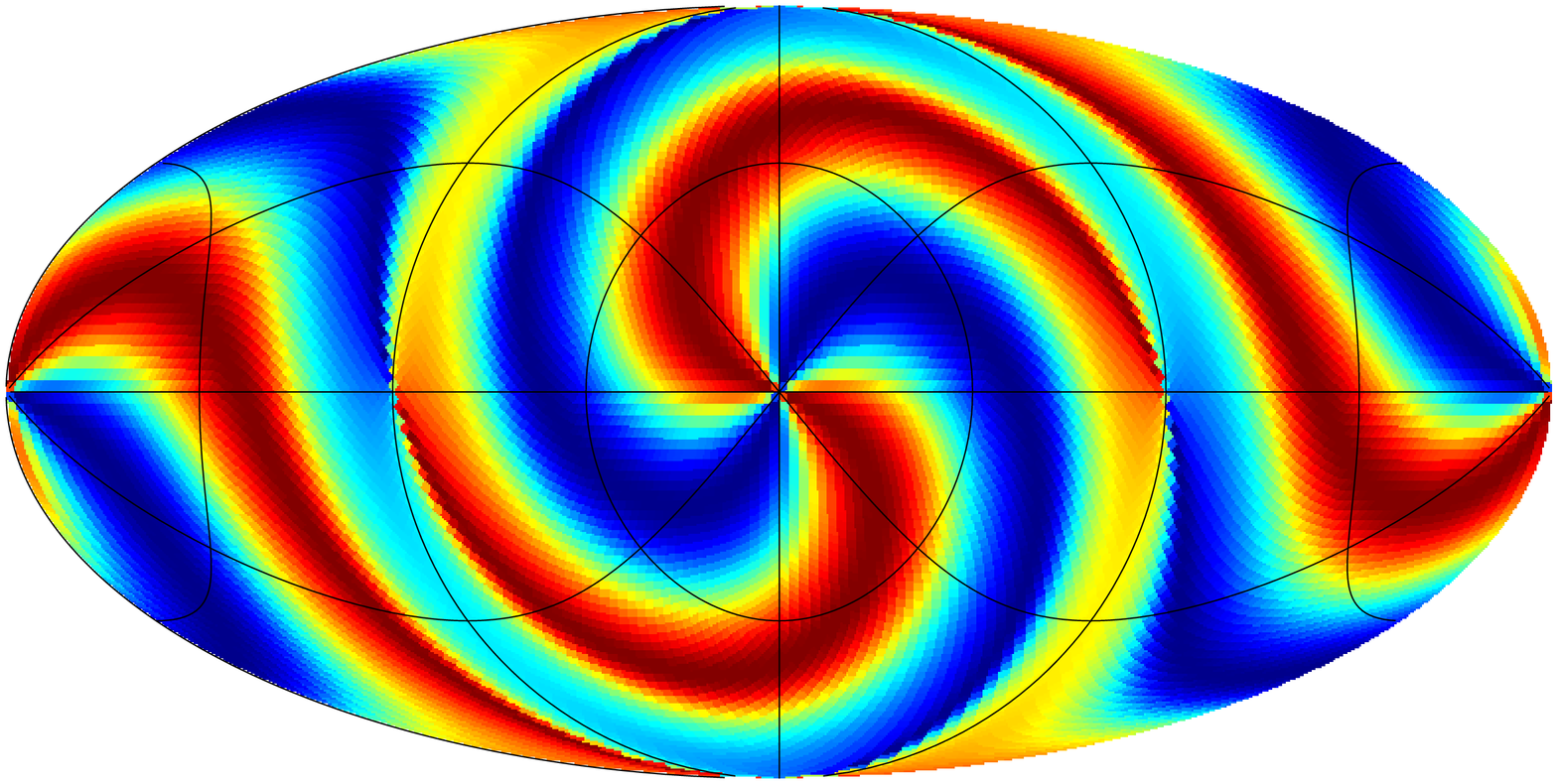}\\
  \includegraphics[scale=0.15,angle=90]{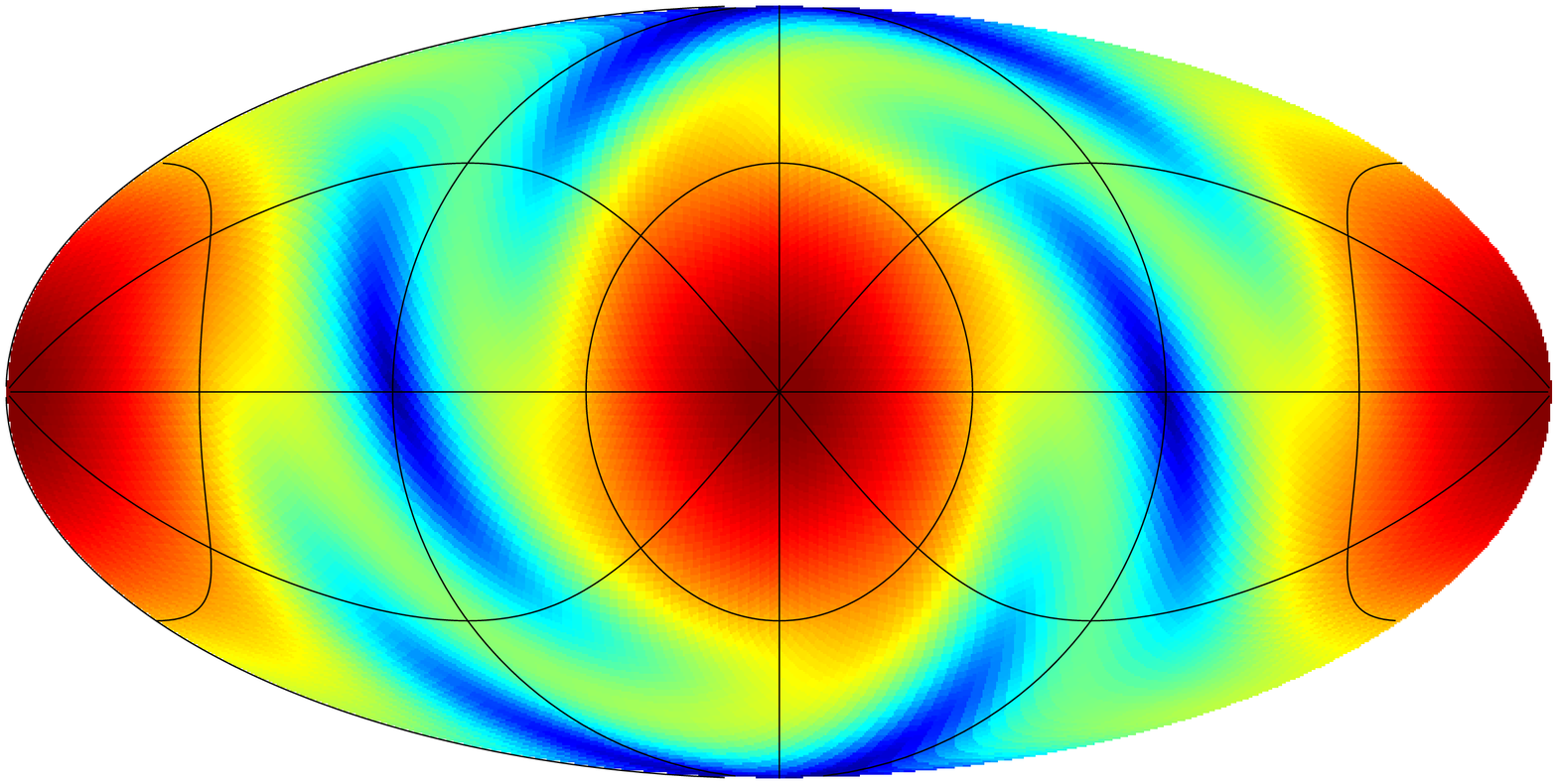}
    \includegraphics[scale=0.15,angle=90]{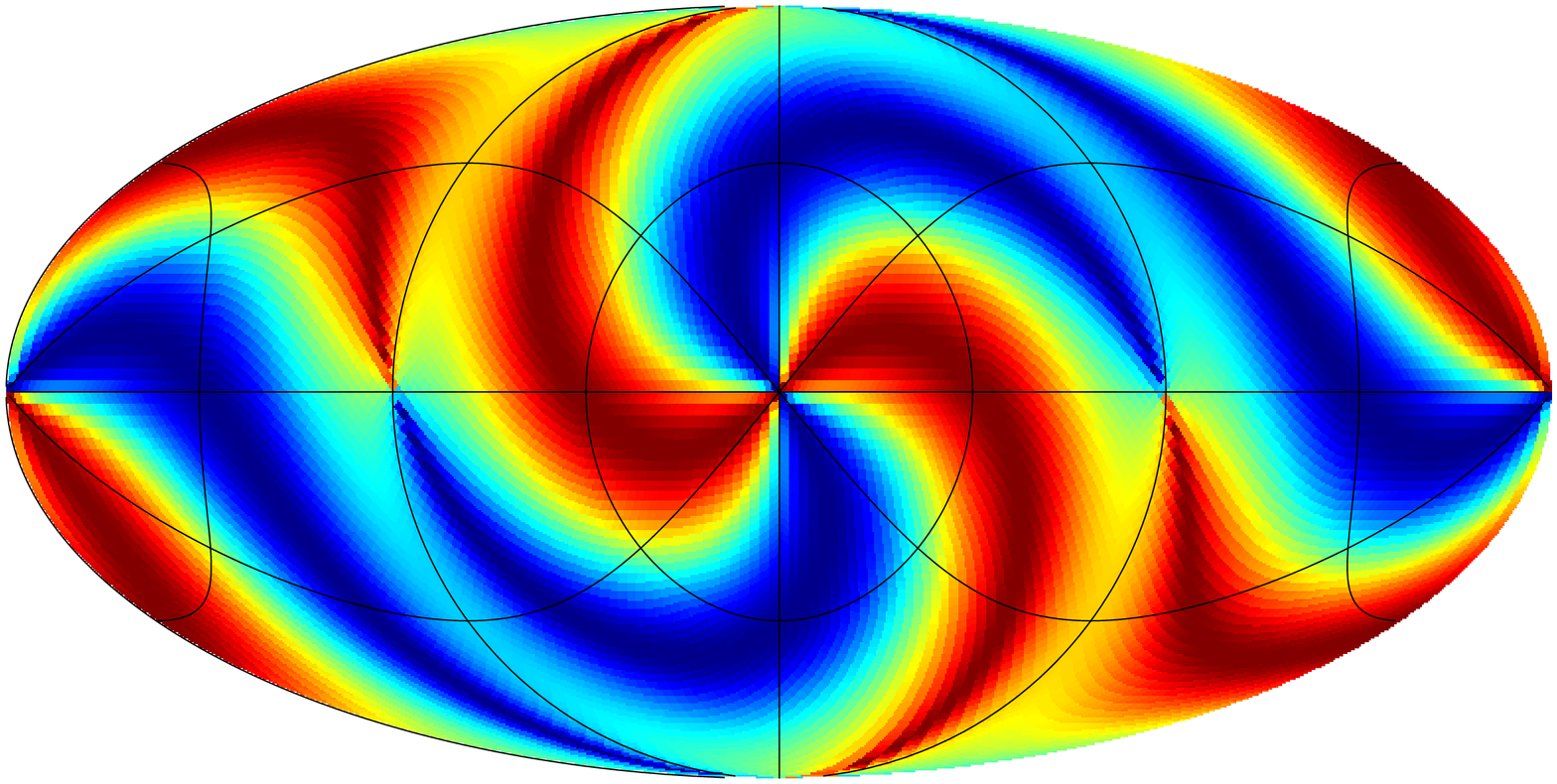}
   \includegraphics[scale=0.15,angle=90]{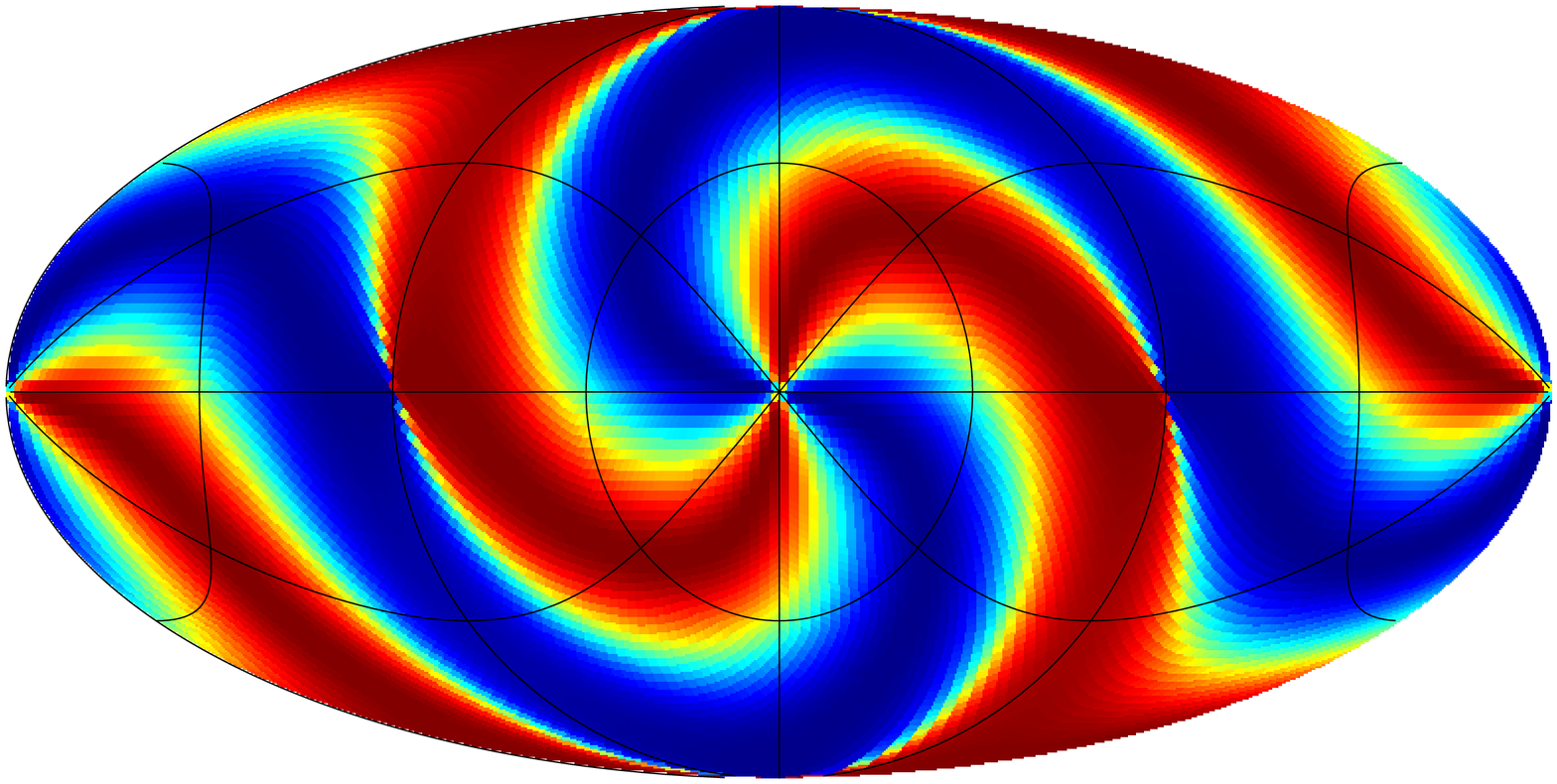}\\
  \includegraphics[scale=0.15,angle=90]{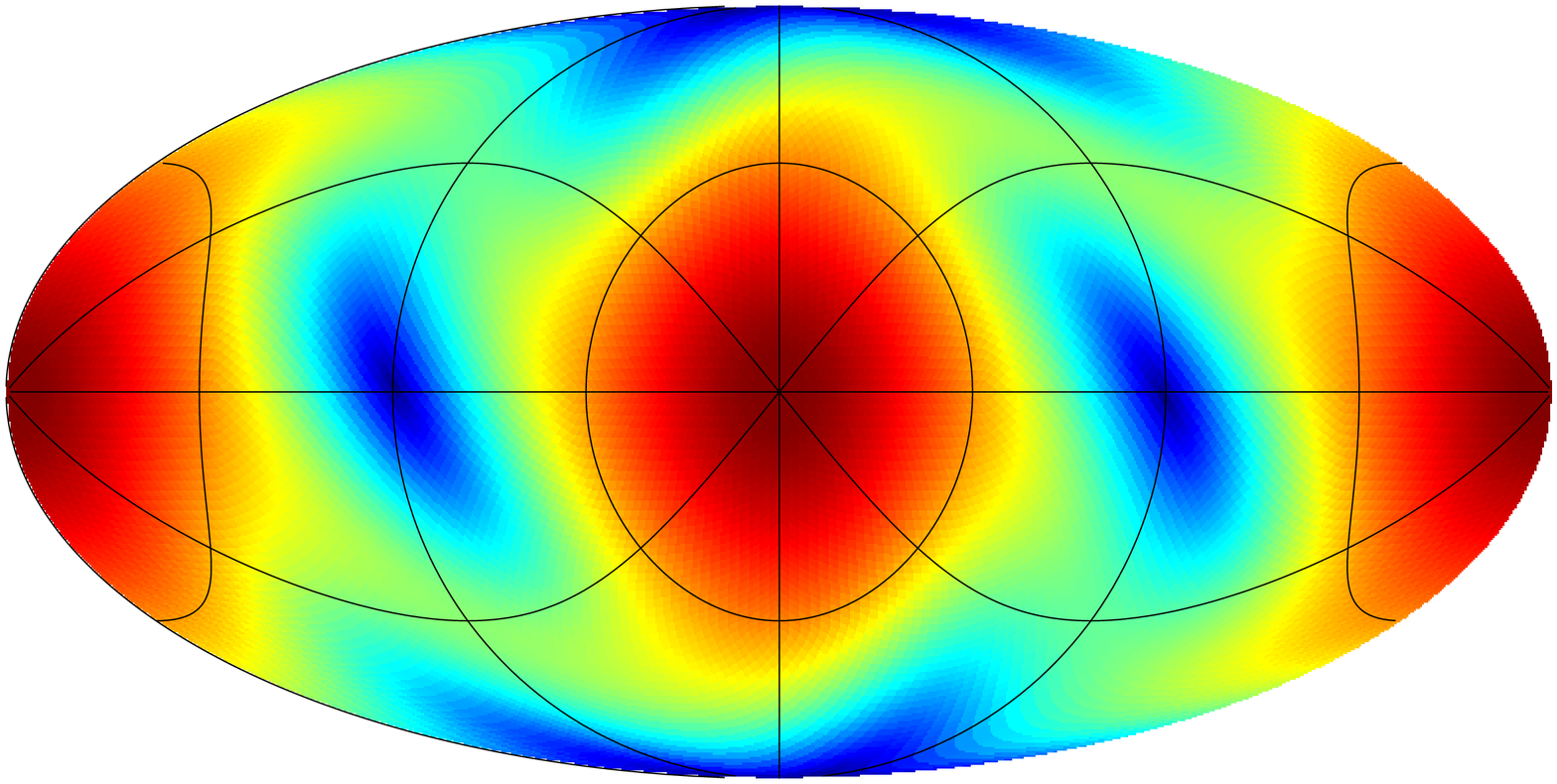}
    \includegraphics[scale=0.15,angle=90]{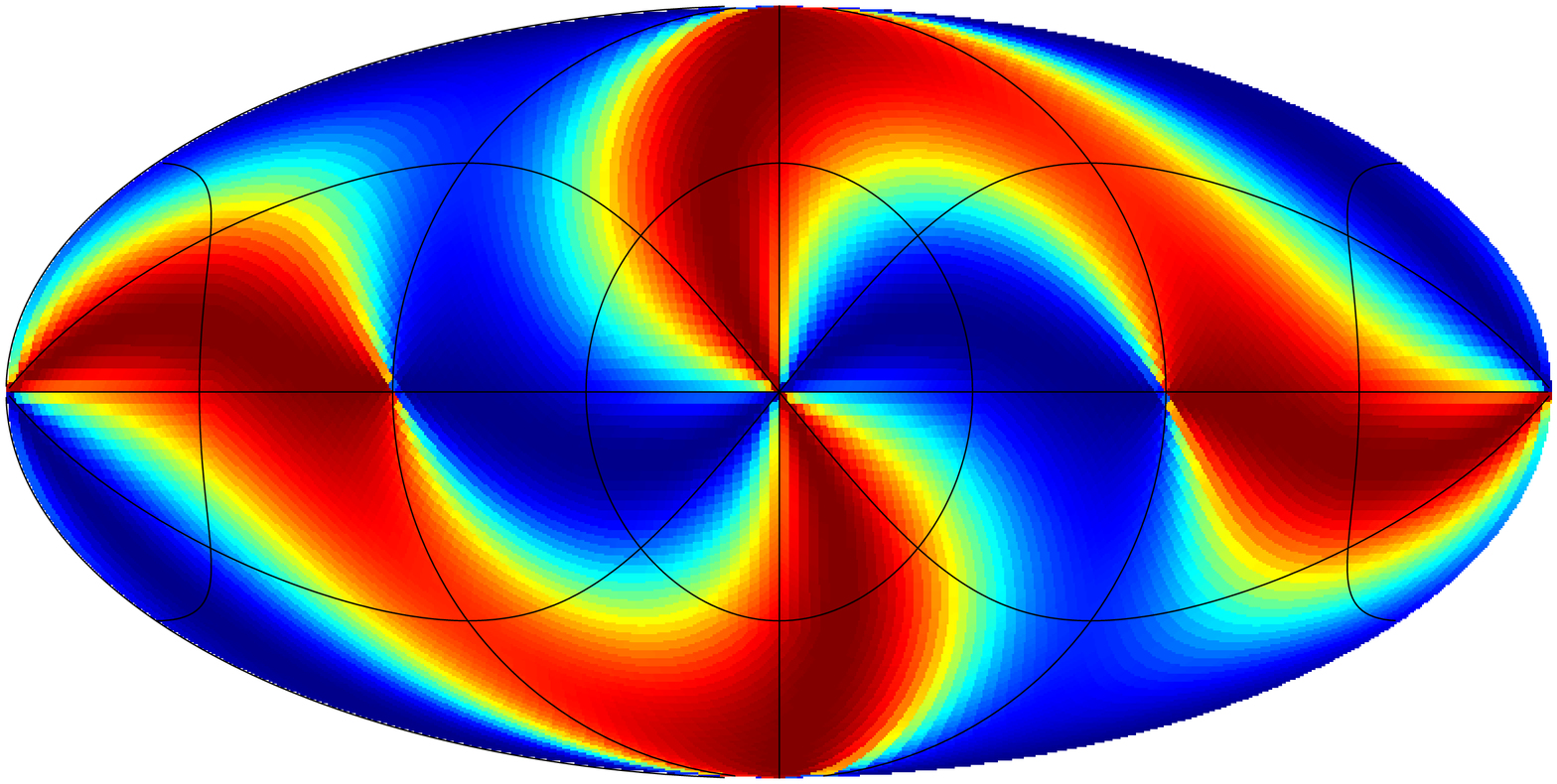}
   \includegraphics[scale=0.15,angle=90]{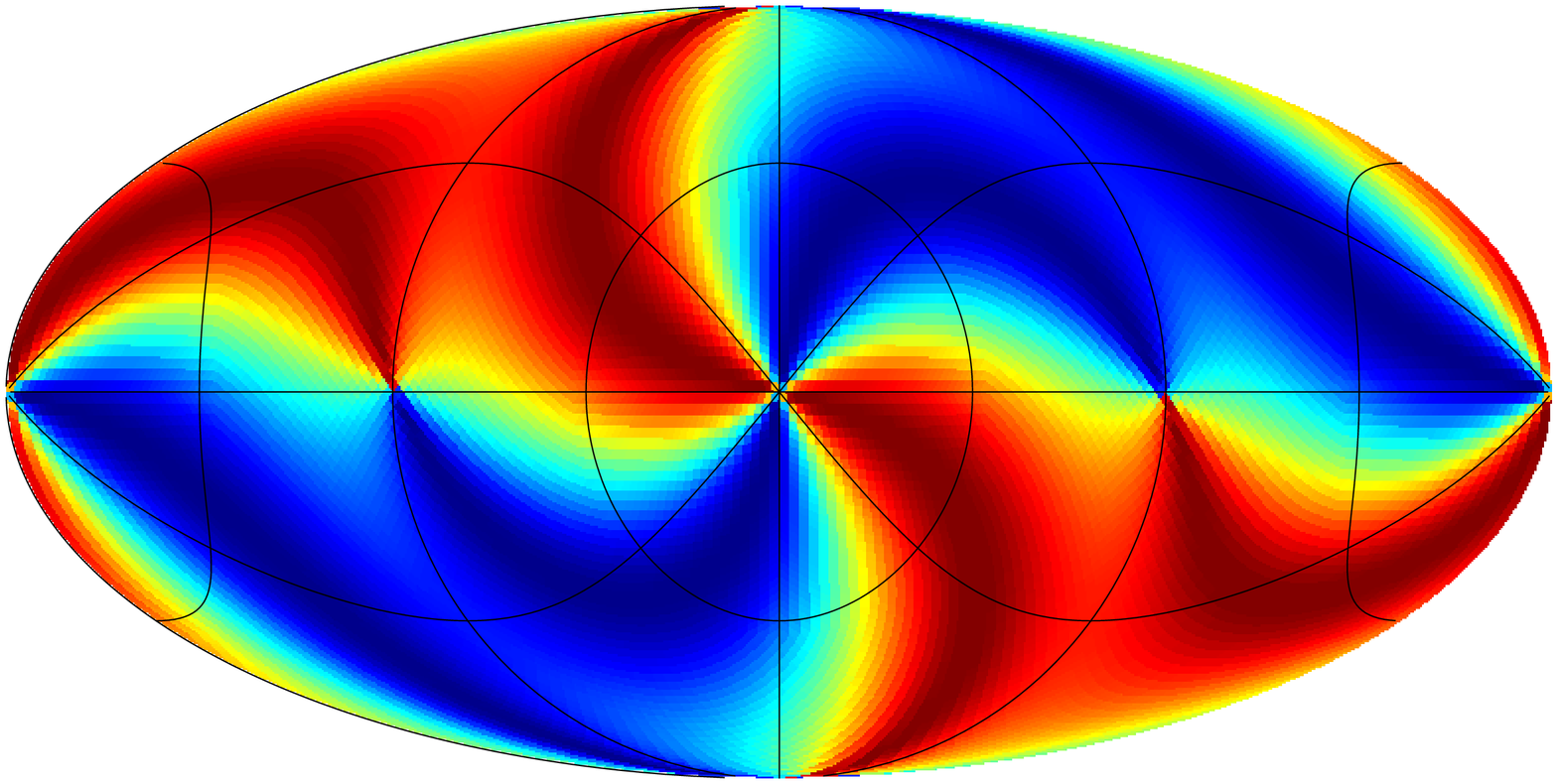}\\
  \includegraphics[scale=0.15,angle=90]{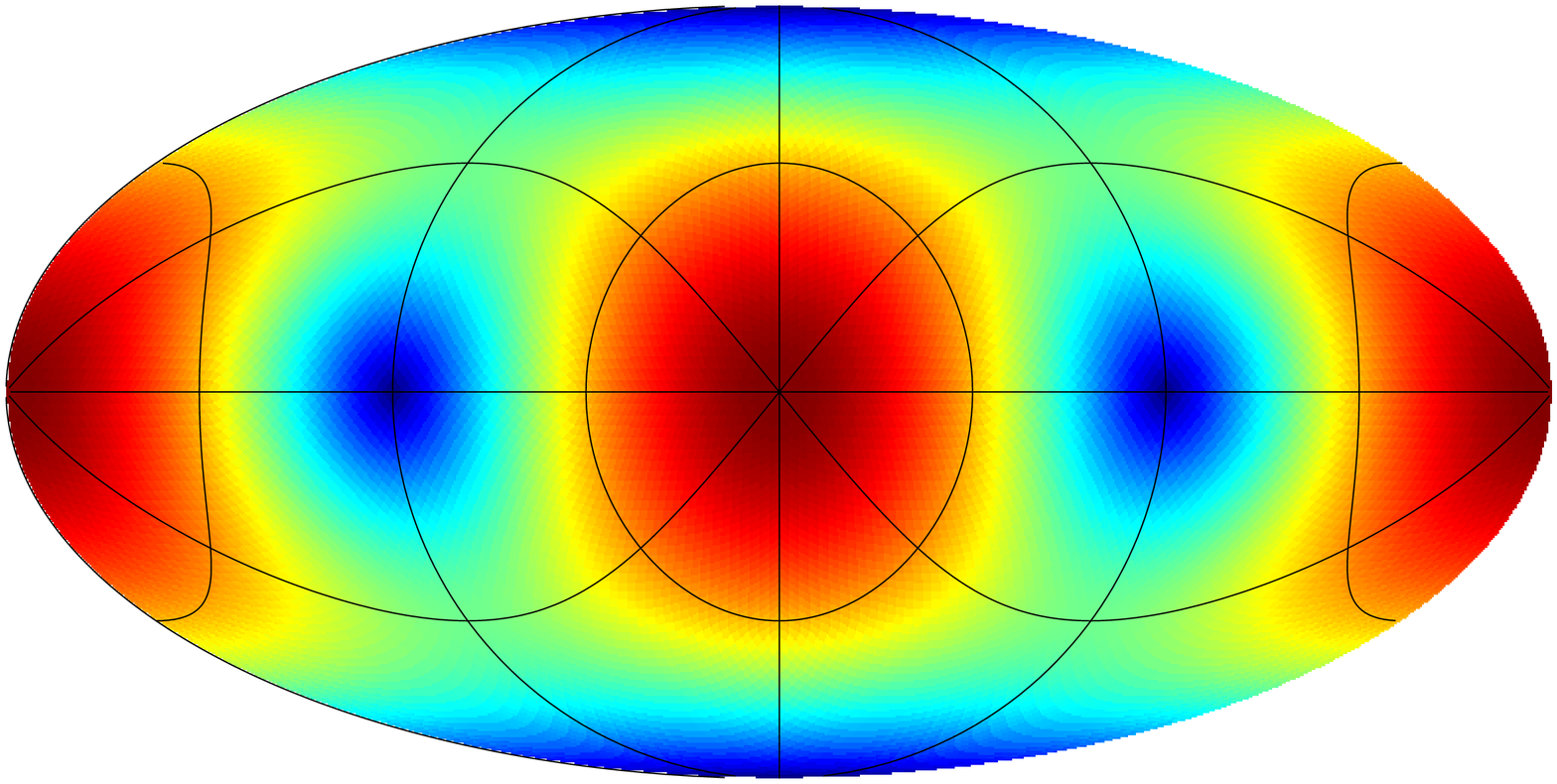}
    \includegraphics[scale=0.15,angle=90]{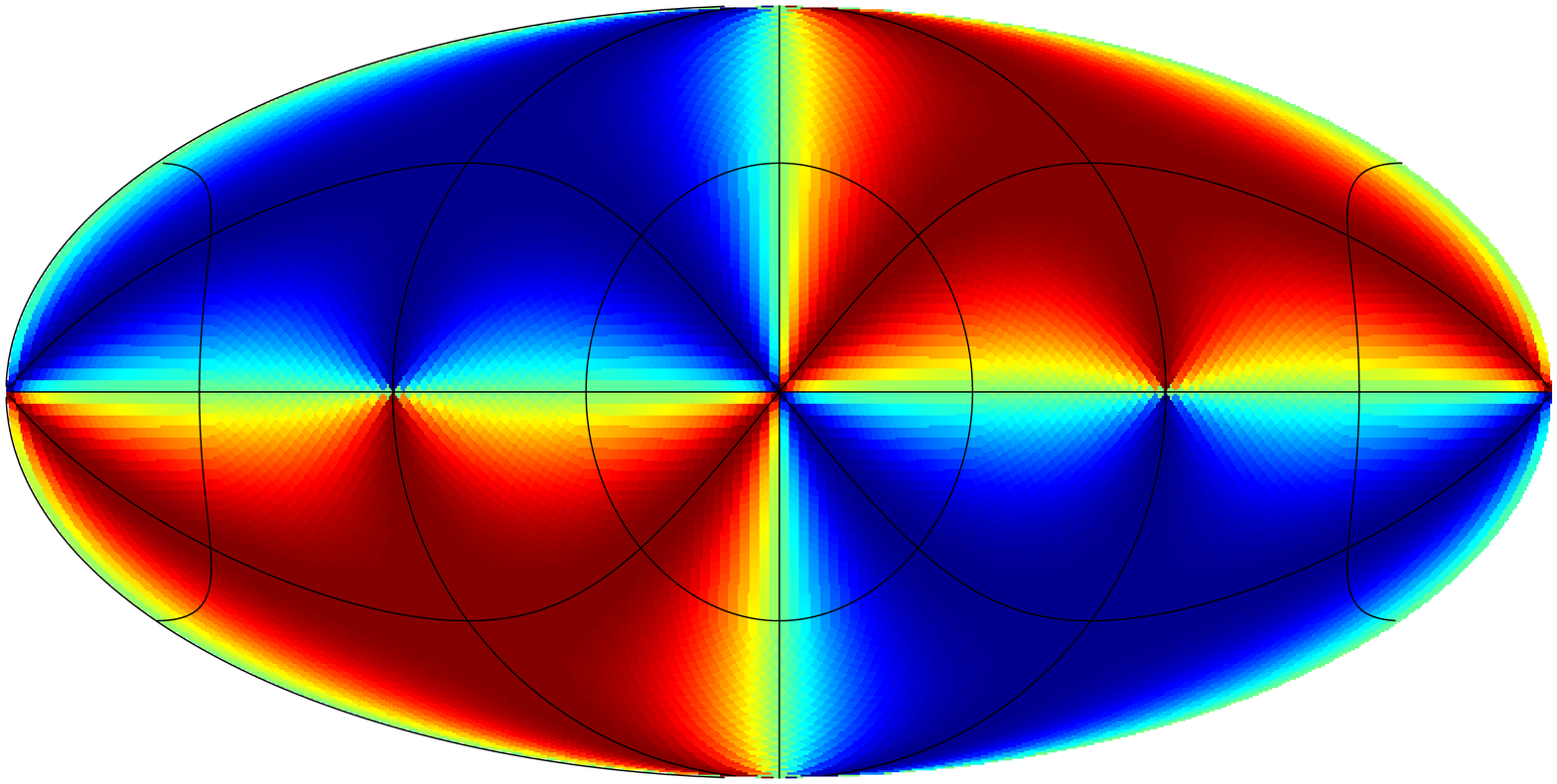}
   \includegraphics[scale=0.15,angle=90]{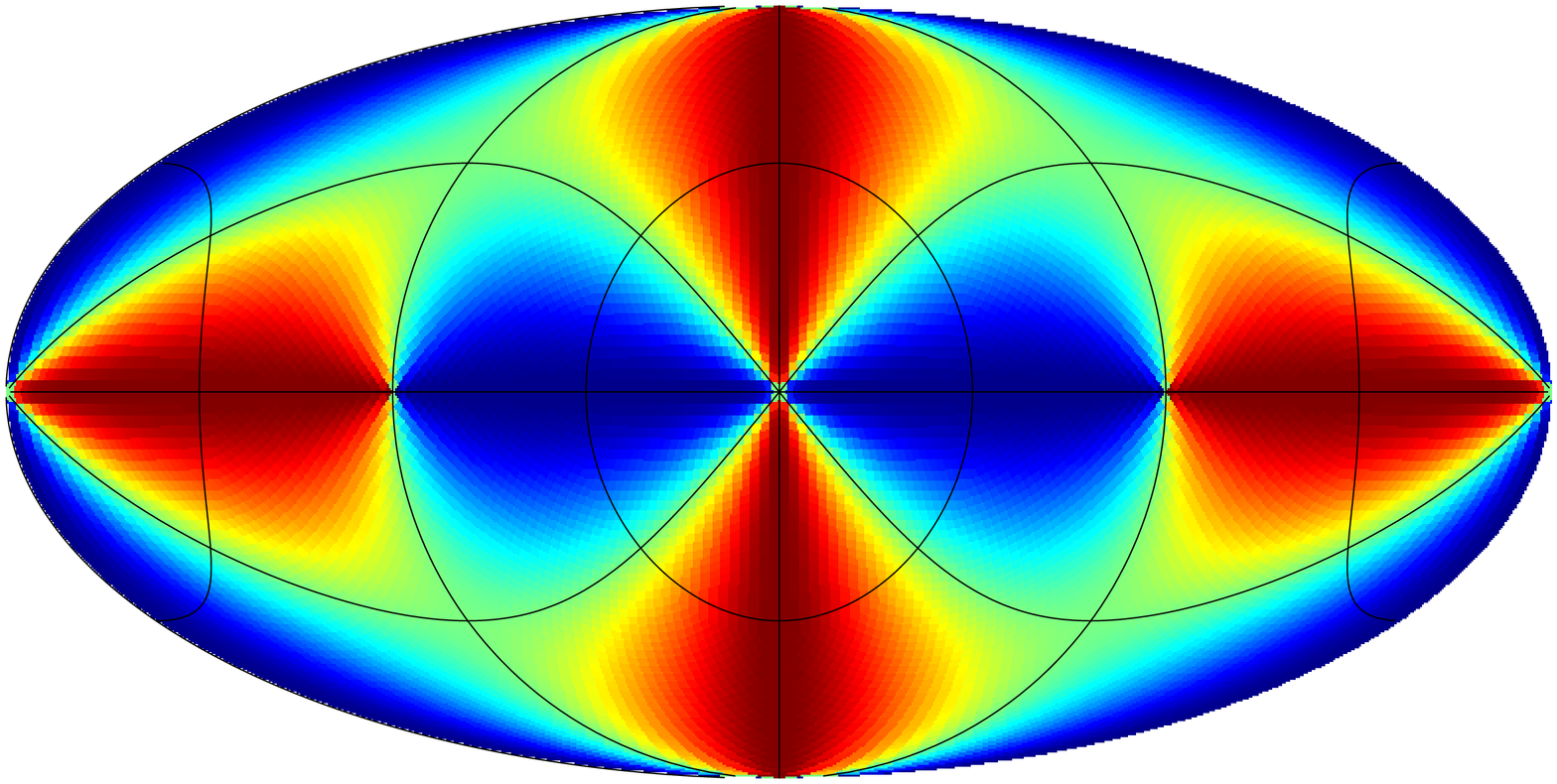}\\
 \end{center}
 \caption{Same as Figure \ref{PVz}, but for  Bianchi type VII$_0$.}
\end{figure}

Finally, for completeness we show results for Bianchi Type IX in
Figure 7. This provides an interesting example of polarization
behaviour because, in terms of polarization degree, the pattern does
change at all with time but the Stokes maps $Q$ and $U$ do evolve.
Bianchi Type IX models have positively curved spatial sections, and
are equivalent in some sense to FRW models with the addition of
circularly-polarized gravitational waves. These cause a rotation of
the polarization angle but do not change the overall magnitude. The
positive curvature does not allow for any focussing effects so the
temperature pattern does not change with time either.
\begin{figure}
\begin{center}
  \includegraphics[scale=0.15,angle=90]{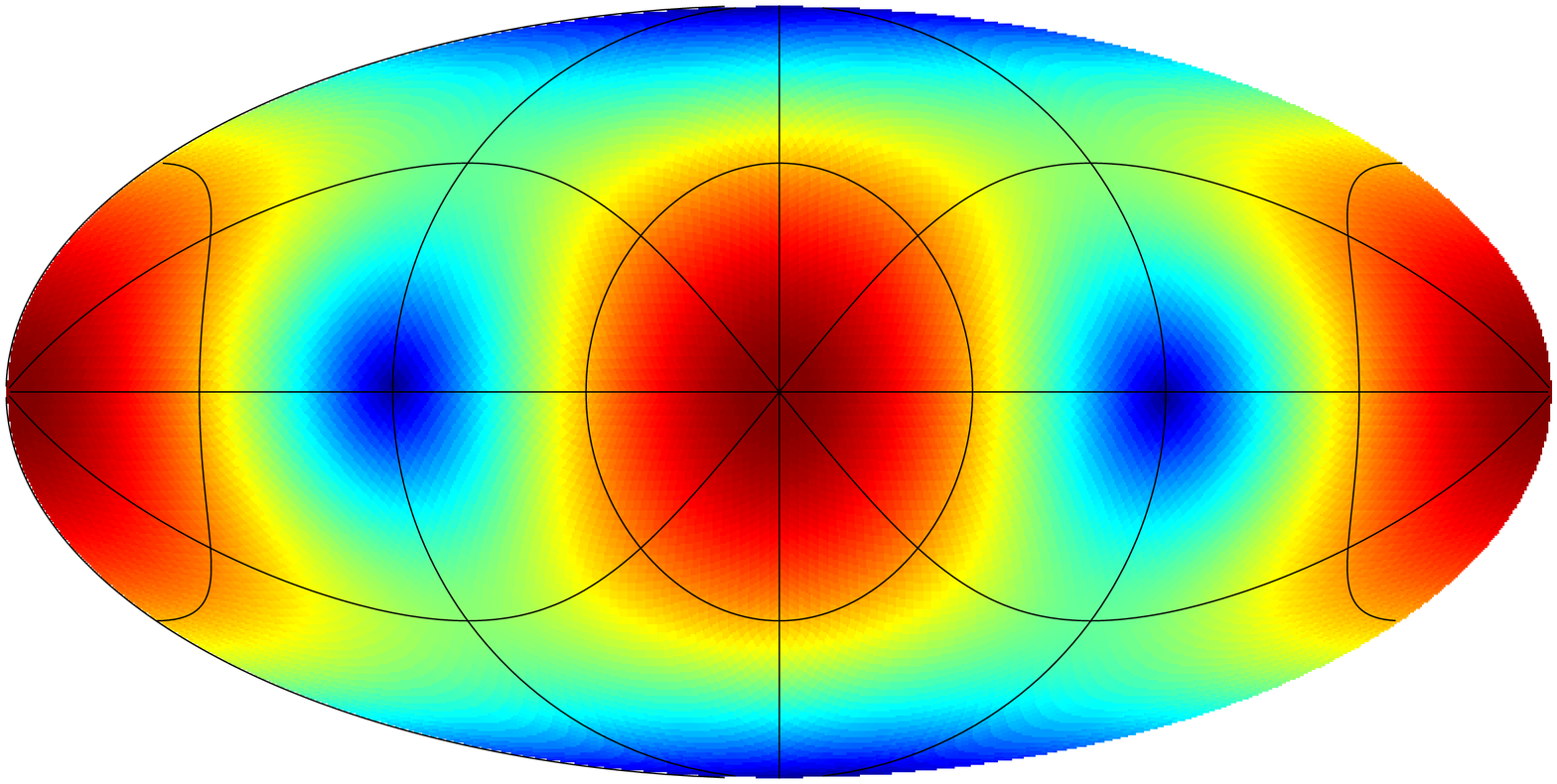}
  \includegraphics[scale=0.15,angle=90]{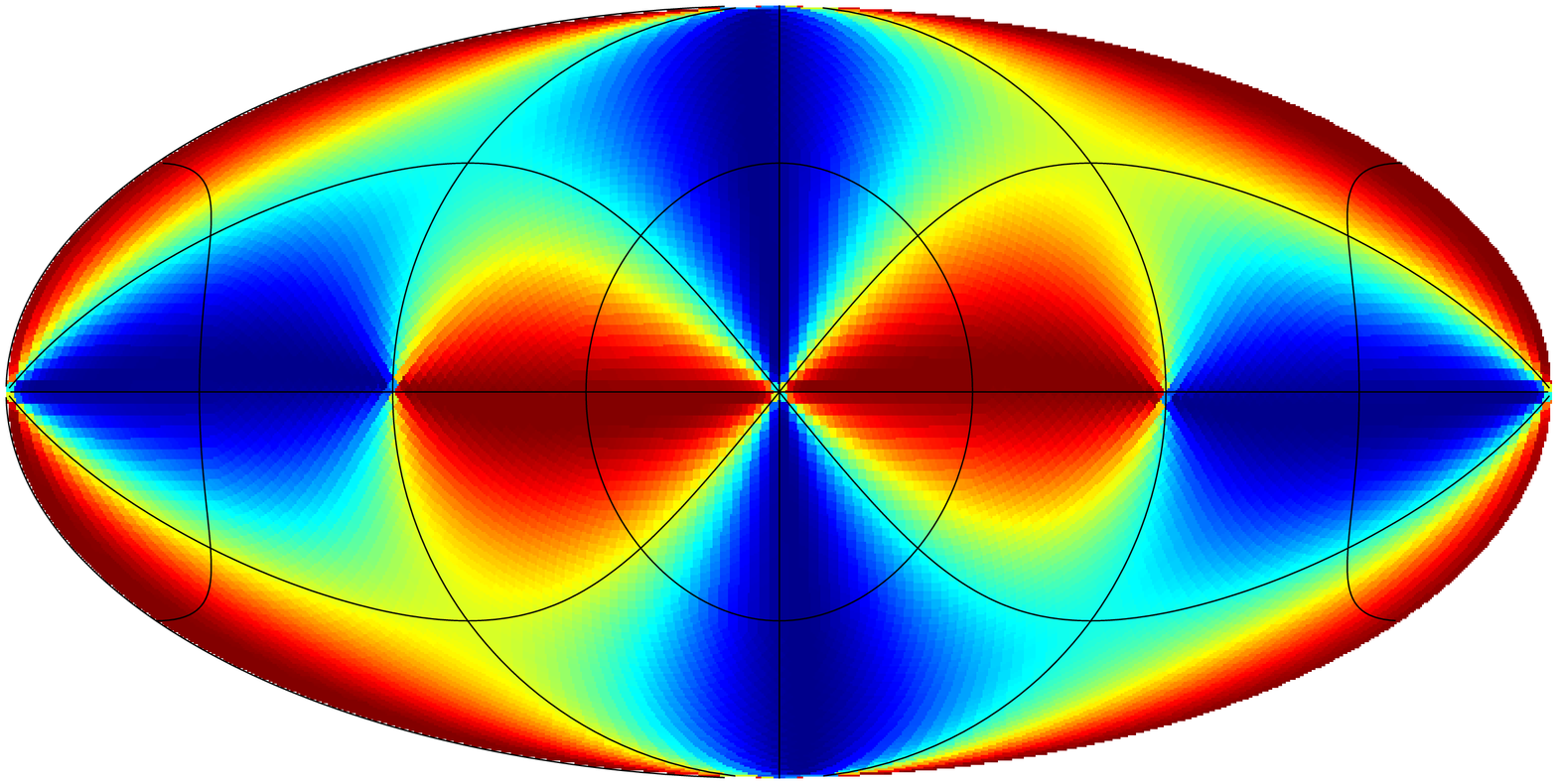}
   \includegraphics[scale=0.15,angle=90]{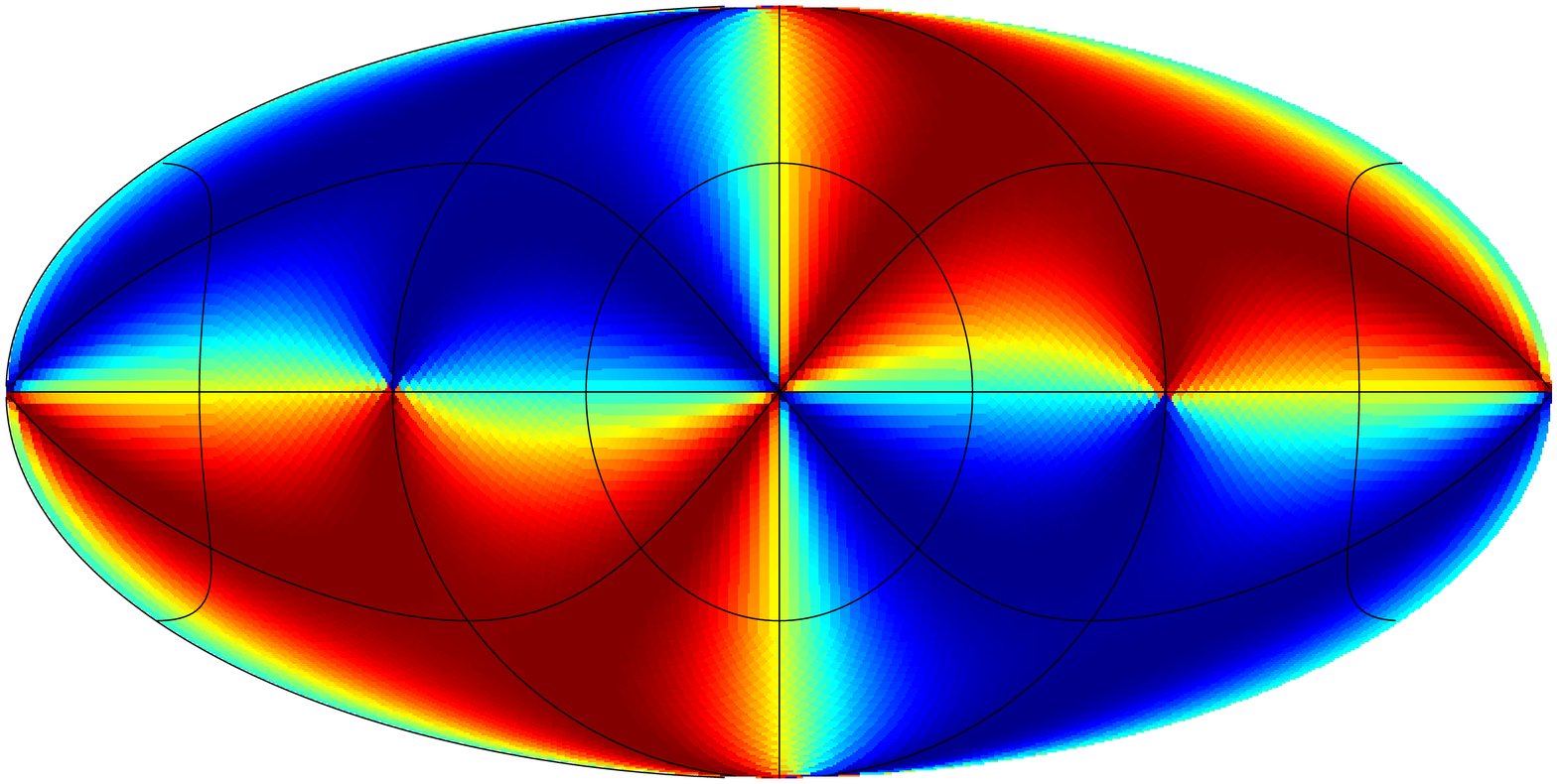}\\
  \includegraphics[scale=0.15,angle=90]{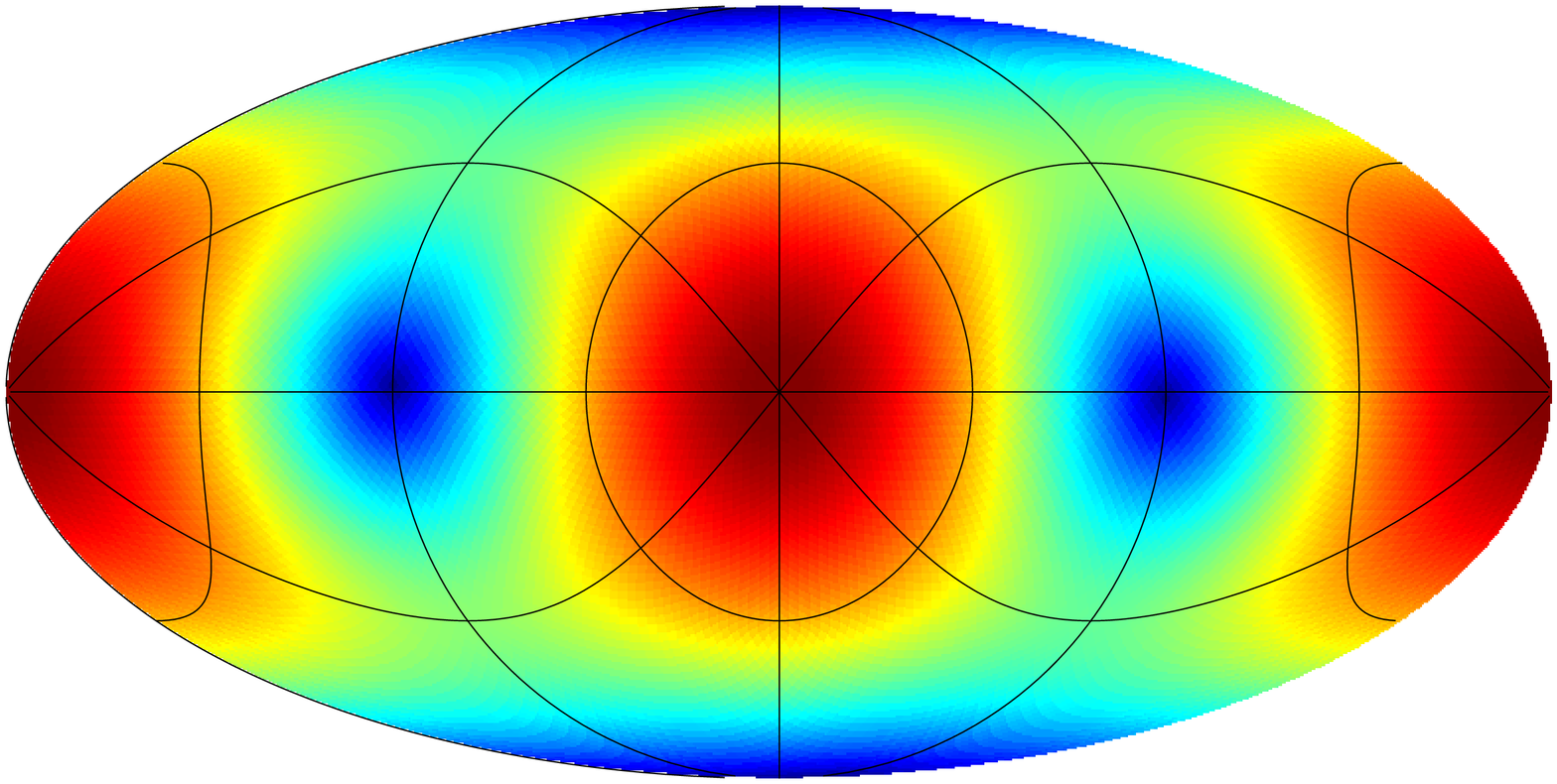}
    \includegraphics[scale=0.15,angle=90]{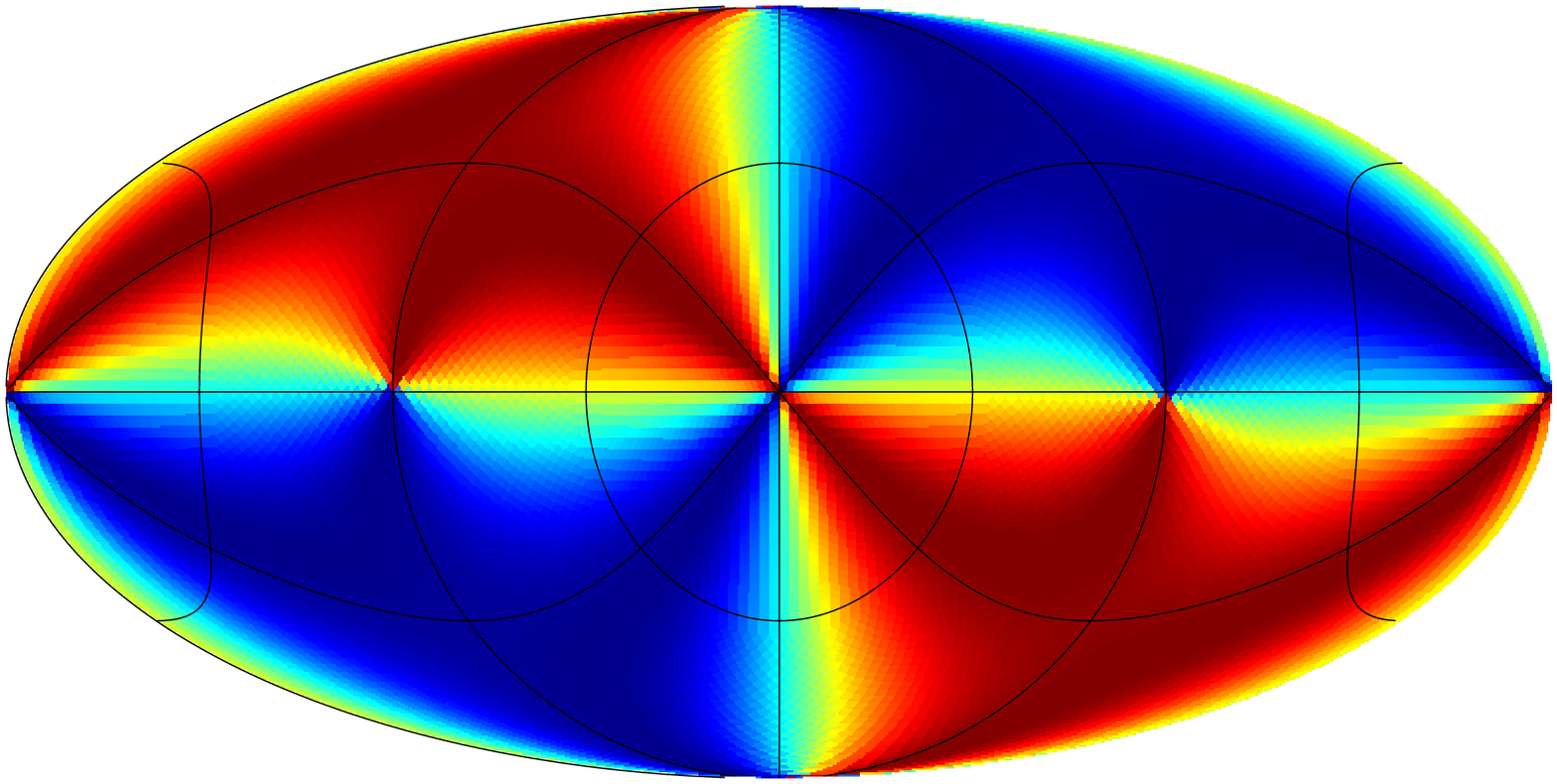}
   \includegraphics[scale=0.15,angle=90]{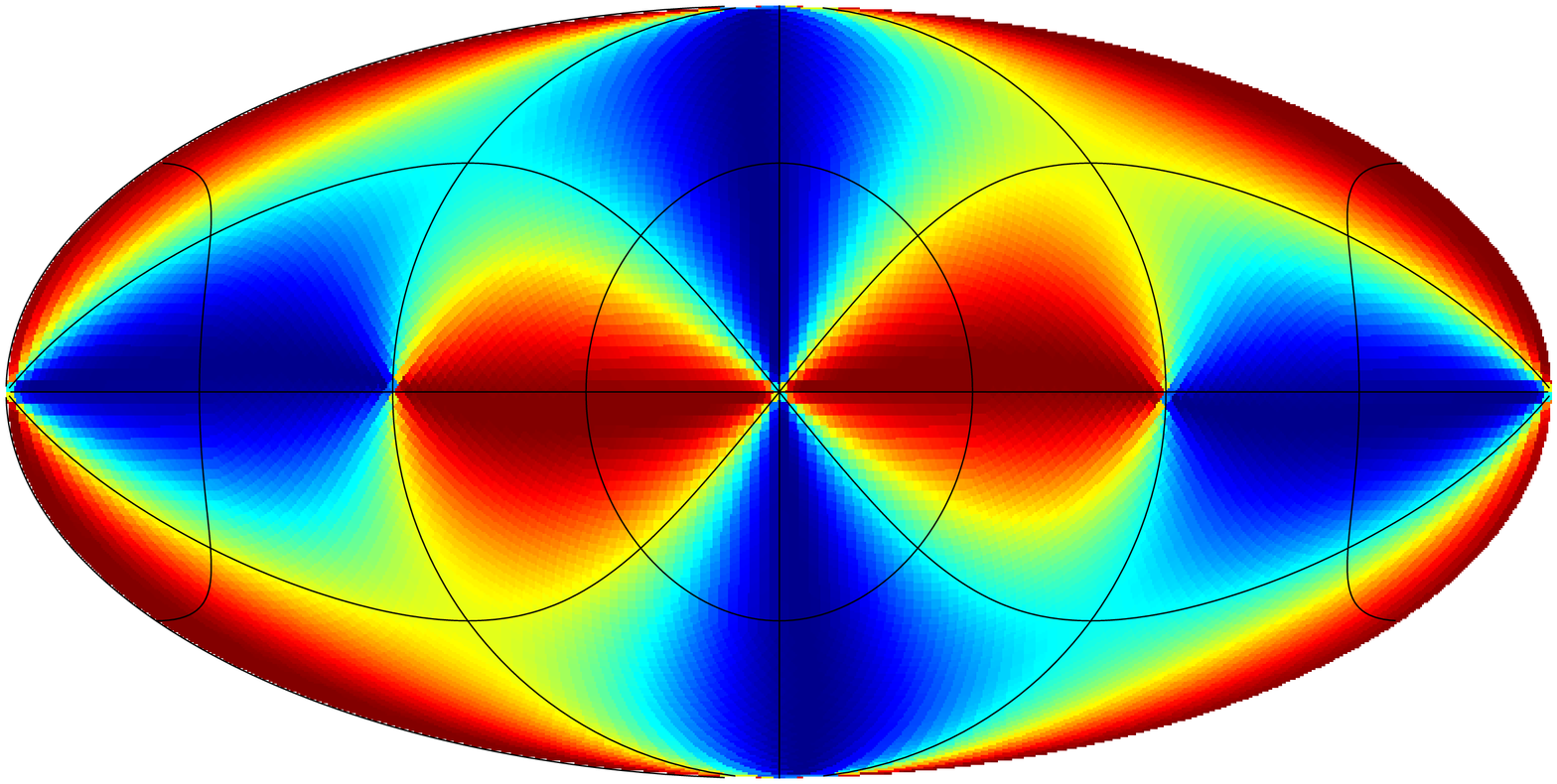}\\
  \includegraphics[scale=0.15,angle=90]{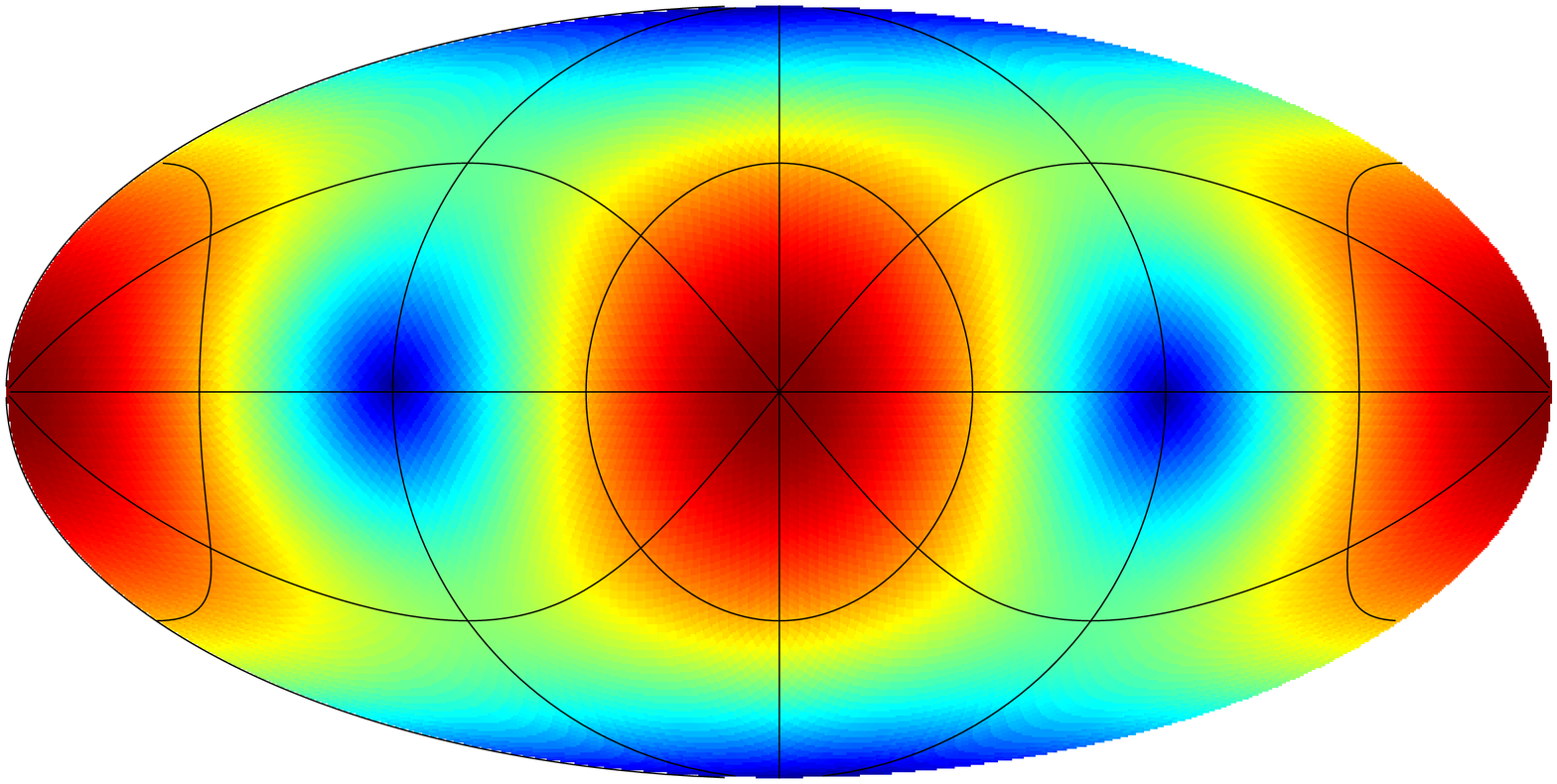}
    \includegraphics[scale=0.15,angle=90]{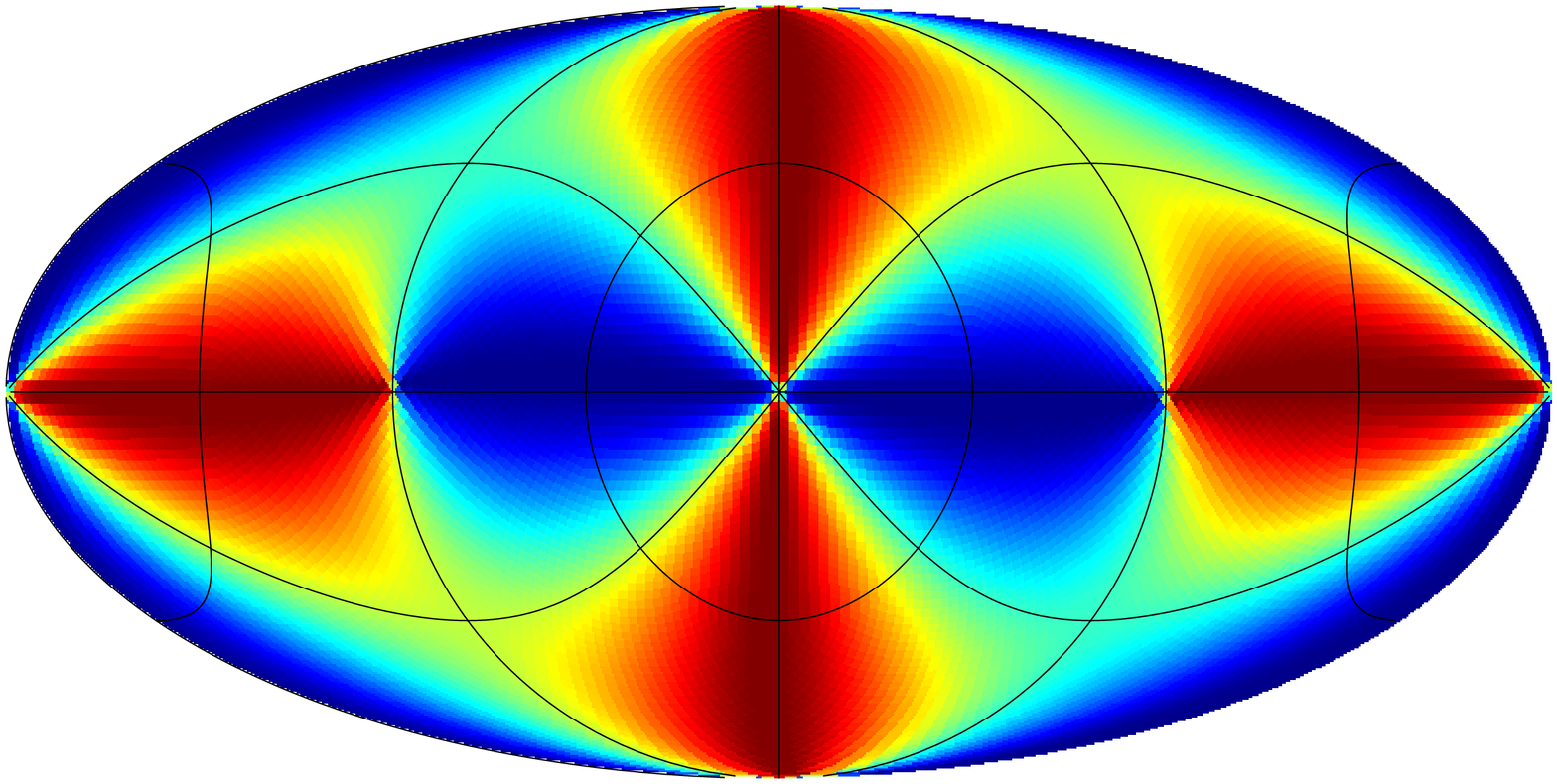}
   \includegraphics[scale=0.15,angle=90]{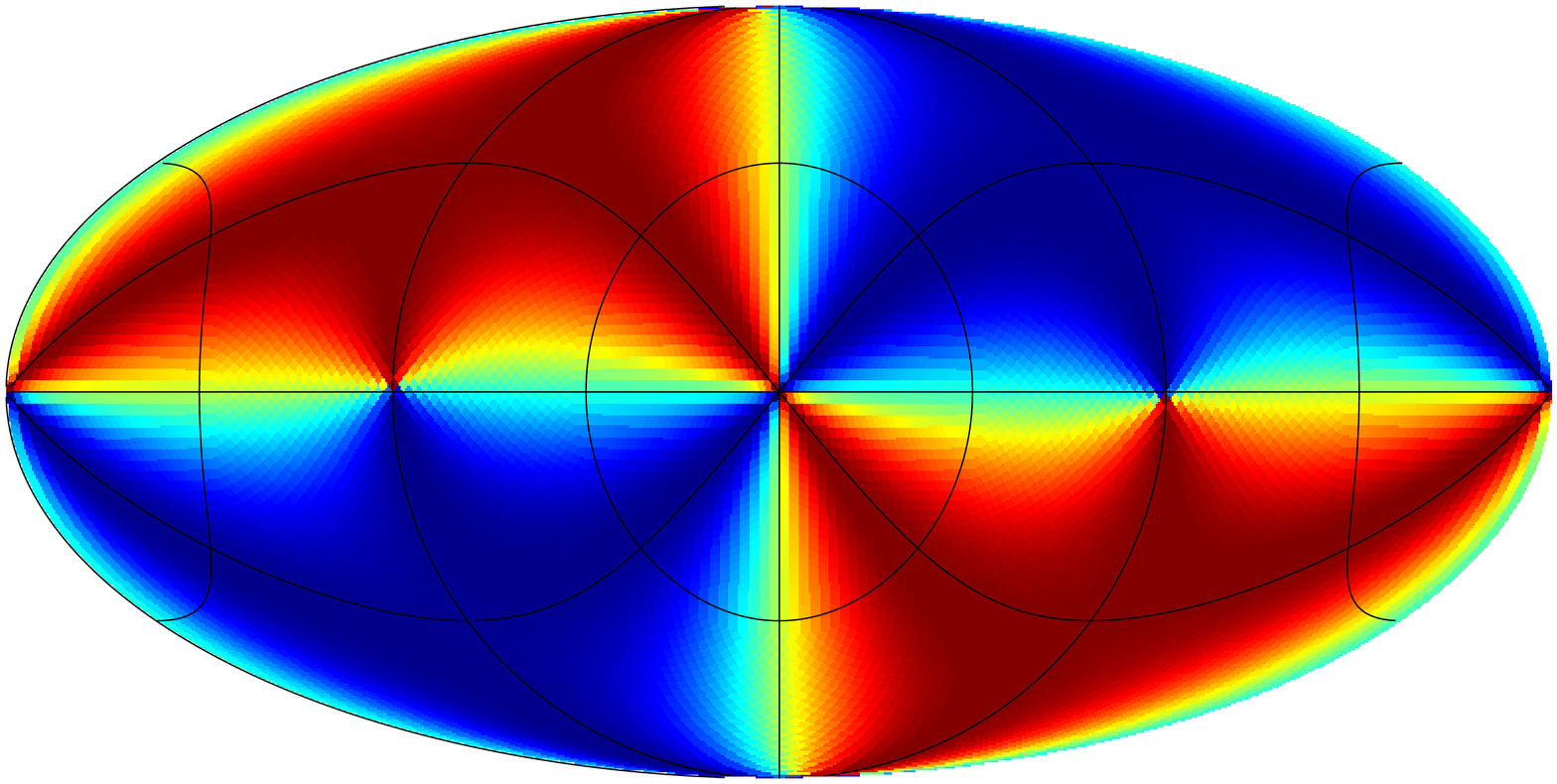}\\
  \includegraphics[scale=0.15,angle=90]{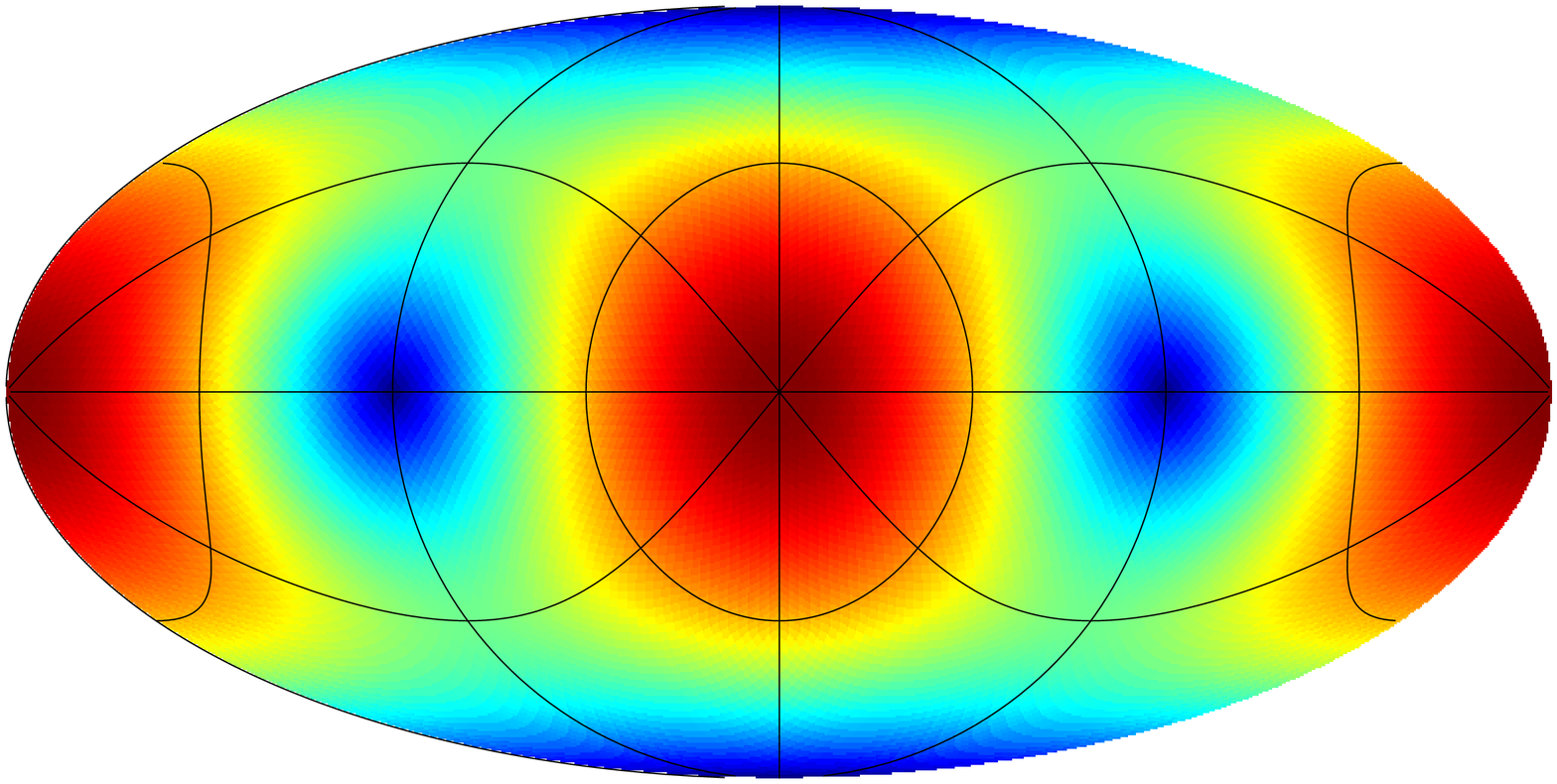}
    \includegraphics[scale=0.15,angle=90]{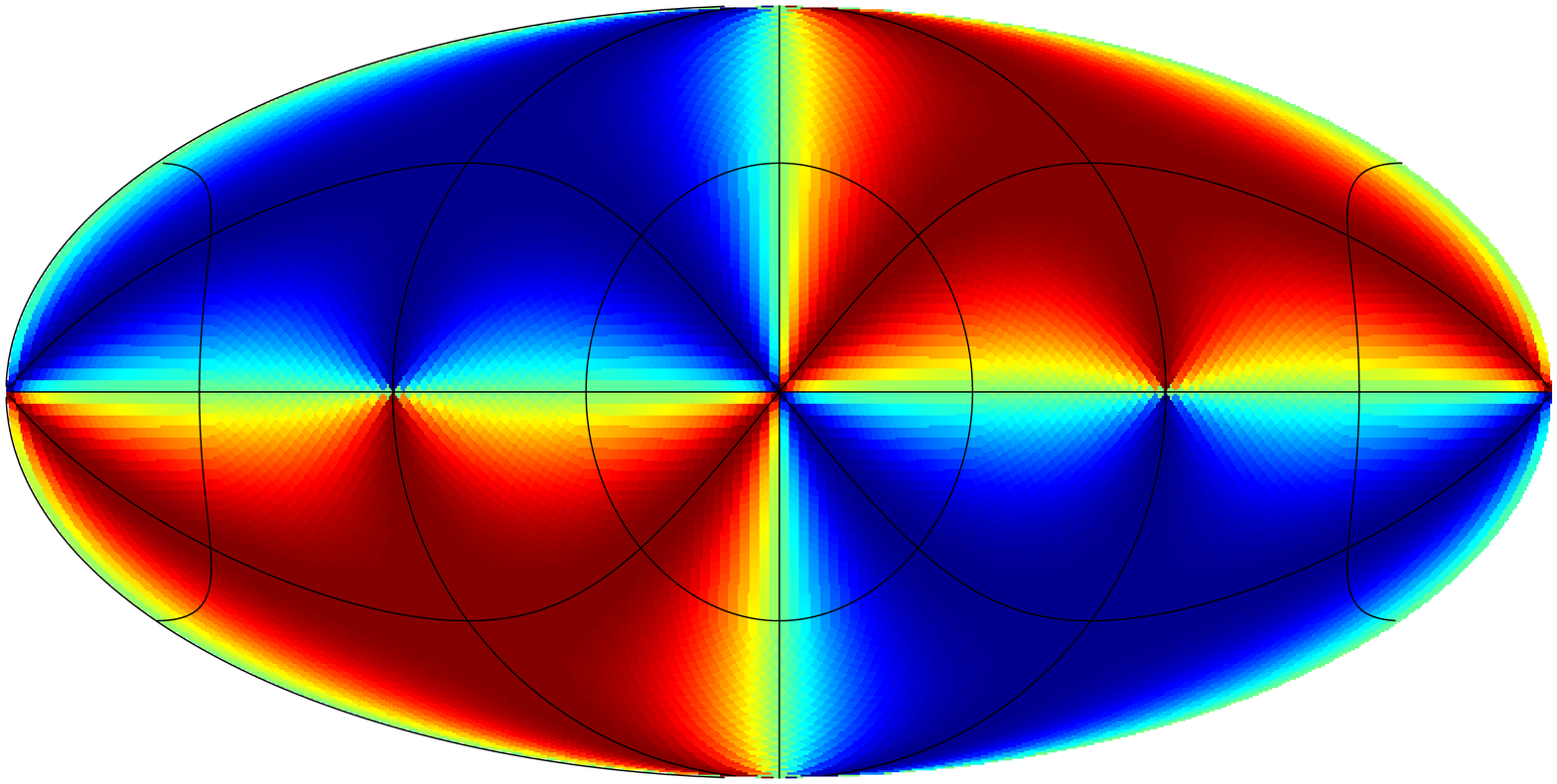}
   \includegraphics[scale=0.15,angle=90]{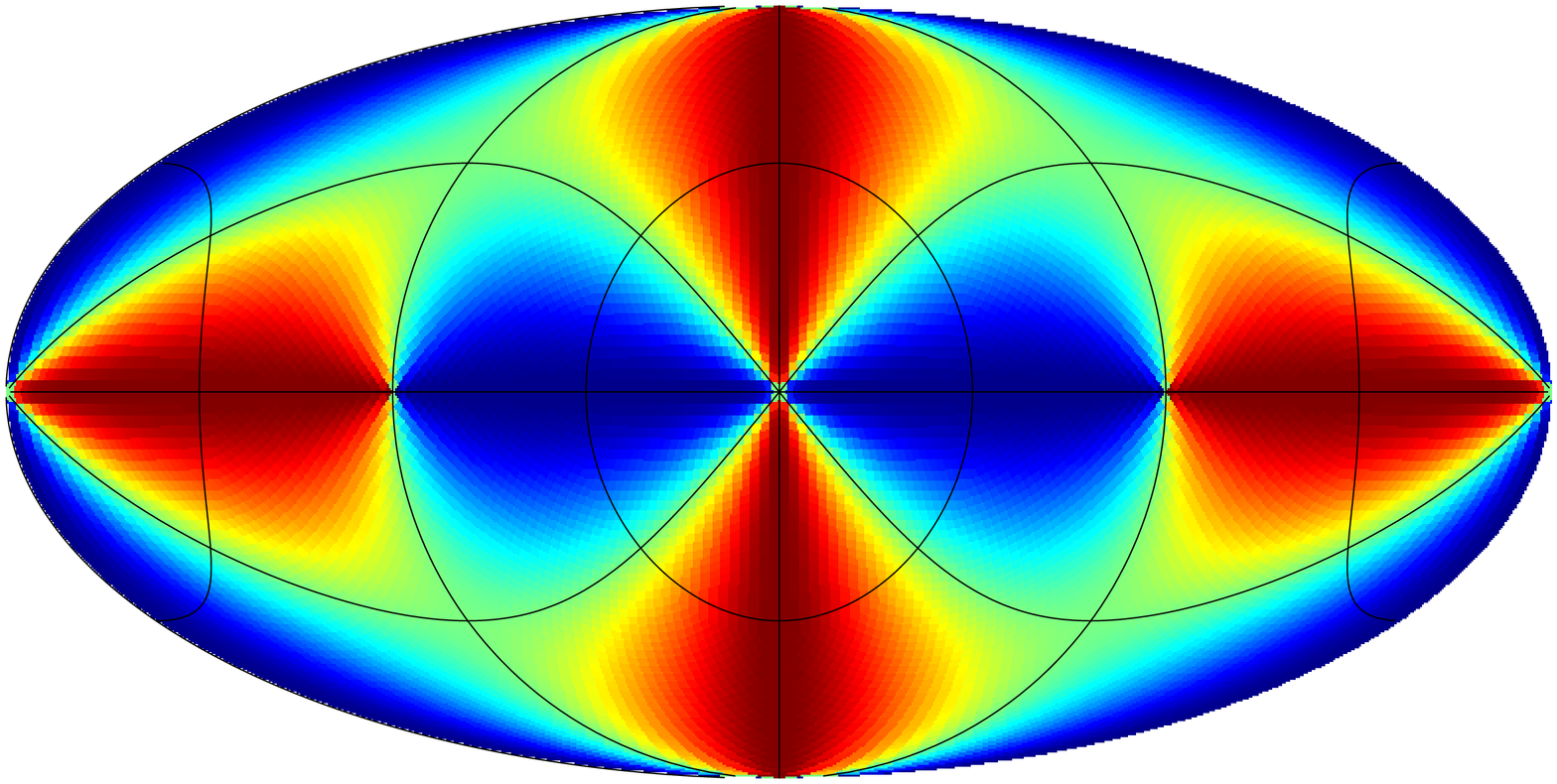}\\
 \caption{Same as Figure \ref{PVz}, but for Bianchi type IX.}
 \end{center}
\end{figure}
\section{Discussion and Conclusions}

The goal of this work was to compute the temperature and
polarization patterns produced in anisotropic relativistic
cosmologies described by various Bianchi types. We focussed on those
types that contain the standard homogeneous and isotropic FRW
background as a limiting case. We constructed an appropriate
description of the radiation field in terms of spin-0 and spin-2
components representing the unpolarized and polarized parts,
respectively. We integrated the equations for the radiation field
numerically, and presented illustrative examples here.

The basic point that emerges from this study is related to the
physical origin of CMB polarization: radiation affected by Thomson
scattering from an electron in a radiation field possessing a
quadrupole anisotropy will inevitably be partially linearly
polarized. In the context of standard cosmological models, the
environment of different electrons varies owing to the presence of
density inhomogeneities and a background of gravitational waves.
These sources of variation are stochastic so the variations in the
polarized component of the radiation field, though correlated with
the temperature variations, are essentially incoherent. In Bianchi
cosmologies, however, global homogeneity requires that each electron
sees {\em the same} quadrupole. The polarized part of the radiation
field is therefore coherent, and is in a fixed relationship to the
temperature variation (once the model is specified).

We have exploited the formalism used to generate the examples shown
here to produce an extensive test-bed of CMB temperature and
polarization maps derived from Bianchi universe models of the types
discussed in this paper. In future work we will analyse these maps
using a variety of statistical measures of anisotropy. It is
important to understand how standard statistical techniques,
designed to be applied to stationary stochastic fluctuations,
perform when applied to patterns which are neither stochastic nor
stationary. Our Bianchi models, for example, yield patterns which
have highly correlated spherical harmonic components in contrast to
the case of a homogeneous Gaussian random field in which the
harmonic modes are independent. It is important to understand how
CMB fluctuations arising from Bianchi-type universes impact on
orthodox analysis procedures and whether they produce characteristic
signatures when analysed in this way, particularly in the presence
of additional stochastic fluctuations. Higher-order statistics will
be necessary to provide a full characterization of the coherent
anisotropic fluctuations we have studied here. Moreover, present and
future CMB experiments are largely aimed at improving the precision
of polarization measurements in order to find evidence of a
stochastic background of primordial gravitational waves. Techniques
advocated to measure polarization pattern over the celestial sphere
are therefore also generally tuned to detect signals of an
incoherent nature. The presence of coherent signals in the CMB could
be an indication of physics beyond the standard cosmological
framework. This will hopefully lead us to better ways of searching
for departures from the concordance cosmology using the next
generation of datasets.

\section*{Acknowledgments}

We thank Jason McEwen, Sasha Polnarev, Andrew Pontzen, Antony
Challinor and Leonid Grishchuk for interesting and helpful
discussions. Rockhee Sung acknowledges an Overseas Scholarship from
the Korean government. Two recent papers \cite{Pontzen1,pontzen}
have also addressed the questions we discuss here, but we have used
a different formalism. Our work can provides independent
confirmation of their results.


\section*{References}

\begin{thebibliography}{99}
\bibitem{WMAP1} Bennett C L et al 2003 {\it Astrophys. J. Supp.} {\bf 148} 1
\bibitem{WMAP2} Hinshaw G et al 2009 {\it Astrophys. J. Supp.}
{\bf 180} 225
\bibitem{Coles} Coles P 2005 {\it Nature} {\bf 433} 248
\bibitem{Guth1982} Guth A H and Pi S Y 1982 {\it Phys. Rev. Lett.} {\bf 49} 1110
\bibitem{Starobinskij1982} Starobinskij A A 1982 {\it Phys. Lett. B.}, {\bf 117} 175
\bibitem{Bardeen1983} Bardeen J M, Steinhardt P J and  Turner M S 1983 {\it Phys. Rev. D.} {\bf 28} 679
\bibitem{Yadav2008} Yadav A P S and  Wandelt B D 2008 {\it Phys. Rev. Lett.} {\bf 100} 181301
\bibitem{Vielva} Vielva P, Martinez-Gonzalez E, Barreiro R B, Sanz J
L and Cayon L 2004 {\it Astrophys. J.} {\bf 609} 22
\bibitem{Cruz1} Cruz M, Martinez-Gonzalez E, Vielva P and Cayon L
2005 {\it Mon. Not. R. astr. Soc.} {\bf 356} 29
\bibitem{Cruz2} Cruz M, Tucci M, Martinez-Gonzalez E and Vielva P
2006 {\it Mon. Not. R. astr. Soc.} {\bf 369} 57
\bibitem{Cruz3} Cruz M, Cayon L, Martinez-Gonzalez E, Vielva P and
Jin J 2007 {\it Astrophys. J.} {\bf 655}, 11
\bibitem{Cruz4} Cruz M, Martinez-Gonzalez E, Vielva P,  Diego J M, Hobson M and Turok
N 2008 arXiv:0804.2904
\bibitem{Cayon} Cayon L, Jin J and Treaster A 2005 {\it Mon. Not. R.
astr. Soc.} {\bf 362} 826
\bibitem{Cspot} Naselsky P D, Christensen P R, Coles P, Verkhodanov
O, Novikov D and Kim J 2009, arXiv/0712.1118
\bibitem{Schwarz2004} Schwarz D J, Starkman G D, Huterer D and Copi C J 2004 {\it Phys. Rev. Lett.} {\bf 93} 221301
\bibitem{Copi2004} Copi C J, Huterer D and  Starkman G D 2004 {\it Phys. Rev. D.} {\bf 70} 043515
\bibitem{Katz2004} Katz G and  Weeks J 2004  {\it Phys. Rev. D.} {\bf 70} 063527
\bibitem{Land2005a} Land K and Magueijo J 2005  {\it Mon. Not. R. astr. Soc.} {\bf 357} 994
\bibitem{Land2005b} Land K and Magueijo J 2005 {\it Mon. Not. R. astr. Soc.} {\bf 362} L16
\bibitem{Land2005c} Land K and Magueijo J 2005 {\it Phys. Rev. D.} {\bf 72} 101302(R)
\bibitem{Land2005d} Land K and Magueijo J 2005  {\it Phys. Rev. Lett.}  {\bf 95} 071301
\bibitem{Land2005e} Land K and Magueijo J 2005  {\it Mon. Not. R. astr. Soc.} {\bf 362} 838
\bibitem{Land2007} Land K and Magueijo J 2007 {\it Mon. Not. R. astr. Soc.} {\bf  378} 153
\bibitem{Copi2006} Copi C J, Huterer D, Schwarz D J and  Starkman  G D 2006 {\it Mon. Not. R. astr. Soc.} {\bf 367} 79
\bibitem{Copi2007} Copi C J, Huterer D, Schwarz D J and  Starkman  G D 2007 {\it Phys. Rev. D.} {\bf 75} 023507
\bibitem{Eriksen2004a} Eriksen H K, Hansen F K,  Banday A J,  G\'{o}rski K M and Lilje P B 2004 {\it Astrophys. J.} {\bf 605} 14
\bibitem{Park2004} Park C 2004 {\it Mon. Not. R. astr. Soc.} {\bf 349} 313
\bibitem{Eriksen2007} Eriksen H K, Banday A J, G\'{o}rski K M, Hansen F K and Lilje P B 2007, {\it Astrophys. J.} {\bf 660} L81
\bibitem{Hoftuft2009} Hoftuft J,  Eriksen H K,  Banday A J, G\'{o}rski K M, Hansen F K and Lilje P B 2009
{\it Astrophys. J. Suppl.} {\bf 699} 985
\bibitem{Hansen2009} Hansen F K, Banday A J, G\'{o}rski K M, Eriksen H K and  Lilje P B 2009
{\it Astrophys. J.} {\bf 704} 1448
\bibitem{HL09} Hanson D and Lewis A 2009, arXiv: 0908.0963
\bibitem{GAWE} Groeneboom N E, Ackerman L, Wehus I K and Eriksen H K
2009, arXiv: 0911.0150
\bibitem{HL10} Hanson D, Lewis A and Challinor A 2010, arXiv:
1003.0198
\bibitem{zb10} Zheng H and Bunn E F 2010, arXiv: 1003.5548
\bibitem{ccno07} Chiang L-Y, Coles P, Naselsky P D and Olesen P 2007
{\it J. Cosmol. Astropart. Phys.} 01(2007)021
\bibitem{cnc07} Chiang L-Y, Naselsky P D and Coles P 2007 {\it
Astrophys. J.} {\bf 664} 8
\bibitem{FP09} Francis C L and Peacock J A 2009, arXiv: 0909.2495
\bibitem{sc10} Short J and Coles P 2010 {\it Mon. Not. R. astr.
Soc.} {\bf 401} 2202
\bibitem{Gris} Grishchuk L P, Doroshkevich A G and Novikov I D 1968
Soviet Physics ZETP {\bf 55} 2281
\bibitem{Ellis1} Ellis G F R  and MacCallum M A H 1969
{\it Commun. Math. Phys.} {\bf 12} 108
\bibitem{Collins1} Collins C B and Hawking S W 1973 {\it
Mon. Not. R. astr. Soc.} {\bf 162} 307
\bibitem{Dautcourt1} Dautcourt G and Rose K 1978 {\it Astr. Nachr.}
{\bf 299} 13
\bibitem{Tolman1} Tolman B W and Matzner R A 1984 {\it Proc. R. Soc. Lond.} {\bf A} {\bf 392} 391
\bibitem{Tolman2} Matzner  R A  and Tolman B W 1982 {\it Phys. Rev. D.} {\bf 26} 10
\bibitem{Tolman3} Tolman B W 1985 {\bf 290} 1
\bibitem{Barrow1} Barrow J D,  Juszkiewicz R
and Sonoda D H 1985  {\it Mon. Not. R. astr. Soc.} {\bf 213} 917
\bibitem{BFS96} Bunn E F, Ferreira P G and Silk J 1996 {\it Phys.
Rev. Lett.} {\bf 77} 2883
\bibitem{khb97} Kogut A, Hinshaw G and Banday A J 1997 {\it Phys.
Rev. D.} {\bf 55} 1901
\bibitem{Jaffe1} Jaffe T R, Banday A J, Eriksen H K, G\'{o}rski
K M  and Hansen F K 2005 {\it Astrophys. J.} {\bf 629} L1
\bibitem{Jaffe2} Jaffe T R, Hervik S, Banday A J and G\'{o}rski
K M 2006 {\it Astrophys. J.} {\bf 644} 701
\bibitem{McEwan1} McEwen J D, Hobson M P, Lasenby A N and Mortlock D
J 2005 {\it Mon. Not. R. astr. Soc.} {\bf 369} 1583
\bibitem{McEwan2} McEwen J D, Hobson M P, Lasenby A N and Mortlock D
J 2006 {\it Mon. Not. R. astr. Soc.} {\bf 371} L50
\bibitem{Bridge} Bridges M, McEwen J D, Lasenby A N and Hobson M P
2007 {\it Mon. Not. R. astr. Soc.} {\bf 377} 1473
\bibitem{sung0} Sung R 2010 PhD thesis, Cardiff University
\bibitem{sung1} Sung R and Coles P 2009 {\it Class. Quantum Grav.}
{\bf 26} 172001
\bibitem{Pontzen1} Pontzen A and Challinor A 2007 {\it Mon. Not. R.
astr. Soc.} {\bf 380} 1387
\bibitem{pontzen} Pontzen A 2009 {\it Phys. Rev. D.} {\bf 79} 103518
\bibitem{WMAPPol} Page L et al 2007 {\it Astrophys. J. Supp.} {\bf
170} 335
\bibitem{Kam} Kamionkowski M, Kosowsky A and Stebbins A 1997 {\it
Phys. Rev. D.} {\bf 55} 7368
\bibitem{Hu1} Hu W and White M D 1997 {\it Phys. Rev. D.} {\bf 56}
596
\bibitem{FE10} Frommert M and En\ss lin T A 2010 {\it Mon. Not. R.
astr. Soc.} {\bf 403} 1739
\bibitem{rees} Rees M J 1968 {\it Astrophys. J.} {\bf 153} L1
\bibitem{nanos} Nanos G P 1979 {\it Astrophys. J.} {\bf 232} 341
\bibitem{NS80} Negroponte J and Silk J 1980 {\it Phys. Rev. Lett.}
{\bf 44} 1433
\bibitem{BP80} Basko M M and Polnarev A G 1980 {\it Sov. Astr.} {\bf
24} 268
\bibitem{Pol85} Polnarev A G 1985 {\it Sov. Astron.}
{\bf 29} 607
\bibitem{fpc94} Frewin R A, Polnarev A G and Coles P {\it Mon. Not.
R. astr. Soc.} {\bf 266} L21
\bibitem{bgen} Barrow J D 1986 in {\em Gravitation in Astrophysics}, eds Carter B and
Hartle J B, {\em Proceedings of NATO ASI Series B}, {\bf 156} 239
\bibitem{cl2} Coles P and Lucchin F 2002 {\it Cosmology: The Origin
and Evolution of Cosmic Structure}, 2nd Edition, John Wiley \& Sons
\bibitem{Ellis2} Ellis G F R 1967 {\it J. Math. Phys.} {\bf 8} 1171
\bibitem{Ellis3} MacCallum M A H  and  Ellis G F R  1970 {\it Commun. Math.
Phys.} {\bf 19} 31
\bibitem{tilt} King A R and Ellis G F R 1973 {\it Commun. Math. Phys.}
{\bf 31} 209
\bibitem{Ellis4} Ellis G F R  2006 {\it Gen. Rel. Grav.} {\bf 38} 1003
\bibitem{Kasner} Kasner E 1921 {\it Trans. Amer. Math. Soc.} {\bf 43} 217
\bibitem{healpix} G\'{o}rski, K M, Hivon E, Banday A J, Wandelt B D, Hansen F K, Reinecke E M and
Bartelmann M 2005 {\em Astrophys. J.} {\bf 622}, 759
\bibitem{BianchiI} Campanelli L, Cea P and Tedesco L 2006 {\it Phys.
Rev. Lett.} {\bf 97} 209903
\bibitem{BianchiIb} Campanelli L, Cea P and Tedesco L 2007 {\it Phys.
Rev. D.} {\bf 76} 063007






\end{thebibliography}
\end{document}